\documentclass[%
 reprint,
 superscriptaddress,
 amsmath,amssymb,
 aps,
 prc,
]{revtex4-2}

\usepackage{graphicx}% Include figure files
\usepackage{dcolumn}% Align table columns on decimal point
\usepackage{bm}% bold math
\usepackage{hyperref}% add hypertext capabilities
\usepackage{comment}
\usepackage{footnote}
\usepackage{multirow}
\begin{document}

%\preprint{APS/123-QED}

\title{Measurement of production branching ratio \\after muon nuclear capture reaction of Al and Si isotopes}% Force line breaks with \\

\author{R.~Mizuno}
    %\email{mizuno@nex.phys.s.u-tokyo.ac.jp}
    \email{rmizuno@triumf.ca}
 \affiliation{Department of Physics, Graduate School of Science, The University of Tokyo, Bunkyo, Tokyo 113-0033, Japan.}
\author{M.~Niikura}
\affiliation{RIKEN Nishina Center, RIKEN, Wako, Saitama 351-0198, Japan.}
 \affiliation{Department of Physics, Graduate School of Science, The University of Tokyo, Bunkyo, Tokyo 113-0033, Japan.}
\author{T.Y.~Saito}
\affiliation{Atomic, Molecular, and Optical Physics Laboratory, RIKEN, Wako, Saitama 351-0198, Japan.}
\author{T.~Matsuzaki}
\affiliation{RIKEN Nishina Center, RIKEN, Wako, Saitama 351-0198, Japan.}
\author{S.~Abe}
\affiliation{Japan Atomic Energy Agency (JAEA), Tokai, Ibaraki 319-1195, Japan}
\author{H.~Fukuda}
\affiliation{Interdisciplinary Graduate School of Engineering Sciences, Kyushu University, Kasuga, Fukuoka 816-8580, Japan}
\author{M.~Hashimoto}
\affiliation{Graduate School of Informatics, Kyoto University, Kyoto, Kyoto 606-8501, Japan}
\author{A.~D.~Hillier}
%\affiliation{STFC Rutherford Appleton Laboratory, Didcot, Oxfordshire OX11 0QX, United Kingdom.}
\affiliation{ISIS Neutron and Muon Facility, STFC Rutherford Appleton Laboratory, Didcot, Oxfordshire OX11 0QX, United Kingdom.}
\author{K.~Ishida}
\affiliation{High Energy Accelerator Research Organization (KEK), Tsukuba, Ibaraki 305-0801, Japan.}
\author{N.~Kawamura}
\affiliation{High Energy Accelerator Research Organization (KEK), Tsukuba, Ibaraki 305-0801, Japan.}
\author{S.~Kawase}
\affiliation{Interdisciplinary Graduate School of Engineering Sciences, Kyushu University, Kasuga, Fukuoka 816-8580, Japan}
\author{T.~Kawata}
\affiliation{Interdisciplinary Graduate School of Engineering Sciences, Kyushu University, Kasuga, Fukuoka 816-8580, Japan}
\author{K.~Kitafuji}
\affiliation{Interdisciplinary Graduate School of Engineering Sciences, Kyushu University, Kasuga, Fukuoka 816-8580, Japan}
\author{F.~Minato}
\affiliation{Department of Physics, Kyushu University, Fukuoka, Fukuoka 891-0395, Japan}
\author{M.~Oishi}
\affiliation{Interdisciplinary Graduate School of Engineering Sciences, Kyushu University, Kasuga, Fukuoka 816-8580, Japan}
\author{A.~Sato}
\affiliation{Department of Physics, Osaka University, Toyonaka, Osaka 560-0043, Japan}
\author{K.~Shimomura}
\affiliation{High Energy Accelerator Research Organization (KEK), Tsukuba, Ibaraki 305-0801, Japan.}
\author{P.~Strasser}
\affiliation{High Energy Accelerator Research Organization (KEK), Tsukuba, Ibaraki 305-0801, Japan.}
\author{S.~Takeshita}
\affiliation{High Energy Accelerator Research Organization (KEK), Tsukuba, Ibaraki 305-0801, Japan.}
\author{D.~Tomono}
\affiliation{Research Center for Nuclear Physics, Osaka University, Mihogaoka 10-1, Ibaraki, Osaka 567-0047, Japan}
\affiliation{High Energy Accelerator Research Organization (KEK), Tsukuba, Ibaraki 305-0801, Japan.}
%\author{I.~Umegaki}
%\affiliation{High Energy Accelerator Research Organization (KEK), Tsukuba, Ibaraki 305-0801, Japan.}
\author{Y.~Watanabe}
\affiliation{Interdisciplinary Graduate School of Engineering Sciences, Kyushu University, Kasuga, Fukuoka 816-8580, Japan}
%\author{Y.~Yamaguchi}
%\affiliation{Japan Atomic Energy Agency (JAEA), Tokai, Ibaraki 319-1195, Japan}

%\altaffiliation[Also at ]{Department of Physics, the University of Tokyo}%Lines break automatically or can be forced with \\
%\affiliation{%


\date{\today}% It is always \today, today,

\begin{abstract}
%About 5\% of article length and $<$ 500 words
%An article usually includes an abstract, a concise summary of the work covered at length in the main body of the article. 
%The muon nuclear capture reaction and subsequent particle emission in $^{27}$Al and $^{28,29,30}$Si were studied by measuring the production branching ratios of the nuclei produced after particle emission. 
\begin{description}
%You may use the \texttt{description} environment to structure your abstract; use the optional argument of the \verb+\item+ command to give the category of each item. 
\item[Background]
Muon nuclear capture is a reaction between a muon and a proton inside a nucleus through weak interactions. This reaction results in the formation of an excited nucleus, which subsequently de-excites by emitting several particles. Examination of the excited state allows for an investigation of the properties of nuclear excitation and particle emission in highly excited nuclei. 
\item[Purpose]
This study investigates muon nuclear capture of $^{27}$Al and $^{28,29,30}$Si, focusing on determining the absolute production branching ratio (BR) following muon nuclear capture and subsequent particle emissions.
By measuring the absolute production BR, we can collect valuable information on the excitation energy distribution of muon nuclear capture. %, allowing for an investigation of the properties of excitation and particle emission in highly excited nuclei. 
\item[Methods]
Measurements were conducted using the in-beam activation method at two pulsed muon facilities: RIKEN-RAL beamline at Rutherford Appleton Laboratory and Materials and Life Science Experimental Facility at Japan Proton Accelerator Research Complex.
Absolute BRs were determined by measuring the number of muons irradiating the target. using a plastic scintillator and the $\beta$-delayed $\gamma$-rays emitted from the produced nuclei using germanium detectors. 
\item[Results]
The absolute production branching ratios of muon nuclear capture on $^{27}$Al and $^{28,29,30}$Si were obtained with the highest accuracy to date.
Predominant neutron emissions, even-odd atomic number dependence of particle emission probabilities, and influence of the neutron excess were observed.
These results were compared with previous measurements and theoretical models and discussed regarding the excitation energy distribution, particle emission mechanism, and nuclear properties, such as resonance in the isovector transition.
\item[Conclusion]
This study emphasizes the importance of considering nuclear structure eﬀects, even-odd eﬀects of proton and neutron numbers, neutron excess, nucleon pairing effect, and particle emission mechanisms, in the context of the muon nuclear capture reaction.

\end{description}
\end{abstract}

\maketitle

\section{Introduction}
%ミューオン原子核捕獲反応については、原子核構造・集団運動を反映した現象だけど、実験データの不足から確立したモデルが未だ存在していない。
A muonic atom is an atomic-bound state comprising %consisting of
a nucleus and a negative muon. 
%After the muon cascades down to its 1s atomic orbit, some of the muons are captured by a proton via weak interaction, called muon nuclear capture. 
After the muon cascades to its 1s atomic orbit, some of the muons are captured by a proton through a weak interaction, a process known as muon nuclear capture. 
The reaction involving a nucleus with $(A, Z)$ can be expressed as $\mu^- +(A, Z)\to (A, Z-1)^* + \nu_\mu$, where $A$ and $Z$ represent the mass and atomic numbers of the nucleus, respectively. 
The reaction produces an excited $(A, Z-1)$ nucleus owing to the large mass of the muon, which is 105.6 MeV/c$^2$. Consequently, these highly excited nuclei can emit several particles, such as neutrons, charged particles, and $\gamma$-rays. 
%Muon nuclear capture is one of the charge exchange reactions which is similar to negative pion capture or electron capture. The differences between these reactions are that muon nuclear capture does not involve a strong interaction, unlike a pion capture, and muon capture can occur in all nuclei because their masses are large enough, while electron capture requires an energy higher than the reaction threshold energy. %Q value.
Muon nuclear capture is classified as a charge exchange reaction, similar to negative pion or electron capture. However, unlike pion capture, muon nuclear capture does not involve a strong interaction. Additionally, muon capture can occur in all nuclei owing to muon's significant masses, whereas electron capture necessitates energy exceeding that of the reaction threshold.
%Another characteristic of the muon nuclear capture reaction is that it populates a highly excited state with small angular momentum transfer; the initial state is a static atomic state with 1s muon 
Another distinguishing feature of muon nuclear capture is its ability to populate a highly excited state with minimal angular momentum transfer. The initial state comprises a static atomic state with a 1s muon (spin 1/2 and zero orbital angular momentum) and a proton within the nucleus (possessing orbital angular momentum).
In addition, the muon nuclear capture occurs through a weak interaction, and the highly excited state generated by this process has the potential to provide less model-dependent insight into giant resonance, particularly isovector transition. This complements charge exchange reactions, such as (n, p) reaction.
%However, there is no established model of muon nuclear capture and particle emission yet due to the lack of experimental data.
However, muon nuclear capture and particle emission have no established model owing to the lack of experimental data.
The distribution of excitation energy cannot be directly measured owing to the challenges associated with measuring the energy of escaped $\nu_\mu$.
Consequently, %we investigate 
our research focuses on the muon nuclear capture reaction and subsequent particle emission, leveraging our current knowledge of nuclear reactions, advanced measurement techniques, and intense muon beams.

The structure of the excitation function, the excitation energy distribution, is reflected in particle emissions.
Based on previous measurements of the emitted particle energies, the typical average excitation energy is approximately 10--20 MeV, with the excitation energy reaching up to approximately 100 MeV.
%Since direct measurement of the emitted particle is restricted from the threshold energy due to the energy deposit inside the target material, we measured the production branching ratio (BR) of the left nuclei after the particle emission to investigate the particle emission probabilities.
However, direct measurement of emitted particles is limited by the particle separation energy and energy deposition within the target material. To overcome this limitation, we have measured the production branching ratio (BR) of the remaining nuclei after particle emission to assess particle emission probabilities.
%A few nuclei exist in which the production BR after the muon nuclear capture reaction has been measured owing to experimental limitations. 
Production BRs after muon nuclear capture have only been measured for a limited number of nuclei because of experimental limitations.
Recently, a method referred to as the in-beam activation method capable of measuring short-lived nuclei using the activation method has been developed, and production BRs of muonic palladium isotopes have been comprehensively measured~\cite{Niikura2024-ck}.
To accurately measure the absolute BR, the number of stopped muons must be counted. %to measure the absolute BR using the activation method. 
However, quantifying the number of muons within a beam pulse at a high-intensity muon beam facility presents a significant challenge.
This study conducted step-by-step measurements to assess the absolute number of stopped muons using a high-intensity muon beam, and the in-beam activation method was extended to measure the absolute production BR.

%励起関数の概形
$^{27}$Al and $^{28,29,30}$Si were selected as the target because they were optimal for measuring the production BR and exploring excitated states.
$^{28}$Si, in particular, was selected owing to the %wealth of previous studies of 
extensive prior research on muon nuclear capture~\cite{Measday2001-mw}.
%However, previous measurements were vague and showed some discrepancies, and they are worth re-evaluating.
However, previous measurements in this atomic mass region have been somewhat ambiguous and have displayed discrepancies, warranting a reevaluation.
Furthermore, the emission of charged particles, in addition to the neutron emission, is anticipated in this mass region. 
Recent studies have focused on charged particle emission~\cite{AlCap_Collaboration2022-ne, Manabe2023-lc, Autran2024-mp, Minato2023-wu}.
The production BR of charged particles is expected to provide insights into the characteristics of particle emission from highly excited nuclei and the emission probabilities of light nuclei. These insights are crucial for understanding the cluster properties within the nucleus.
%which are related to the cluster properties within the nucleus.
In this study, both natural abundance ($^\mathrm{nat}$Si) and isotopically enriched silicons ($^{28,29,30}$Si) were measured. 
Additionally, $^{27}$Al was measured owing to its monoisotopic nature %isotopic enrichment with natural abundance 
and proximity in atomic number to silicon isotopes, making it ideal for investigating the even-odd effect of atomic number. 
This mass region facilitates a comprehensive discussion based on both the shell structure and collective properties of the nucleus.

%目次
The remainder of this paper is organized as follows. The experimental method and the details of the experiments are described in Sects.~\ref{sec:method} and \ref{sec:experiment}, respectively.
The analysis procedure is presented in Sect.~\ref{sec:analysis} and the result of the deduced BRs for all the targets is summarized in Sect.~\ref{sec:result}.
In Sect.~\ref{sec:Discussion}, the obtained BRs are discussed by comparing them with previous results and model calculations, and %regarding the systematics of mass and atomic number.
by considering the systematics with mass and atomic number.
Finally, Sect.~\ref{sec:summary} summarizes and concludes the paper.
\section{Method}\label{sec:method}
In this study, the production BR ($b$) is defined as 
\begin{equation} \label{eq:defineBR}
    b \equiv \frac{N_\mathrm{prod}}{N_\mathrm{cap}},
    %b = \frac{N_\gamma /\epsilon_\gamma}{N_\mathrm{cap} P_\mathrm{decay} I_\gamma},
\end{equation}
where $N_\mathrm{prod}$ represents the number of produced nuclei and $N_\mathrm{cap}$ represents the number of muon nuclear capture reactions.
$N_\mathrm{prod}$ was determined using the in-beam activation method, which is a technique developed to measure short-lifetime isotopes with an activation method that utilizes the time structure of the pulse beam~\cite{Niikura2024-ck}.
%The in-beam activation method was adopted in the present measurement because most of the reaction residues have a lifetime shorter than a few minutes for silicon isotopes.
This method was chosen for the present measurement owing to the short lifetimes of most reaction residues for silicon isotopes.
%The present paper will use the same notation as the Ref.~\cite{Niikura2024-ck} hereafter.
Throughout this study, the same notation as in Ref.~\cite{Niikura2024-ck} will be used.

$N_\mathrm{cap}$ is expressed with the total number of muons irradiating the target ($N_\mathrm{muon}$), muon nuclear capture probabilities ($P_\mathrm{cap}$), and the beam stopping rate in the target ($\epsilon_\mathrm{stop}$), that is %i.e.
\begin{equation} \label{eq:Ncap}
    N_\mathrm{cap} = N_\mathrm{muon}P_\mathrm{cap}\epsilon_\mathrm{stop}
    %= \int I_\mathrm{beam} dt P_\mathrm{cap}\epsilon_\mathrm{stop},
    = \Sigma n_\mathrm{beam}~P_\mathrm{cap}\epsilon_\mathrm{stop},
\end{equation}
where $n_\mathrm{beam}$ represents the number of muons in one pulse. %during the beam irradiation.
$\epsilon_\mathrm{stop}$ was set to be equal to one by adjusting the beam momentum and target thickness as outlined in Sect.~\ref{sec:experiment}. $P_\mathrm{cap}$ of aluminum and silicon isotopes were obtained using the following equation:
\begin{equation}
    %P_\mathrm{cap} = \frac{\Lambda_\mathrm{cap}}{\Lambda_\mathrm{cap}+Q/\tau_{\mu^+}},
    %P_\mathrm{cap} = \frac{\Lambda_\mathrm{cap}}{\Lambda_\mathrm{total}},
    P_\mathrm{cap} = \Lambda_\mathrm{cap}\tau_\mathrm{total},
\end{equation}
where $\Lambda_\mathrm{cap}$ represents the muon capture rate and $\tau_\mathrm{total}$ represents the lifetime of the muonic atom referred from Ref.~\cite{Mizuno2024}.
Counting the number of muons in each pulse accurately is generally a challenging task when dealing with a high-intensity beam.
Therefore, to obtain the absolute BR by measuring $n_\mathrm{beam}$, %the present measurements were conducted with the three steps below:
a series of measurements were performed following the steps outlined below:
1) The absolute BR measurement for main channels was conducted using a countable low-intensity pulsed muon beam.
2) The beam irradiation number was then calibrated with a high-intensity pulsed muon beam using the measured absolute BR and Eqs.~(\ref{eq:defineBR}) and (\ref{eq:Ncap}).
3) Finally, the high-statistics BR measurement was conducted using the calibrated high-intensity beam, especially for isotopes having small production BR.

Experiments were performed at two pulsed muon beam facilities: part 1) at the ISIS Neutron and Muon Source at Rutherford Appleton Laboratory (RIKEN-RAL beamlines)~\cite{Matsuzaki2001-pg, Hillier2018-vd} and part 2) and 3) at the Materials and Life Science Experimental Facility (MLF) at the Japan Proton Accelerator Research Complex (J-PARC)~\cite{Higemoto2017-qv}.
The intensity of the muon beam at RIKEN-RAL beamlines can be reduced to a countable number and provide less electron contamination. On the other hand, the high-intensity beam at J-PARC %has the advantage of obtaining 
provided high statistical data and was ideal for comprehensive measurement, including those involving low-probability branches. Additionally, the high-intensity beam offered a high signal-to-noise (S/N) ratio because the typical background of the activation measurement is natural background radiation.

\section{Experiment}\label{sec:experiment}
Two experiments were conducted at RIKEN-RAL and J-PARC, which are referred to as the RAL experiment and the J-PARC experiment, respectively, hereinafter.
The RAL experiment was conducted at Port 4 of RIKEN-RAL at ISIS. 
A proton beam of 800 MeV, accelerated in the ISIS RCS synchrotron, irradiated a graphite target at Target Station 1 (TS1) and produced pions. The ISIS synchrotron output has a double-pulse structure with a 50 Hz repetition rate, with four out of five pulses directed to TS1. The negative muon beam, resulting from the decay of the negative pions, was then transported and delivered to Port 4. 
The J-PARC experiment was conducted at the D1 area at MLF, J-PARC. 
A graphite pion production target was irradiated with a proton beam accelerated to 3 GeV by an RCS synchrotron, operating at an intensity of approximately 730~kW.
The negative muon beam was obtained from the decay of pions to the D1 area.
The muon beam had a double-pulse structure with a 25 Hz repetition rate.

\begin{figure}[htbp]
    \centering
    %\begin{minipage}{1.0\textwidth}
    %\hspace{-9.5cm}
    \includegraphics[width=0.35\textwidth]{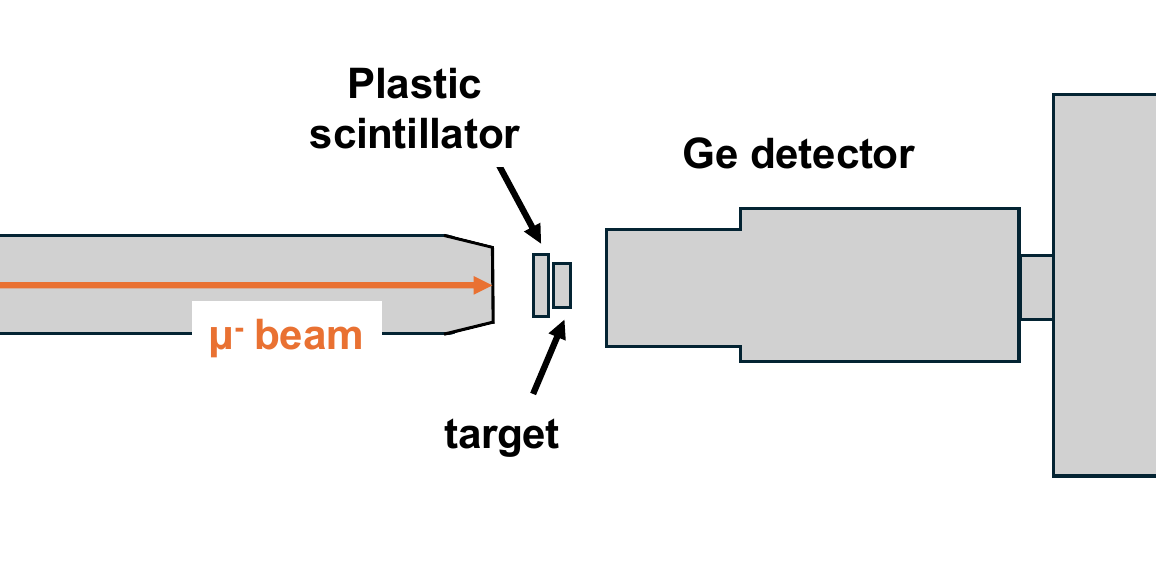} \\
    \small{(a) RAL experiment at Port 4}
    %\end{minipage}
    %\begin{minipage}{1.0\textwidth}
    %\hspace{-9.5cm}
    \includegraphics[width=0.45\textwidth]{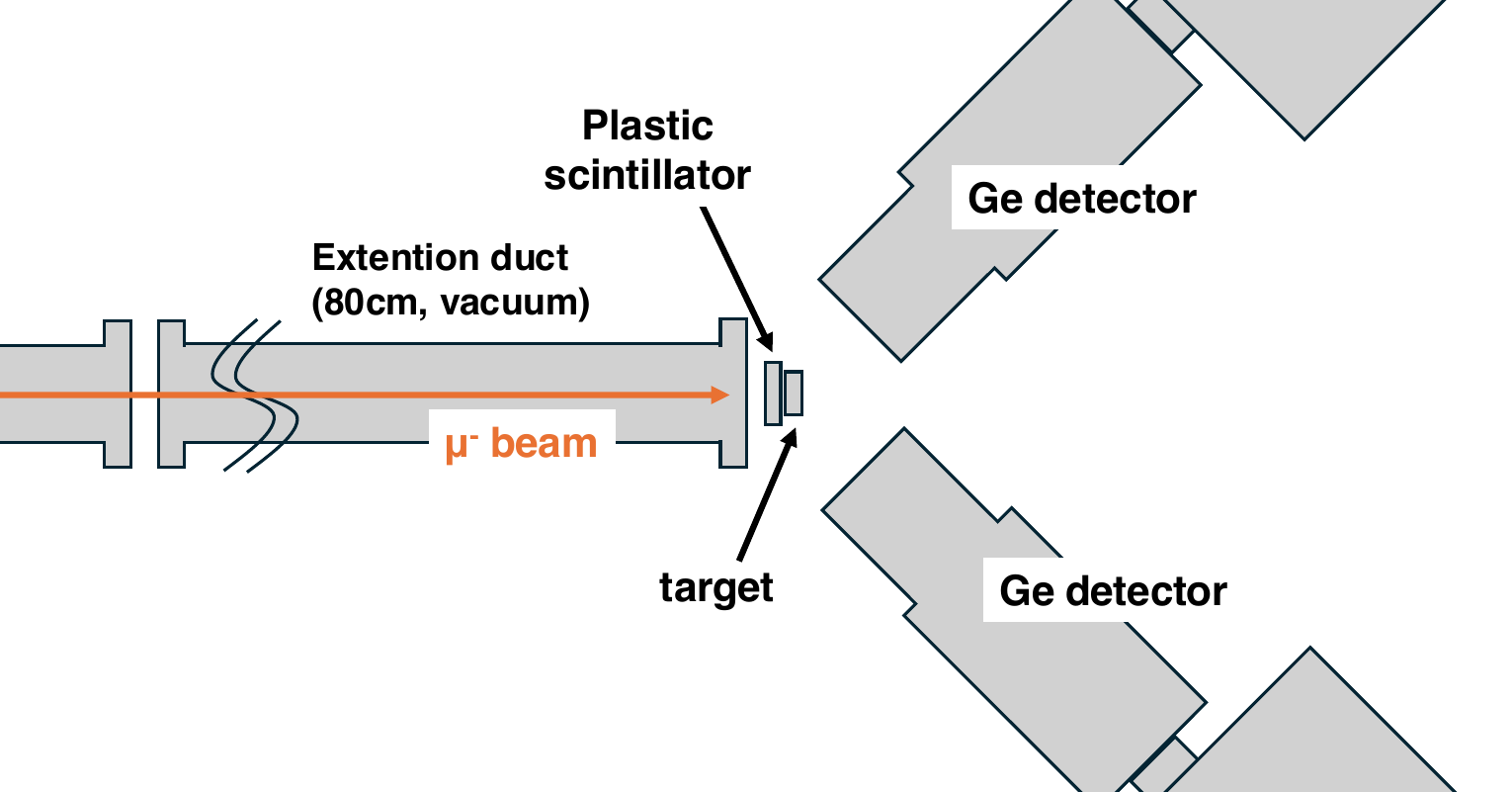}\\
    \small{(b) J-PARC experiment at D1 area}
    %\end{minipage}
    \caption{Schematic of the experimental setup at (a) ISIS, RAL, and (b) MLF, J-PARC (not to scale). In the J-PARC experiment, the beam duct was extended by 80~cm, with the detectors set downstream to a $\mu$SR spectrometer.}
    \label{fig:setup_RALMLF}
\end{figure}

The schematics of the experimental setups at (a)RAL and (b)J-PARC are shown in Fig.~\ref{fig:setup_RALMLF}. The concept of the setup was consistent in both experiments. 
The number of muons irradiating targets was counted using a plastic scintillator installed upstream of the target. 
In the RAL experiment, the plastic scintillator was set approximately 35~mm away from the end of the beam collimator, whereas in the J-PARC experiment, it was placed 30~mm from the end of the beam duct.
The plastic scintillator had a thickness of 0.5 mm, collimated by a 10~mm thick acrylic plate with a diameter of 20 mm. %a diameter of 20 mm and 
Photomultipliers (PMT) manufactured by Hamamatsu (H11934-100-10 and H11934-100 in the RAL and J-PARC experiments, respectively) were utilized to read out the signal from the plastic scintillator. Two PMTs were employed in the RAL experiment on both sides with an applied voltage of 800~V, whereas in the J-PARC experiment, one PMT was utilized with a voltage of 720~V to prevent output signal saturation.

\begin{figure}[htbp]
    \centering
    \includegraphics[width=0.45\textwidth]{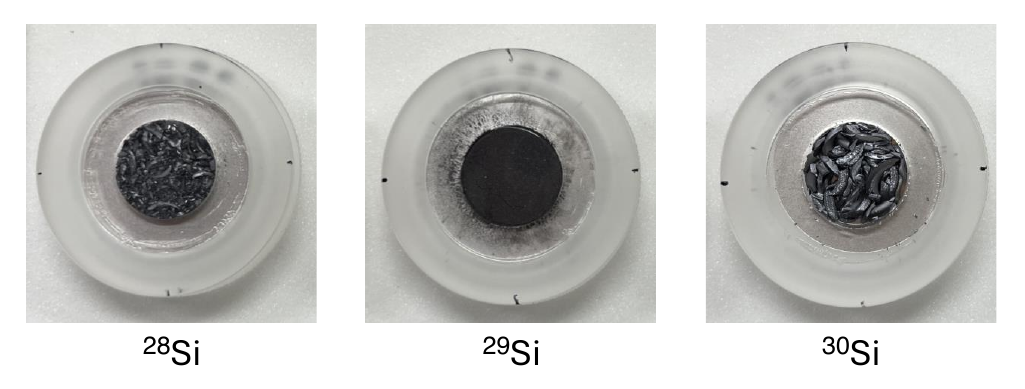}
    \caption{Photographs of isotopically enriched silicon targets. Target materials were contained in a hole of the acrylic case with a diameter and thickness of 15~mm and 2.8~mm, respectively.}
    \label{fig:target_picture}
\end{figure}

%Target
The activation of five targets was measured, including aluminum and silicon targets with both natural abundance and isotopically enriched compositions ($^{28,29,30}$Si).
The aluminum and natural abundance silicon targets were in the form of metal plates, whereas the enriched targets were in powder and small piece form, as shown in Fig.~\ref{fig:target_picture}. The enriched targets were encapsulated in acrylic target cases.
For the background reference, an empty acrylic case was also irradiated.
Details regarding the form, size, and enrichments of each target are listed in Table~\ref{tab:targets_info_measure}.
The probability of the muon nuclear capture reaction for each target is also provided in Table~\ref{tab:targets_info_measure}~\cite{Mizuno2024}.
The targets were positioned 1 mm away from the plastic scintillator to ensure that all muons passing through the plastic scintillator stopped in the target, particularly important for metal plate targets.

%Beam
The momentum of the muon beam ($p$) was 36~$\mathrm{MeV}/c$ ($\Delta p/p=4$\%) at RAL and 41 $\mathrm{MeV}/c$ ($\Delta p/p=3$\%) at J-PARC %, respectively, which was chosen to ensure all muons stopped at the target.
guarantee that all muons stopped at the plate target.
%設定値：J-PRAC:38MeV, RAL: 35MeV
A plastic scintillator with a size of 50$\times$50$\times$5 mm$^3$ was positioned behind the targets prior to the measurement and served as a veto counter to validate that muons did not pass through the target. 
The isotopically enriched targets were in powder form, so some of the irradiated muons may have passed through them. Additionally, the radius of the acrylic target case was smaller than that of the plastic scintillator. To determine the $\epsilon_\mathrm{stop}$ of the enriched targets, the natural abundance silicon target was utilized for decomposition analysis in Sect.~\ref{subsec:muon_irrad_num}. 
The difference in the momentum setting between RAL and J-PARC was attributed to the presence of an additional beam duct in the J-PARC experiment. 
In the J-PARC experiment, the beam transport line was extended by 80 cm with a vacuum duct after the muon beam passed through the air. The extended beam duct was introduced to place detectors downstream of a $\mu$SR spectrometer in the D1 area. The momentum of the muon beam was degraded with the Kapton foil at both ends of the extended beam duct.

%Ge
High-purity p-type coaxial germanium detectors were used for the decay $\gamma$-ray measurement.
Details regarding the detector performance and the data acquisition system of the germanium detectors were summarized in Ref.~\cite{Mizuno2024-br}. 
In the RAL experiment, a Canberra GX5019 was placed at 0 degrees relative to the muon beamline and 58 mm away from the target position. 
The dynamic range of the measured $\gamma$-ray obtained with the germanium detector was set to 0.020--2.6~MeV. The typical count rate of the germanium detector was approximately 40-60 counts per second (cps) during beam irradiation and approximately 15 cps without the beam.
In the J-PARC experiment, two Canberra-manufactured germanium detectors, GX5019 and GC3018, were utilized. They were positioned at 45 degrees to the muon beamline and 55 mm away from the target. Owing to the higher intensity and larger electron contamination in J-PARC, the germanium detectors could not be set at 0 degrees to the beamline.
The dynamic ranges of the germanium detectors were 0.070--10.4~MeV with GX5019 and 0.015--2.1~MeV with GC3018. 
%The typical count rate of each germanium detector was approximately 100-200 cps during the beam irradiation and approximately 20-30 cps without the beam. 
During beam irradiation, the typical count rate for each germanium detector was approximately 100-200 cps, dropping to 20-30 cps without the beam. 
%The germanium detectors were surrounded by lead blocks to reduce the natural background radiation in both experiments. 
To minimize natural background radiation, the detectors were shielded by lead blocks in both experiments.

%DAQ
All the signals from the detectors were acquired using 500-MS/s 14-bit waveform digitizers, CAEN V1730B and V1730SD~\cite{CAENV1730}, in both experiments. 
The firmwares of the digitizer boards were equipped with digital pulse processing for pulse height analysis (DPP-PHA) and DPP for charge integration and pulse shape discriminator (DPP-PSD), respectively. 
The energy and timestamp data from the germanium detectors were processed using DPP-PHA.
Additionally, the waveforms of all the PMT signals from plastic scintillators were recorded. 
A signal from the germanium detectors was acquired and saved in the self-trigger mode, whereas signals from the plastic scintillator were obtained based on a 50 Hz or 25 Hz trigger condition from the synchrotron in the RAL and J-PARC experiments, respectively. The timing signal from the Cerenkov counter installed near the pion production target was also obtained and served as a precise timing reference for muon beam irradiation in the RAL experiment. In the J-PARC experiment, the RCS kicker signal was used as a precise timing reference instead of a Cherenkov signal.

%Run summary
The measurement conditions for each target are listed in Table~\ref{tab:targets_info_measure}.
$\beta$-delayed $\gamma$-rays were measured both during the beam irradiation and after the irradiation (decay measurement) because certain $\gamma$-ray energy peaks from the long-lived isotopes produced demonstrate a higher S/N ratio in the off-beam condition.

\begin{table*}[htbp]
    \centering
    \caption{Summary of the composition and shape information of targets. The probability of the muon nuclear capture reaction ($P_\mathrm{cap}$) is listed in the sixth column~\cite{Mizuno2024}. Furthermore, the measurement conditions of each target were summarized. Beam irradiation and decay measurement times are listed in the first and second rows of each experiment, and with the stopping rate $\epsilon_{\mathrm{stop}}$ summarized for each target.}
    \begin{ruledtabular}
    \begin{tabular}{cccccccccccc}%\hline\hline
     & & & & & &\multicolumn{3}{c}{RAL} & \multicolumn{3}{c}{J-PARC}\\
    Target    & Form  & Size(mm$^3$) & Weight(g) & Enrichment & $P_\mathrm{cap}$(\%) & Irradiation & Decay & $\epsilon_{\mathrm{stop}}$ & Irradiation & Decay & $\epsilon_{\mathrm{stop}}$\\\hline
    %$\bar n_\mu$ 
    $^{27}$Al  & metal plate   & 50$\times$50$\times$2 & 13.6  &  99$+$\% & 60.96(4) & 20.0h & 0.3h  & 1.0 & 0.7h & 0.0h & 1.0\\
    $^\mathrm{nat}$Si      & metal plate   & $\phi$50$\times$1.96  & 9.00  & %$>$99.99999\%
     - & 66.07(5) & 8.3h  & 0.4h  & 1.0 & 7.3h & 6.7h & 1.0\\
    $^{28}$Si  & metal powder  & $\phi$15$\times$2.8   & 0.500 & 99.93\%\footnotemark[1] & 66.41(23) & 14.8h & 10.0h & 0.407(14)  & 4.7h & 0.25h & 0.298(9)\\%0.409(21)&0.299(15)
    $^{29}$Si  & metal powder  & $\phi$15$\times$2.8   & 0.500 & 99.25\%\footnotemark[2] & 64.14(16) & 5.4h  & 3.0h  & 0.652(169) & 4.7h & 0.25h & 0.260(63)\\%0.624(189)&0.252(72)
    $^{30}$Si  & small pieces  & $\phi$15$\times$2.8   & 0.500 & 99.64\%\footnotemark[3] & 61.21(12) & 23.7h & 29.2h  & 0.349(26)  & 4.5h & 0.25h & 0.261(19)\\%0.352(29)&0.263(21)%\hline\hline
    empty      & acrylic case & - & - & - & - & 4.0h  & 0.0h  & - & 1.5h & 0.0h & - \\%\hline\hline
    \end{tabular}
    \end{ruledtabular}
    %\footnotetext[1]{cite by Suzuki et. al.~\cite{Suzuki1968}}
    %\footnotetext[1]{from Ref.~\cite{Mizuno2024}}
    \footnotetext[1]{The other composition of $^{28}$Si is 0.065\% of $^{29}$Si and 0.005\% of $^{30}$Si.}
    \footnotetext[2]{The other composition of $^{29}$Si is 0.20\% of $^{28}$Si and 0.55\% of $^{30}$Si.}
    \footnotetext[3]{The other composition of $^{30}$Si is 0.32\% of $^{28}$Si and 0.04\% of $^{29}$Si.}
    \label{tab:targets_info_measure}
\end{table*}
\section{Anaysis}\label{sec:analysis}
Figure~\ref{fig:Ge_spectrum} shows the segment of the $\gamma$-ray energy spectra of each target obtained with a germanium detector at the RAL experiment. 
The activation method %was performed with these energy spectra, and the absolute BRs were calculated by deducing 
utilized these energy spectra to calculate the absolute BRs by inferring 
the muon irradiation number from a plastic scintillator signal.
%In this section, 
%This section delves into the data evaluation, activation analysis, absolute production BR analysis using a low-intensity beam, and the calibration method of the plastic scintillator with a high-intensity beam.
In this section, data evaluation, activation analysis, absolute production BR analysis using a low-intensity beam, and the calibration method of the plastic scintillator with a high-intensity beam are described. 

\subsection{Activation analysis and data evaluation}
The number of reaction residues can be determined through $\beta$-delayed $\gamma$-ray spectroscopy, leveraging the well-established $\beta$-decay schemes of the reaction residues.
The number of produced nuclei ($N_\mathrm{prod}$) was determined using the in-beam activation method, and is expressed as follows:
\begin{equation} \label{eq:Nprod}
    N_\mathrm{prod} = \frac{N_\gamma/\epsilon_\gamma \epsilon_\mathrm{LT}}{ P_\mathrm{decay} I_\gamma} - N_\mathrm{decay},
\end{equation}
where $N_\gamma$ represents the number of the detected $\beta$-delayed $\gamma$-ray, $\epsilon_\gamma$ represents the $\gamma$-ray detection efficiency of the detectors, 
$\epsilon_\mathrm{LT}$ represents the analysis live-time ratio, 
$P_\mathrm{decay}$ represents the decay probability during the measurement period, and $I_\gamma$ represents the $\gamma$-ray intensity per decay of the subject nuclei.
In this study, $N_\mathrm{prod}$ is defined as the direct production number, obtained by subtracting decay components from $\beta$-decay of other nuclei
and isomeric decay ($N_\mathrm{decay}$).

\begin{figure}[htbp]
    \centering
    \includegraphics[width=0.98\linewidth]{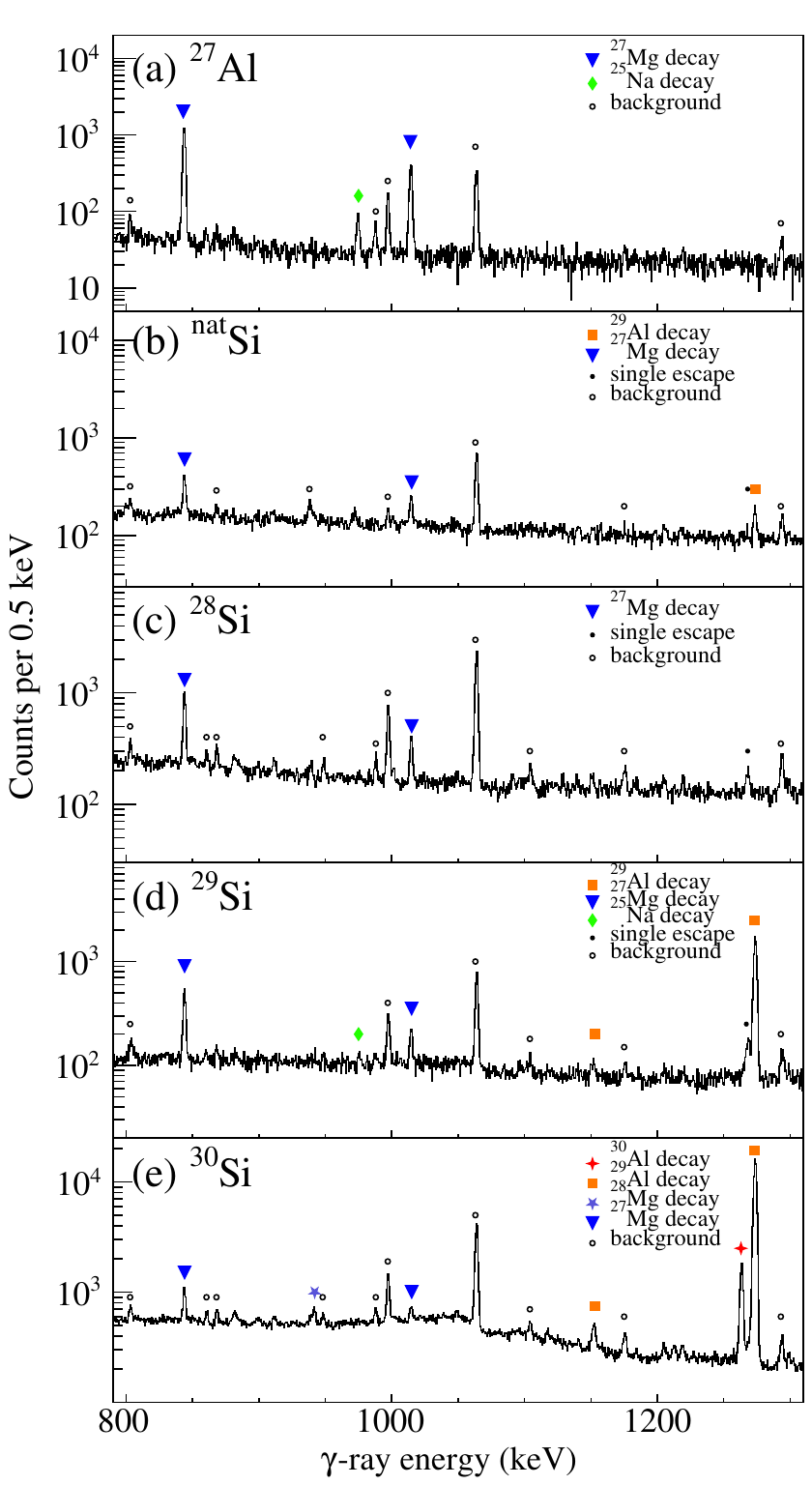}
    \caption{Part of the $\gamma$-ray energy spectra of $^{27}$Al, $^\mathrm{nat}$Si, and $^{28,29,30}$Si target obtained with a germanium detector at the RAL experiment. $\beta$-delayed $\gamma$-ray energy peaks are denoted with each symbol, background peaks are denoted with open circles, and peaks originating from single escapes of other peaks are denoted with filled circles.}
    \label{fig:Ge_spectrum}
\end{figure}

The $N_\gamma$ values were determined based on the measured energy peaks of $\beta$-delayed $\gamma$-ray, some of which are indicated by closed symbols in Fig.~\ref{fig:Ge_spectrum}. 
The identification of these $\gamma$-rays was achieved through their energy, with additional confirmation provided by half-life for some $\gamma$-rays. %and the origin of some $\gamma$-rays was also confirmed using lifetime. 
%The germanium detectors utilized in this study was calibrated using standard $\gamma$-ray sources.
The energy calibration of germanium detectors was conducted with standard $\gamma$-ray sources.
%sources. 60Co, 133Ba, and 137Cs were used in the RAL experiment, and 60Co and 152Eu were used in the J-PARC experiment, respectively
To account for any shifts in energy peaks obtained during the measurement, a correction for gain drift was applied using typical $\gamma$-ray energy peaks. 
Additionally, potential overlaps with single and double escape peaks from the higher energy peaks were examined for all the observed peaks.
%Ge efficiency, self-absorption, gain shift, dead time, muon gate, background
The background component of each $\gamma$-ray energy peak was subtracted, if present.
The sources of this background included environmental background and beam irradiation on materials other than the target. 
For example, the creation of background $^{28}$Al through the neutron capture reaction, $^{27}$Al(n, $\gamma$)$^{28}$Al was attributed to neutron presence in the muon beamlines and the aluminum components of the germanium detector's cryostat case. %muon facility environment
%Especially for the J-PARC experiment, the background component from the beam irradiation on the beam duct, which was made of stainless steel, was critical for the analysis because stainless steel contained silicon approximately 1\%.
In the J-PARC experiment, particular attention was paid to the background contribution from beam irradiation on the stainless steel beam duct, which contained silicon approximately 1\%.
%The amount of environmental background was evaluated with background measurement without beam irradiation and normalized with measurement time.
The environmental background level was assessed through background measurements conducted without beam irradiation and then normalized based on the measurement time. Additionally, the background component associated with beam irradiation was determined through empty target measurements and normalized based on the muon irradiation number during each measurement.

The peak detection efficiency of the germanium detector, $\epsilon_\gamma$, was calibrated using standard $\gamma$-ray sources in the RAL experiment and evaluated through GEANT4 simulation~\cite{Agostinelli2003-eh, Allison2006-ho, Allison2016-zr} for the J-PARC experiment. The uncertainty of the GEANT4 simulation was comparable with that of the standard $\gamma$-ray sources, approximately 3\%~\cite{Mizuno2024-br}.
Furthermore, the impact of self-absorption of emitted $\gamma$-ray within the target and target case was estimated using the GEANT4 simulation. 
To validate the accuracy of the self-absorption estimation using GEANT4, the efficiency was compared with and without a 5.0 mm-thick acrylic plate placed in front of the standard $\gamma$-ray sources.

Owing to the close proximity of the detector to the target, some X-rays and $\gamma$-rays were detected simultaneously at the prompt timing of the beam arrival.
This resulted in the signals from the germanium detector becoming saturated for a few milliseconds after beam irradiation. %because detectors were placed at a closed distance to the target, 
Therefore, signals within 0.17~ms (RAL experiment) and 2.0~ms (J-PARC experiment) following beam irradiation were excluded from the analysis, with the effect being accounted for through the use of a live-time ratio, $\epsilon_\mathrm{LT}$, in the analysis process.

$P_\mathrm{decay}$ was calculated using the half-life of the produced nuclei ($T_{1/2}=\mathrm{ln}(2)/\lambda$) and the muon irradiation number as follows:
%$\tau=1/\lambda$
\begin{equation}
    P_\mathrm{decay} \equiv \frac{\int\lambda n_\mathrm{nucl}(t) dt}{N_\mathrm{muon}}.
\end{equation}
%where $n_\mathrm{nucl}(t)$ represents the number of radioactive reaction products at time $t$ assuming all the irradiated muon produced the given nucleus ($P_\mathrm{cap}\epsilon_\mathrm{stop}b=1$)~\cite{Niikura2024-ck}. 
%%the number of produced nuclei at time $t$
%The numerator represents the total decay number during the measurement period while assuming $P_\mathrm{cap}\epsilon_\mathrm{stop}b=1$ as well. 
Here, the numerator represents the total decay number during the 
measurement period assuming all the irradiated muon produced the given 
nucleus ($P_\mathrm{cap}\epsilon_\mathrm{stop}b=1$)~\cite{Niikura2024-ck}, where $n_\mathrm{nucl}(t)$ represents the number of radioactive reaction products at time $t$ while assuming $P_\mathrm{cap}\epsilon_\mathrm{stop}b=1$ as well.
The half-life of the produced nuclei and $I_\gamma$ were obtained from Evaluated and Compiled Nuclear Structure Data (ENSDF)~\cite{Shamsuzzoha-Basunia2012-jo,Shamsuzzoha-Basunia2013-wi,Shamsuzzoha-Basunia2011-nq,Basunia2016-mn,Firestone2009-aw,Basunia2022-zw,Shamsuzzoha-Basunia2021-sj,Basunia2015-ab,Firestone2015-jd,Tilley1998-sf,Tilley1995-zp,Alburger1969-ic}. %Basunia2024:A=30

The $N_\mathrm{prod}$ was calculated with Eq.~(\ref{eq:Nprod}) using $N_\gamma$ with and without beam irradiation.
%In this study, $N_\mathrm{prod}$ is defined as the direct production number, obtained by subtracting decay components from $\beta$-decay of other nuclei and isomeric decay. 
%as will be described in Sect.~\ref{sec:result}. 
%The subtraction was conducted by solving the Bateman equation~\cite{Bateman1910-sl, Niikura2024-ck}.
The subtraction of %the decay components from $\beta$-decay of other nuclei and isomeric decay 
$N_\mathrm{decay}$ was conducted by solving the Bateman equation~\cite{Bateman1910-sl, Niikura2024-ck}.
The direct production number of isomeric states was determined independently from the ground state production.

\subsection{Number of muons irradiating the target}\label{subsec:muon_irrad_num}
\begin{figure}[htbp]
    \centering
    \includegraphics[width=0.4\textwidth]{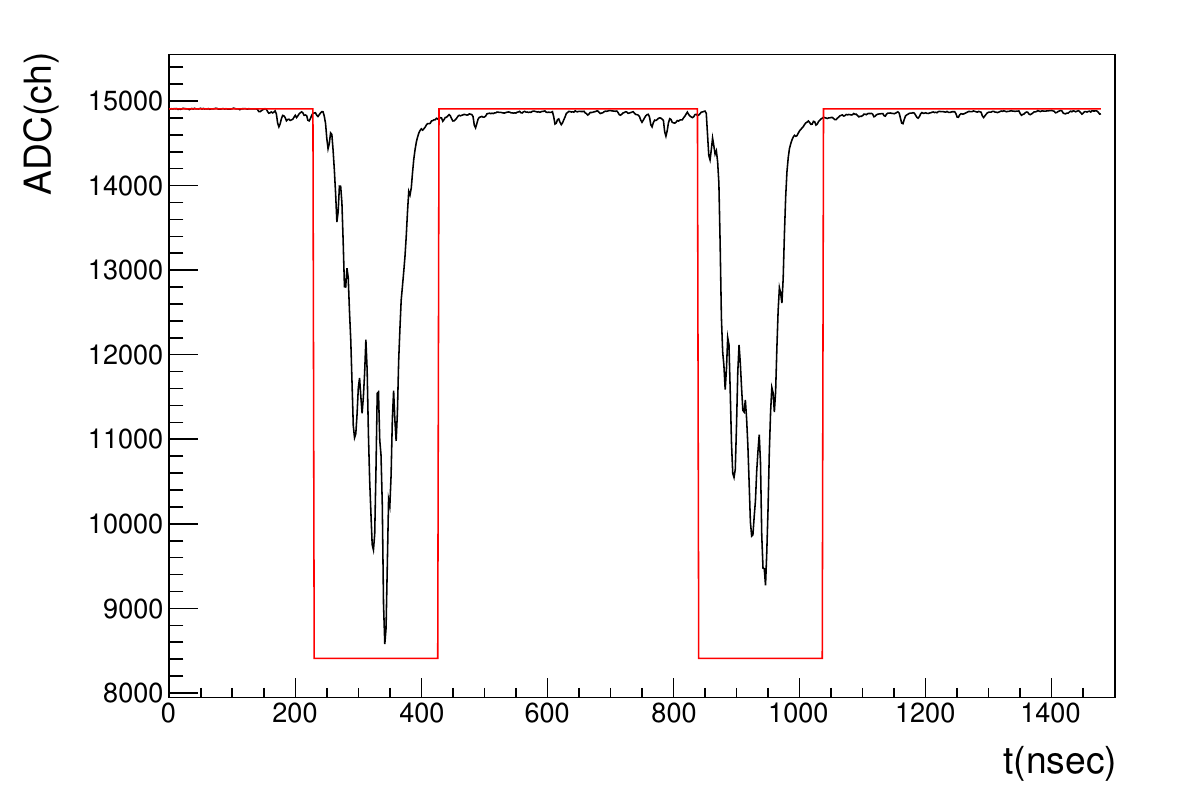}
    \caption{Raw signal of the plastic scintillator from PMT in one pulse obtained at J-PARC. 
    The time distance from the kicker signal is shown on the horizontal axis.
    %in nanoseconds with some offset
    The red line represents the charge integral gate of the muon irradiation. The gate width was chosen to be 150 ns and 200 ns in the RAL and J-PARC experiments, respectively.}
    \label{fig:plastic_rawsignal}
\end{figure}

\begin{figure}[htbp]
    \centering
    %\begin{minipage}[b]{0.49\linewidth}
    %\centering
    %\includegraphics[width=0.4\textwidth]{charge_integration_a.pdf}\\ %1.0
    \includegraphics[width=0.4\textwidth]{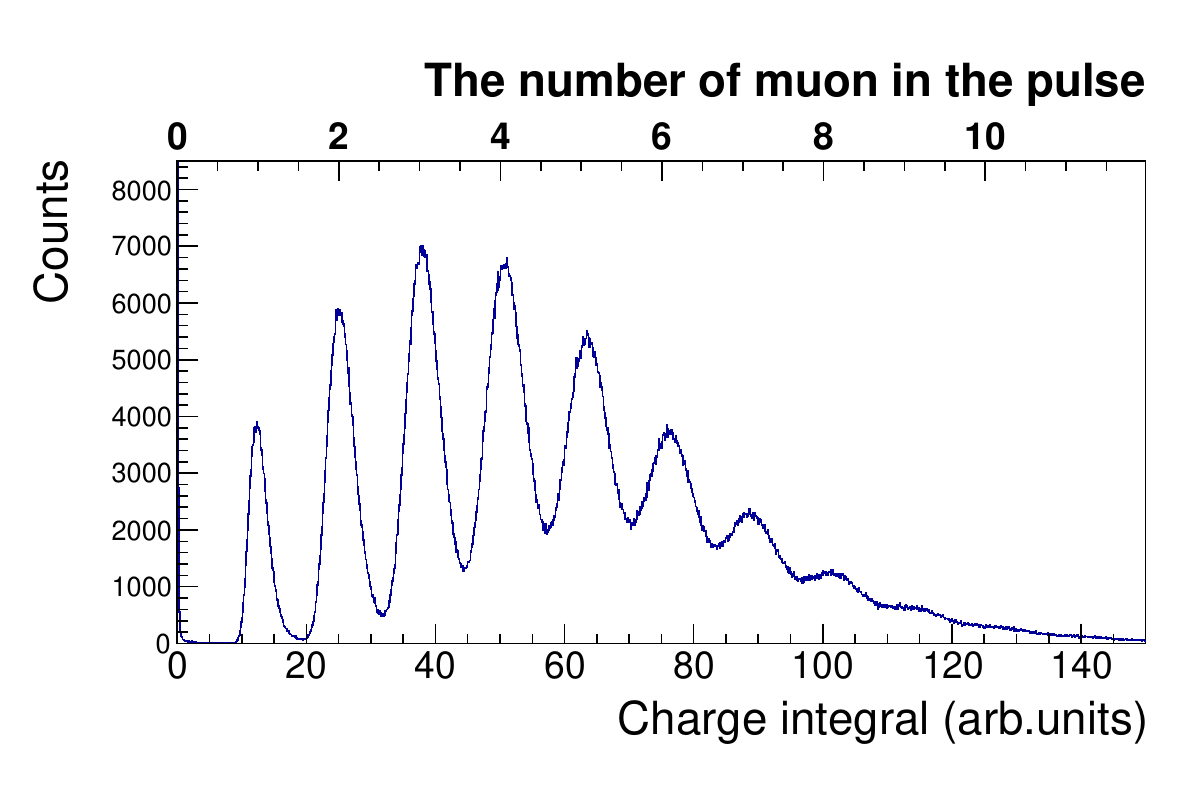}\\
    \small{(a) Muon beam with 35 MeV/c at RAL. The beam intensity was reduced to countable numbers.}
    %\end{minipage}
    %\begin{minipage}[b]{0.49\linewidth}
    %\centering
    \includegraphics[width=0.4\textwidth]{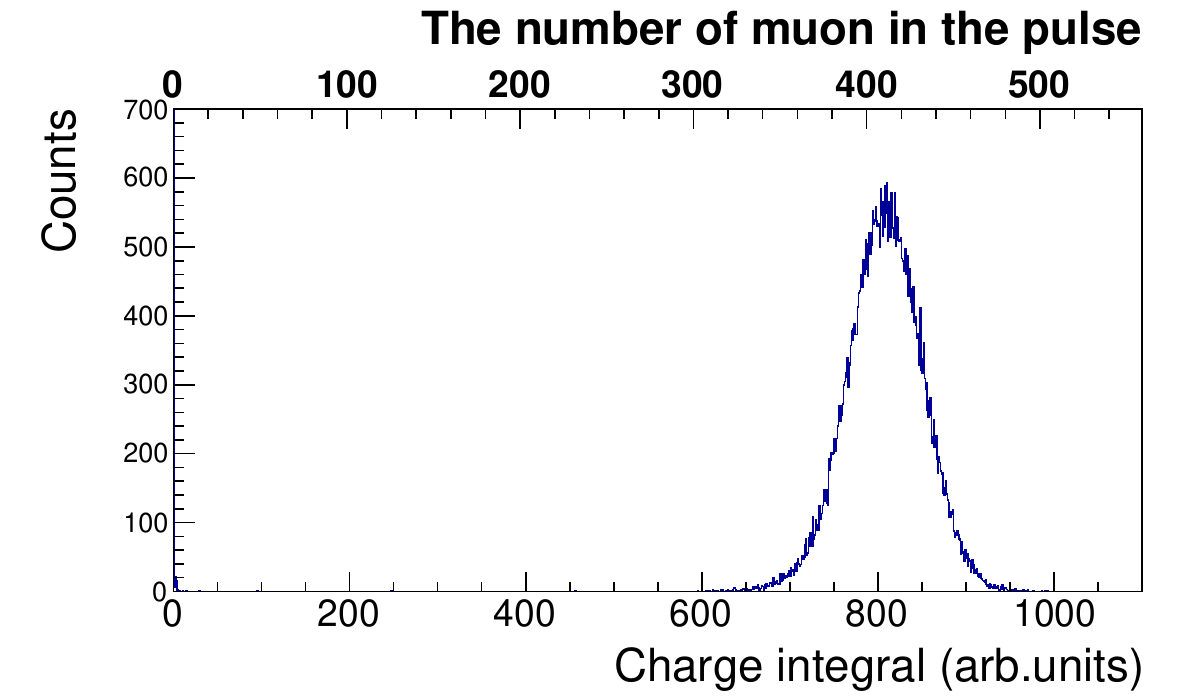}\\%1.0
    \small{(b) Muon beam with 38 MeV/c at J-PARC.}
    %\end{minipage}
    \caption{Spectra of the charge integral obtained by utilizing the plastic scintillator at (a) RAL and (b) J-PARC. The bottom horizontal axis represents the charge integral with arbitrary units, whereas the vertical axis represents the count. The upper horizontal axis represents the absolute muon number in the pulse.} 
    \label{fig:muon_plastic}
\end{figure}

The number of muons irradiating the target ($N_\mathrm{muon}$ in Eq.~(\ref{eq:Ncap})) was measured by utilizing a plastic scintillator. 
The energy deposit of the muons through the plastic scintillator is proportional to the number of muons irradiating the target. 
The raw signal from the PMT in one pulse obtained at J-PARC is shown in Fig.~\ref{fig:plastic_rawsignal}. 
The smaller signal detected prior to muon irradiation corresponds to electron contamination in the beam.
The signals from the plastic scintillator underwent waveform analysis. 
The charge integral of the signals was determined by integrating the signal height within the muon irradiation timing gate indicated by the red line. 
The charge integral gate for muon irradiation was carefully selected to ensure that all muons within the pulse were included while minimizing electron contamination. In the RAL experiment, the gate width was set at 150 ns with a 330 ns distance, whereas in the J-PARC experiment, it was at 200 ns with a 610 ns distance.
The spectrum of the charge integral obtained through waveform analysis is shown in Fig.~\ref{fig:muon_plastic}.

%\subsection{Absolute BR with countable muon beam at RAL}\label{subsec:analysis-RAL}
%As shown in Fig.~\ref{fig:muon_plastic} (a), the peaks of the charge integral of the plastic scintillator were clearly distinguished, 
%%separated and corresponded to the number of muons in one pulse at RAL.
%indicating the number of muons present in each pulse at the RAL.
%%The irradiation number of muons during the measurement was countable for all beam pulses with such a low-intensity muon beam.
%%The measurement process 
%A low-intensity beam allowed for the accurate quantification of the muon irradiation during the experiment. %, even with a low-intensity beam.
%%The muon beam intensity at RAL was tuned to be a countable intensity.
%The muon beam intensity at RAL was meticulously adjusted to ensure precise measurements. 
The muon beam intensity at RAL was meticulously adjusted to ensure precise quantification of the muon irradiation during the experiment. 
As shown in Fig.~\ref{fig:muon_plastic} (a), the peaks of the charge integral of the plastic scintillator were clearly distinguished, indicating the number of muons present in each pulse at the RAL.
The charge integral of the plastic scintillator was calibrated to the number of muons by fitting each peak in the histogram. %to correlate with the number of muons present. 
Utilizing the derived muon count in each pulse, the absolute production BR was calculated using the Eqs.~(\ref{eq:defineBR}), (\ref{eq:Ncap}), and (\ref{eq:Nprod}). 

%enrichedのdecompositionについて
In this experiment, the enriched silicon targets were in powder form, as shown in Fig.~\ref{fig:target_picture}. Some muons were not fully stopped in the target, $\epsilon_\mathrm{stop}\neq$1, for these powder targets.
Therefore, the determined BR values are relative and expressed as follows: $b_\mathrm{rel}~=~\epsilon_\mathrm{stop}~b$. 
The absolute BR of each enriched target can be determined by decomposing utilizing the result of the natural abundance silicon target as 
%\begin{equation}
%    P_\mathrm{cap}^\mathrm{natSi} b_i =  
%    A^{^{28}\mathrm{Si}} P_\mathrm{cap}^{^{28}\mathrm{Si}} b_i^{^{28}\mathrm{Si}} + 
%    A^{^{29}\mathrm{Si}} P_\mathrm{cap}^{^{29}\mathrm{Si}} b_i^{^{29}\mathrm{Si}} + 
%    A^{^{30}\mathrm{Si}} P_\mathrm{cap}^{^{30}\mathrm{Si}} b_i^{^{30}\mathrm{Si}}
%\end{equation}
\begin{equation}\label{eq:decomposition}
    P_\mathrm{cap}^\mathrm{^{nat}Si} b_i^\mathrm{^{nat}Si} =  
    \Sigma_j A^j P_\mathrm{cap}^j \frac{1}{\epsilon_\mathrm{stop}^j} b_{i~\mathrm{rel}}^j,
\end{equation}
where $i$ represents produced nuclei, $j$ corresponds to each isotope: $j=^{28,29,30}$Si, and $A^j$ represents the abundance of each isotope ($A^{^{28}\mathrm{Si}}=0.9223, A^{^{29}\mathrm{Si}}=0.0467, A^{^{30}\mathrm{Si}}=0.0310$).
$\epsilon_\mathrm{stop}^j$ were calculated using least square method with Eq.~(\ref{eq:decomposition}) and production BRs of four produced nuclei, $^{30}$Al, $^{29}$Al, $^{28}$Al, and $^{27}$Mg from $^\mathrm{nat}$Si and $^{28,29,30}$Si.
Furthermore, for the enriched targets $^{28,29,30}$Si, the impurity of the target listed in Table~\ref{tab:targets_info_measure} was simultaneously considered in the decomposition process.
The determined stopping rates of the various targets are listed in Table~\ref{tab:targets_info_measure}.
The discrepancy in the $\epsilon_\mathrm{stop}$ values between the two experiments may be attributed to variations in the momentum bite and the focus of the beam spot.
Through this analysis, the absolute BRs of aluminum and silicon isotopes were determined using measurable muon irradiation at RAL.

%\subsection{Calibration of plastic scintillator at J-PARC}
As the beam intensity becomes higher, the spectrum of the charge integral of the plastic scintillator becomes continuous, as shown in Fig.~\ref{fig:muon_plastic} (b). This figure displays the spectrum of the charge integral obtained with the high-intensity muon beam containing approximately 400 particles per pulse (ppp) at J-PARC.
The linearity, ranging from the countable muons less than approximately ten ppp to high-intensity muon beams, was not guaranteed owing to the potential quenching effect of the plastic scintillator by a high-density pulse beam.
Therefore, the calibration of the charge integral to the muon number was conducted in the J-PARC experiment. 

The absolute BR of the muon nuclear capture reaction of the $^{27}$Al target was utilized for this calibration.
Assuming linearity between the charge integral of the PMT signal and the number of muons passing through the plastic scintillator within the range of beam fluctuation, the number of muons can be expressed as $n_\mathrm{beam} = A \times Q$, where $Q$ represents the charge integral of the PMT signals, and $A$ represents the calibration parameter to convert $Q$ to the number of muons.
%Using this expression, 
The calibration parameter $A$ was determined through in-beam activation measurement of $^{27}$Al using Eqs.~(\ref{eq:defineBR}), (\ref{eq:Ncap}), and the absolute BR of the $^{27}$Al($\mu$, $\nu_\mu$)$^{27}$Mg reaction, which was 9.90(33)\% as determined in the RAL experiment. 
Notably, the effect of the electron contamination in the muon charge integral gate at J-PARC was included in the calibration parameter $Q$.
%otably, the calibration parameter accounted for the impact of electron contamination at J-PARC.

After the calibration process was conducted, 
%Following the calibration process, 
the absolute production BR was calculated using the same methodology as in the RAL experiment. 
This approach enabled a high statistical absolute BR measurement with the high-intensity pulsed muon beam at J-PARC.

\subsection{Uncertainty}
%stat., Ig, tau, stopping rate, Pcap, Nmu, eff\\

%%%%%%%%%%%%% D論 %%%%%%%%%%%%%%%%
%考慮された誤差の紹介ができれば良いことにする。コンパイルの仕方で誤差評価だいぶ面倒だが、それは記載しない。RALとMLFそれぞれの結果が出せるところまで。
%誤差の候補
The following items were considered as uncertainty: statistical uncertainty (including uncertainties of the background component when it was subtracted), uncertainty in $I_\gamma$, uncertainty in half-lifes ($T_{1/2}$), uncertainty in the peak detection efficiency of the germanium detectors ($\epsilon_\gamma$), systematic uncertainty in the irradiated muon number ($N_\mathrm{beam}$), uncertainty in the stopping rate ($\epsilon_\mathrm{stop}$), and uncertainty in the capture probability ($P_\mathrm{cap}$).

%%Ig
%RALとJ-PARC共通事項
The uncertainty in $I_\gamma$ and half-life value was cited from the ENSDF database, and the uncertainty in capture rates was determined based on the uncertainty in the referenced lifetime of muonic atoms~\cite{Mizuno2024}.
The uncertainties stemming from the lifetime and muon capture rate were negligible compared with other uncertainties.
For $I_\gamma$, the uncertainties in the relative intensity of each energy $\gamma$-rays ($\Delta I_\gamma^{\textit{rel}}$) and absolute intensity ($\Delta I_\gamma^{\textit{abs}}$) were treated separately, following the approach outlined in the Ref.~\cite{Niikura2024-ck}. $\Delta I_\gamma^{\textit{rel}}$ and $\Delta I_\gamma^{\textit{abs}}$ were sourced from the uncertainty of each $\gamma$-ray column and that of the normalization factor in the ENSDF database, respectively.
These uncertainties were distinctly labeled in Tables~\ref{tab:Result_27Al}--\ref{tab:Result_30Si}.
The uncertainty of the stopping rate was determined during the decomposition process.

%%Ge eff, muon count, stopping rate
%RALとJ-PARCで独立に扱う内容
In the RAL experiment, as the absolute BR value is determined using the yield of $\gamma$-ray and the measured number of muons, the uncertainty in the efficiency of the germanium detectors is considered as an absolute uncertainty. 
The uncertainty in the number of muon irradiations in the RAL experiment was negligible owing to the clear separation of peaks, as shown in Fig.~\ref{fig:muon_plastic} (a).
The peak detection efficiency of the germanium detectors had a systematic uncertainty of 3\%, originating from the uncertainty in the activities of standard $\gamma$-ray sources.
At J-PARC, only the relative values for each target were measured and normalized using the BR of $^{27}$Al($\mu^-$, $\nu_\mu$)$^{27}$Mg during the calibration process.
The uncertainty in the relative efficiency of the germanium detectors was deemed negligible; hence, the germanium efficiency uncertainty was not considered independently.
The systematic uncertainty in the number of muons in the J-PARC experiment was attributed to the calibration uncertainty of the plastic scintillator, primarily resulting from the uncertainty in the absolute BR of $^{27}$Mg produced from muonic $^{27}$Al. The uncertainty in the BR of $^{27}$Mg from $^{27}$Al was dominated by the germanium efficiency uncertainty at the RAL experiment. %mainly influenced by

\section{Result}\label{sec:result}
\begin{table*}[htbp]
    \centering \small
    \caption{Absolute production BR of each isotope produced from muonic $^{27}$Al. $P_\mathrm{cap}=60.96(4)\%$ was utilized as the muon capture probability. The parent nucleus of the decay (Nucleus), spin-parity of the decaying state (State), decay
mode (Decay), half-life ($T_{1/2}$), $\gamma$-ray energy ($E_\gamma$), $\gamma$-ray intensity ($I_\gamma$), BR determined using each $\gamma$-ray intensity from the RAL ($b_\gamma^{\tt{RAL}}$) and J-PARC ($b_\gamma^{\tt{J-PARC}}$) experiments are listed in the table. The weighted average values of BR of each experiment are listed below the result of each $\gamma$-ray peak, with the compiled BR for each decaying state ($b$) listed in the last column. The decay properties are obtained from ENSDF~\cite{Shamsuzzoha-Basunia2012-jo,Shamsuzzoha-Basunia2013-wi,Shamsuzzoha-Basunia2011-nq,Basunia2016-mn,Firestone2009-aw,Basunia2022-zw,Shamsuzzoha-Basunia2021-sj,Basunia2015-ab,Firestone2015-jd,Tilley1998-sf,Tilley1995-zp,Alburger1969-ic}.}
    \hspace*{-0.0cm}
    \begin{ruledtabular}
    \begin{tabular}{cccccclll}%\hline
    Nucleus   & State & Decay & $T_{1/2}$ & $E_\gamma$ (keV) & $I_\gamma$ (\%) & \multicolumn{1}{c}{$b_\gamma^{\tt{RAL}}$ (\%)} & \multicolumn{1}{c}{$b_\gamma^{\tt{J-PARC}}$ (\%)} & \multicolumn{1}{c}{$b$ (\%)} \\\hline
    $^{27}$Mg & 1/2$^+$ & $\beta^-$ & 9.458(12) min & 170.82 & 0.860(20) & - & -\footnotemark[4]\\
              &         &           &               & 843.76 & 71.800(20)& 9.97(17) & -\footnotemark[4]\\
              &         &           &               & 1014.52& 28.200(20)& 9.65(32) & -\footnotemark[4]\\
              %&         &           &               &        & average   & 9.90(15) &  \\
              &     & & & & $\Delta I_\gamma^{abs}/I_\gamma^{abs}$=0.11\% & 9.90(15)(30) & -\footnotemark[4] & 9.90(15)(30)\\
    $^{23}$Mg & 3/2$^+$ & $\epsilon$& 11.3046(45) s &  440.5 & 7.85(11)  & 3.02(41) & 3.42(26)\\
              &     & & & & $\Delta I_\gamma^{abs}/I_\gamma^{abs}$=incl. & 3.02(41)(11) & 3.42(26)(13) & $<$3.31(22)(12)\footnotemark[5]\\
    $^{26}$Na & 3$^+$   & $\beta^-$ & 1.07128(25) s& 1808.71& 99.08(50)\footnotemark[1] & 0.79(7) & 0.78(7)\\%0.787(71)&0.779(70)
              &     & & & & $\Delta I_\gamma^{abs}/I_\gamma^{abs}$=0.03\% & 0.79(7)(3) & 0.78(7)(3) & 0.783(50)(28)\\%0.787(73)(28) & 0.779(72)(28) & 0.783(50)(28)\\
    $^{25}$Na & 5/2$^+$ & $\beta^-$ & 59.1(6) s    & 389.710& 12.68(22)& 2.52(21) & 2.55(8)\\
              &         &           &               & 585.028& 13.00(18)& - & 2.39(10)\\
              &         &           &               & 974.742& 14.95(22)& 2.90(24) & 2.49(10)\\
              &         &           &               &1611.716& 9.48(14) & - & 2.57(15)\\
              %&         &           &               &        & average   & 2.68(21) & 2.50(8)\\>2.59, 2.46
              &     & & & & $\Delta I_\gamma^{abs}/I_\gamma^{abs}$=5.4\% & 2.68(26)(10) & 2.50(16)(9) & 2.52(16)(9)\footnotemark[6]\\
    $^{24}$Na & 4$^+$   & $\beta^-$ & 14.956(3) h  & 1368.625& 99.994(2)& 0.67(18) & -\\
              %&         &           &               & 2754.007& 99.8550(50)& - & \\
              &     & & & & $\Delta I_\gamma^{abs}/I_\gamma^{abs}$=incl. & 0.67(18)(2)$^{\ddag}$ & & 0.67(18)(2)\\%0.674(181)(24)& &0.674(189)(24)
              & 1$^+$   & IT        & 20.18(10) ms  & 472.2023& 99.95(5)\footnotemark[2]   & 0.937(37)& 1.005(27)\\%>0.679
              &     & & & & $\Delta I_\gamma^{abs}/I_\gamma^{abs}$=0.0\% & 0.937(37)(34)& 1.005(27)(37)& 0.980(22)(35)\footnotemark[7]\\
    $^{25}$Ne & 1/2$^+$ & $\beta^-$ & 602(8) ms     & 89.53  & 95.4(8)   & $<$ 0.074 & $<$0.035\\
    $^{24}$Ne & 0$^+$   & $\beta^-$ & 3.38(2) min   & 472.2023 & 100.0(2)& & $<$0.275\\
              &         &           &               & 874.420  & 7.9(2)  & $<$ 0.255 & & \\%$<$0.255\footnotemark[7]\\
              %&     & & & & $\Delta I_\gamma^{abs}/I_\gamma^{abs}$=0.0\% & & & $^{\#4}$\\
    $^{23}$Ne & 5/2$^+$ & $\beta^-$ & 37.25(10) s   & 440.5   & 32.9(10) & 0.72(10) & 0.83(6)\\%0.721(97) & 0.830(62)\\
              &     & & & & $\Delta I_\gamma^{abs}/I_\gamma^{abs}$=incl. & 0.72(10)(3) & 0.83(7)(3)& $<$0.80(6)(3)\footnotemark[5]\\%0.721(99)(26)&0.830(67)(30)&$<$0.798(58)(29)
    $^{21}$Na & 3/2$^+$ & $\beta^-$ & 22.49(4) s  & 350.725 & 5.07(13) & 1.7(7) & -\\%1.70(66)
              &     & & && $\Delta I_\gamma^{abs}/I_\gamma^{abs}$=incl. & 1.70(66)(6) & & $<$1.70(66)(6)\footnotemark[8]\\
    $^{21}$F  & 5/2$^+$ & $\beta^-$ & 4.158(20) s & 350.725 & 89.6(18)\footnotemark[1] & 0.096(37) & - \\%Ig:89.55(180)
              &     & & && $\Delta I_\gamma^{abs}/I_\gamma^{abs}$=3.32\% & 0.096(37)(3) & & $<$0.096(37)(3)\footnotemark[8]\\%0.0957(372)(34)&&0.0957(372)(34)
    $^{20}$Na & 2$^+$   & $\epsilon$& 447.9(23) ms  & 1633.602 & 79.3(11)& - & 0.128(26)\\
              &     & & && $\Delta I_\gamma^{abs}/I_\gamma^{abs}$=incl. & & 0.128(26)(5) & $<$0.128(26)(5)\footnotemark[9]\\%0.1279(260)(47)
    $^{20}$F  & 2$^+$   & $\beta^-$ & 11.163(8) s   & 1633.602 & 99.9995(10)\footnotemark[3] & - & 0.102(21)\\ 
              &     & & && $\Delta I_\gamma^{abs}/I_\gamma^{abs}$=incl. & & 0.102(21)(4) & $<$0.102(21)(4)\footnotemark[9]\\%0.1019(207)(37)
    $^{19}$O  & 5/2$^+$ & $\beta^-$ & 26.88(5) s    %& 197.142 & 95.9(21) & ** & \\
                                                    & 1356.843 & 50.4(11)& $<$ 0.037 & $<$ 0.073\\
    %\hline
    \end{tabular}
    \end{ruledtabular}
    \footnotetext[0]{* The uncertainties in the $b_\gamma^{\tt{RAL}}$ and $b_\gamma^{\tt{J-PARC}}$ columns in each $\gamma$-ray row include the statistical uncertainty from the $\gamma$-ray peak count and uncertainty from $I_\gamma^{rel}$. The first set of parentheses in the weighted average row indicates the relative uncertainty, encompassing the statistical uncertainty, uncertainties from $I_\gamma$ (relative and absolute), and the lifetime of the nuclei. The second set of parentheses includes the uncertainties of the efficiency of the germanium detector, muon capture rate, and number of muons counted with the plastic scintillator, excluding the relative uncertainty. The uncertainties from the lifetime and muon capture rate were negligible compared with other uncertainties.}
    \footnotetext[0]{** $I_\gamma$ uncertainty includes both $\Delta I_\gamma^{rel}$ and $\Delta I_\gamma^{abs}$ for the nuclei denoted as $\Delta I_\gamma^{abs}/I_\gamma^{abs}$=incl.}
    \footnotetext[0]{$\ddag$ Decay component from the mother nuclei was subtracted.} 
    \footnotetext[1]{$\Delta I_\gamma^{rel}$ of these peaks are not provided in the ENSDF database and were estimated from other $\Delta I_\gamma$.}
    \footnotetext[2]{$\Delta I_\gamma^{rel}$ is not provided in the ENSDF database. 
    The uncertainty originating from the probability of beta decay from $^{24m}$Na (0.05\%) was estimated as 0.05\%.}
    %The uncertainty of $I_\gamma$ is 0.05\% even if the probability of beta decay from $^{24m}$Na has 100\% uncertainty, which is negligible compared with the statistical uncertainty.}
    \footnotetext[3]{$\Delta I_\gamma^{rel}$ of this peak is not provided in the ENSDF database and was estimated from the probability of the beta decay to the ground state.}
    \footnotetext[4]{BR of $^{27}$Mg was utilized as the calibration reference in the analysis of the J-PARC experiment.}
    \footnotetext[5]{A peak at 440.5 keV originated from $^{23}$Mg and $^{23}$Ne. The upper limit of these isotopes could not be restricted by the other energy peaks. The production BR of these isotopes both represent the upper limit.}
    \footnotetext[6]{The lower limit of $^{25}$Na is 2.46\% calculated from the upper limit of $^{25}$Ne.}
    \footnotetext[7]{A peak at 472.2 keV originated from $^{24\textit{m}}$Na and $^{24}$Ne. The upper limit of $^{24}$Ne is estimated from the statistics during the decay measurement after stopping the beam irradiation, with the upper limit of the non-detected peak at 874.41 keV. The lower limit of $^{24\textit{m}}$Na is 0.679\% calculated from the upper limit of $^{24}$Ne.}
    \footnotetext[8]{ A peak at 350.7 keV could originate from $^{21}$Na and $^{21}$F. The background component was subtracted since the peak was also observed in the empty target measurement.}
    \footnotetext[9]{A peak at 1633.6 keV could originate from $^{20}$Na and $^{20}$F. The upper limit of these isotopes could not be restricted by the other energy peaks. The production BR of these isotopes both represent the upper limit.}
    \label{tab:Result_27Al}
\end{table*}
%\end{comment}
\begin{table*}[htbp]
    \centering \small
    \caption{Absolute production BR of each isotope produced from muonic natural abundance Si. $P_\mathrm{cap}=66.07(5)\%$ was utilized as the muon capture probability. The same notations as those in Table~\ref{tab:Result_27Al} were utilized.}
    \hspace*{-0cm}
    \begin{ruledtabular}
    \begin{tabular}{cccccclll}%\hline
    %Nucleus   & State & Decay & $T_{1/2}$ & $E_\gamma$ (keV) & $I_\gamma$ (\%) & $b_\gamma^{\tt{RAL}}$ (\%) & $b_\gamma^{\tt{J-PARC}}$ & $b$ (\%) \\\hline
    Nucleus   & State & Decay & $T_{1/2}$ & $E_\gamma$ (keV) & $I_\gamma$ (\%) & \multicolumn{1}{c}{$b_\gamma^{\tt{RAL}}$ (\%)} & \multicolumn{1}{c}{$b_\gamma^{\tt{J-PARC}}$ (\%)} & \multicolumn{1}{c}{$b$ (\%)}\\\hline
    $^{30}$Al & 3$^+$ & $\beta^-$ & 3.62(6) s   & 3498.33 & 32.6(14)\footnotemark[1] & - & 0.396(27) \\
              &     & & & & $\Delta I_\gamma^{abs}/I_\gamma^{abs}$=incl. & & 0.396(28)(15) & 0.396(28)(15)\\
    $^{29}$Al & 5/2$^+$ & $\beta^-$ & 6.56(6) min & 1152.57  & 1.031(27)     & -        & 2.80(50) \\ 
              &         &           &             & 1273.36  & 91.3(9)\footnotemark[2] & -        & 2.56(6) \\%RAL:2.31(15)%Ig:91.26(90)
              &         &           &             & 2028.09  & 3.514(27)     & -        & 2.62(18) \\
              &         &           &             & 2425.73  & 5.23(6)     & -        & 2.78(15) \\%Ig:5.229(55)
              %&         &           &             &          &      average  &  & 2.59(5)\\
              &     & & & & $\Delta I_\gamma^{abs}/I_\gamma^{abs}$=0.07\% & - & 2.57(12)(9)$^\ddag$ & 2.57(12)(9)\\%$\#1$\\%RAL:2.29(15)(17)$\ddag$
    $^{28}$Al & 3$^+$ & $\beta^-$ & 2.245(2) min & 1778.987 & 100(0) & 19.69(45) & 20.37(22)\\
              &     & & & & $\Delta I_\gamma^{abs}/I_\gamma^{abs}$=0.00\% & 19.67(45)(59)$^\ddag$ & 20.33(22)(75)$^\ddag$ & 20.08(28)(60)\\
    $^{29}$Mg & 3/2$^+$ & $\beta^-$ & 1.30(12) s & & & - & - & 0.019(7)\footnotemark[6]\\%0.0189(72)
    $^{28}$Mg & 0$^+$   & $\beta^-$ & 20.915(9) h & & & - & - & 0.125(27)\footnotemark[6]\\
    $^{27}$Mg & 1/2$^+$ & $\beta^-$ & 9.458(12) min & 843.76 & 71.800(20) & 2.95(17)  & 3.01(24)\\
              &         &           &               & 1014.52 & 28.200(20) & 3.03(37) & 2.75(7)\\
              %&         &           &               &        & average & 2.98(16) & 2.90(7)\\
              &     & & & & $\Delta I_\gamma^{abs}/I_\gamma^{abs}$=0.11\% & 2.97(16)(9) & 2.89(19)(11)& 2.94(13)(9)\footnotemark[7]\\
    %$^{28}$Na & 1$^+$   & $\beta^-$ & 30.5(4) ms  & 1473.5 & 37(5) & - & $<$ \\
    $^{26}$Na & 3$^+$ & $\beta^-$ & 1.07128(25) s & 1808.71 & 99.08(50)\footnotemark[2] & $<$0.388 & \\
              &         &           &               & 1128.89 & 5.88(3) &  & $<$0.287  \\
    $^{25}$Na & 5/2$^+$ & $\beta^-$ & 59.1(6) s & 389.710 & 12.68(22) & - & 0.207(31)\footnotemark[5]\\ 
              &         &           &           & 585.028 & 13.00(18) & - & 0.355(40)\\
              &     & & && $\Delta I_\gamma^{abs}/I_\gamma^{abs}$=5.4\% & & 0.355(44)(13) & 0.355(44)(13)\footnotemark[8]\\
    $^{24}$Na & 4$^+$   & $\beta^-$ & 14.956(3) h  & 1368.625& 99.994(2)& - & - & 0.52(10)\footnotemark[6]\\%0.521(98)
              & 1$^+$   & IT        & 20.18(10) ms  & 472.2023& 99.95(5)\footnotemark[3]   & 1.21(10)& 1.158(27)\\%1.213(96)
              &     & & & & $\Delta I_\gamma^{abs}/I_\gamma^{abs}$=0.00\% &1.21(10)(4) & 1.158(27)(42)& 1.164(32)(35)\footnotemark[9]\\%1.213(96)(36)
    $^{25}$Ne & 1/2$^+$ & $\beta^-$ & 602(8) ms     & 89.53  & 95.4(8)   & $<$ 0.48 & $<$0.074\\
    $^{24}$Ne & 0$^+$   & $\beta^-$ & 3.38(2) min   & 472.2023 & 100.0(2)&  & $<$0.173& \\%$<$0.173\footnotemark[9]\\
              %&         &           &               & 874.420  & 7.9(2)  & - & $<$ 1.22\\
    $^{20}$Na & 2$^+$   & $\epsilon$& 447.9(23) ms  & 1633.602 & 79.3(11)& - & 0.119(9)\\
              &     & & & & $\Delta I_\gamma^{abs}/I_\gamma^{abs}$=incl. & & 0.119(9)(4) & $<$0.119(9)(4)\footnotemark[10] \\
    $^{20}$F  & 2$^+$   & $\beta^-$ & 11.163(8) s   & 1633.602 & 99.9995(10)\footnotemark[4] & - &0.094(7) \\
              &     & & & & $\Delta I_\gamma^{abs}/I_\gamma^{abs}$=incl. & & 0.094(7)(3) & $<$0.094(7)(3)\footnotemark[10]\\
    %\hline
    \end{tabular}
    \end{ruledtabular}
    \footnotetext[0]{* The notation of the uncertainties is the same as that in Table~\ref{tab:Result_27Al}.}
    \footnotetext[0]{** $I_\gamma$ uncertainty includes both $\Delta I_\gamma^{rel}$ and $\Delta I_\gamma^{abs}$ for the nuclei denoted as $\Delta I_\gamma^{abs}/I_\gamma^{abs}$=incl.}
    \footnotetext[0]{$\ddag$: The decay component from the mother nuclei was subtracted.}
    \footnotetext[1]{Absolute $I_\gamma$ was not provided in the ENSDF and was calculated from the intensity of beta decay to each energy level and the probability of $\gamma$-ray transition from each state~\cite{Endt1974-tz, Klotz1974-mi, Alburger1974-ie}. The uncertainties encompassed absolute uncertainty.} 
    \footnotetext[2]{$\Delta I_\gamma^{rel}$ of these peaks are not provided in the ENSDF database and were estimated from other $\Delta I_\gamma$.}
    \footnotetext[3]{$\Delta I_\gamma^{rel}$ is not provided in the ENSDF database. The uncertainty originating from the probability of beta decay from $^{24m}$Na (0.05\%) was estimated as 0.05\%.}
    %The uncertainty of $I_\gamma$ is 0.05\% even if the probability of beta decay from $^{24m}$Na has 100\% uncertainty, which is negligible compared with the statistical uncertainty.}
    \footnotetext[4]{$\Delta I_\gamma^{rel}$ of this peak is not provided in the ENSDF database and was estimated from the probability of the beta decay to the ground state. }
    \footnotetext[5]{The BR determined with 389.7 keV was minimal compared with the result of the other peaks obtained at J-PARC. This result was excluded from the weighted average value.}
    \footnotetext[6]{The BRs of these nuclei were calculated from the result of enriched silicon targets.}
    \footnotetext[7]{The lower limit of $^{27}$Mg is 2.93\% calculated from the upper limit of $^{27}$Na (0.01\%), calculated using the result of $^{29,30}$Si.}
    \footnotetext[8]{The lower limit of $^{25}$Na is 0.280\% calculated from the upper limit of $^{25}$Ne.}
    \footnotetext[9]{The peak at 472.2 keV originated from $^{24m}$Na and $^{24}$Ne. The upper limit of $^{24}$Ne was estimated from the statistics obtained during the decay measurement after stopping the beam irradiation. The lower limit of $^{24\textit{m}}$Na was 0.998\% calculated from the upper limit of $^{24}$Ne.}
    \footnotetext[10]{A peak at 1633.6 keV could originate from $^{20}$Na and $^{20}$F. The upper limit of these isotopes could not be limited by the other energy peaks. The production BR of these isotopes both represent the upper limits.}
    \label{tab:Result_natSi}
\end{table*}
\begin{table*}[htbp]
    \centering \small
    \caption{Absolute production BR of each isotope produced from muonic $^{28}$Si. $P_\mathrm{cap}=66.41(23)\%$ was utilzed. The stopping rates were 0.407(14)\% and 0.298(9)\% in the RAL and J-PARC experiments, respectively. The same notations as those in Table~\ref{tab:Result_27Al} were utilized.}
    \hspace*{-0cm}
    \begin{ruledtabular}
    \begin{tabular}{cccccclll}%\hline
    %Nucleus   & State & Decay & $T_{1/2}$ & $E_\gamma$ (keV) & $I_\gamma$ (\%) & $b_\gamma^{\tt{RAL}}$ (\%) & $b_\gamma^{\tt{J-PARC}}$ & $b$ (\%) \\\hline
    Nucleus   & State & Decay & $T_{1/2}$ & $E_\gamma$ (keV) & $I_\gamma$ (\%) & \multicolumn{1}{c}{$b_\gamma^{\tt{RAL}}$ (\%)} & \multicolumn{1}{c}{$b_\gamma^{\tt{J-PARC}}$ (\%)} & \multicolumn{1}{c}{$b$ (\%)}\\\hline
    $^{28}$Al & 3$^+$ & $\beta^-$ & 2.245(2) min & 1778.987 & 100 & 18.94(19) & 18.93(10)\\
              &     & & && $\Delta I_\gamma^{abs}/I_\gamma^{abs}$=0.0\% & 18.92(20)(94)$^\dag$& 18.91(10)(92)$^\dag$& 18.92(10)(91) \\
    $^{27}$Mg & 1/2$^+$ & $\beta^-$ & 9.458(12) min & 843.76  & 71.800(20) & 2.82(6)  & 2.93(10)\\%2.818(65)  & 2.932(97)
              &         &           &               & 1014.52 & 28.200(20) & 3.06(14) & 2.84(9)\\%3.061(142) & 2.838(90)
              %&         &           &               &        & average & 2.847(59) &  2.872(66)\\
              &    & & && $\Delta I_\gamma^{abs}/I_\gamma^{abs}$=0.11\% & 2.86(6)(14)$^\dag$& 2.88(7)(14)$^\dag$ & 2.869(45)(137) \\%2.860(59)(143)& 2.882(66)(140)
    $^{26}$Na & 3$^+$ & $\beta^-$ & 1.07128(25) s & 1808.71 & 99.08(50)\footnotemark[1] & $<$0.185 &  \\
              &         &           &               & 1128.89 & 5.88(3) &  & $<$1.17  \\
    $^{25}$Na & 5/2$^+$ & $\beta^-$ & 59.1(6) s & 585.028 & 13.00(18) & 0.24(17) & - \\%0.244(171)
              %&         &           &               &        & average & 0.243(170)>0.162 & - \\
              &     & & && $\Delta I_\gamma^{abs}/I_\gamma^{abs}$=5.4\% & 0.24(17)(1)$^\dag$ & & 0.24(17)(1)\footnotemark[4]\\%0.243(171)(12)
    $^{24}$Na & 4$^+$   & $\beta^-$ & 14.956(3) h  & 1368.625& 99.994(2)& 0.53(10) & - \\%0.526(104)
              &     & & && $\Delta I_\gamma^{abs}/I_\gamma^{abs}$=incl. & 0.53(10)(3)$^\dag$$^\ddag$& & 0.53(10)(3) \\%0.526(104)(26)
              & 1$^+$   & IT        & 20.18(10) ms  & 472.2023& 99.95(5)\footnotemark[2]   & 1.182(36) & 1.134(48)\\
              %&         &           &               &        & average & >0.988 &  \\
              &     & & && $\Delta I_\gamma^{abs}/I_\gamma^{abs}$=0.0\% & 1.182(36)(59)$^\dag$& 1.134(49)(55)$^\dag$& 1.161(29)(56)\footnotemark[5] \\
    $^{25}$Ne & 1/2$^+$ & $\beta^-$ & 602(8) ms     & 89.53  & 95.4(8)   & $<$0.244 & $<$0.172 &$<$0.080\footnotemark[6] \\
    $^{24}$Ne & 0$^+$   & $\beta^-$ & 3.38(2) min   & 472.2023 & 100.0(2)&  & $<$0.563 & $<$0.187\footnotemark[5]\footnotemark[6]\\
              &         &           &               & 874.420  & 7.9(2)  & $<$ 0.568 &  \\
    $^{20}$Na & 2$^+$   & $\epsilon$& 447.9(23) ms  & 1633.602 & 79.3(11)& 0.144(47) & 0.148(32)\\
              &     & & && $\Delta I_\gamma^{abs}/I_\gamma^{abs}$=incl. & 0.144(47)(7)$^\dag$&0.148(32)(7)$^\dag$ & $<$0.147(26)(7)\footnotemark[7]\\
    $^{20}$F  & 2$^+$   & $\beta^-$ & 11.163(8) s   & 1633.602 & 99.9995(10)\footnotemark[3] & 0.114(37) & 0.117(24)\\
              &     & & && $\Delta I_\gamma^{abs}/I_\gamma^{abs}$=incl. & 0.114(37)(6)$^\dag$& 0.117(24)(6)$^\dag$& $<$0.116(20)(6)\footnotemark[7]\\
    %\hline
    \end{tabular}
    \end{ruledtabular}
    \footnotetext[0]{* The notation of the uncertainties is the same as that in Table~\ref{tab:Result_27Al}. The uncertainty of the stopping rate was included in the uncertainty in the second parenthesis.}
    \footnotetext[0]{** $I_\gamma$ uncertainty includes both $\Delta I_\gamma^{rel}$ and $\Delta I_\gamma^{abs}$ for the nuclei denoted as $\Delta I_\gamma^{abs}/I_\gamma^{abs}$=incl.}
    \footnotetext[0]{$\dag$: The impurity of the enriched target (Table~\ref{tab:targets_info_measure}) was taken into account with the decomposition process.}
    \footnotetext[0]{$\ddag$: The decay component from the mother nuclei was subtracted.}
    \footnotetext[1]{$\Delta I_\gamma^{rel}$ of the peak is not provided in the ENSDF database and was estimated from other $\Delta I_\gamma$.}
    \footnotetext[2]{$\Delta I_\gamma^{rel}$ is not provided in the ENSDF database. The uncertainty originating from the probability of beta decay from $^{24m}$Na (0.05\%) was estimated as 0.05\%.}
    %The uncertainty of $I_\gamma$ is 0.05\% even if the probability of beta decay from $^{24m}$Na has 100\% uncertainty, which is negligible compared with the statistical uncertainty.}
    \footnotetext[3]{$\Delta I_\gamma^{rel}$ of this peak is not provided in the ENSDF database and was estimated from the probability of the beta decay to the ground state. }
    \footnotetext[4]{The lower limit of $^{25}$Na is 0.163\% calculated from the upper limit of $^{25}$Ne.}
    \footnotetext[5]{A peak at 472.2 keV originated from $^{24\textit{m}}$Na and $^{24}$Ne. The upper limit of $^{24}$Ne was estimated from the statistics during the decay measurement after stopping the beam irradiation, with the upper limit of the non-detected peak at 874.41 keV. The lower limit of $^{24\textit{m}}$Na is 0.993\% calculated from the upper limit of $^{24}$Ne.}
    \footnotetext[6]{The upper limit of $^{24, 25}$Ne was evaluated from the result of natural abundance silicon.}
    %\footnotetext[7]{The upper limit of $^{24}$Ne was evaluated from the result of natural abundance silicon.}
    \footnotetext[7]{A peak at 1633.6 keV could originate from $^{20}$Na and $^{20}$F. The upper limit of these isotopes could not be restricted by the other energy peaks. The production BR of these isotopes both represent the upper limit.}
    \label{tab:Result_28Si}
\end{table*}
%\squeezetable
\begin{table*}[htbp]
    \centering \small
    \caption{Absolute production BR of each isotope produced from muonic $^{29}$Si. $P_\mathrm{cap}=64.14(16)\%$ was utilized. The stopping rates were 0.652(169)\% and 0.260(63)\% in the RAL and J-PARC experiments, respectively. The same notations as those in Table~\ref{tab:Result_27Al} were utilized.}
    \hspace*{-0cm}
    \begin{ruledtabular}
    \begin{tabular}{cccccclll}%\hline
    %Nucleus   & State & Decay & $T_{1/2}$ & $E_\gamma$ (keV) & $I_\gamma$ (\%) & $b_\gamma^{\tt{RAL}}$ (\%) & $b_\gamma^{\tt{J-PARC}}$ & $b$ (\%) \\\hline
    Nucleus   & State & Decay & $T_{1/2}$ & $E_\gamma$ (keV) & $I_\gamma$ (\%) & \multicolumn{1}{c}{$b_\gamma^{\tt{RAL}}$ (\%)} & \multicolumn{1}{c}{$b_\gamma^{\tt{J-PARC}}$ (\%)} & \multicolumn{1}{c}{$b$ (\%)}\\\hline
    $^{29}$Al & 5/2$^+$ & $\beta^-$ & 6.56(6) min & 1152.57  & 1.031(27)     & 20.(6) & 15.2(24)\\%20.19(587) & 15.17(244)
              &         &           &             & 2028.09  & 3.514(27)     & 13.8(14) & 15.7(8)\\%13.79(141) & 15.73(84)
              &         &           &             & 2425.73  & 5.23(6)     & 15.3(11) & 15.6(7)\\%15.28(110) & 15.60(71)%Ig:5.229(55)
              %&         &           &             &          &     average   & 15.43(89) & 16.10(54)\\
              &     & & && $\Delta I_\gamma^{abs}/I_\gamma^{abs}$=0.07\% & 14.6(9)(39)$^\dag$ & 15.4(6)(38)$^\dag$ & 15.01(48)(373)\\%14.60(88)(391)$^\dag$ & 15.41(55)(384)
    $^{28}$Al & 3$^+$ & $\beta^-$ & 2.245(2) min & 1778.987 & 100(0) & 49.10(33) & 48.22(16)\\
              &     & & && $\Delta I_\gamma^{abs}/I_\gamma^{abs}$=0.00\% & 49.09(34)(1288)$^\dag$$^\ddag$ & 48.34(17)(1180)$^\dag$$^\ddag$ & 48.68(18)(1188) \\
    $^{28}$Mg & 0$^+$   & $\beta^-$ & 20.915(9) h & 30.6383 & 89.0(45)\footnotemark[1] & 1.56(41) & - \\
              &     & & && $\Delta I_\gamma^{abs}/I_\gamma^{abs}$=incl. & 1.56(41)(41)$^\dag$ & & 1.56(41)(41)\\
    $^{27}$Mg & 1/2$^+$ & $\beta^-$ & 9.458(12) min & 843.76  & 71.800(20) & 2.93(9)  & 2.90(11)\\%2.927(89)&2.900(115)
              &         &           &               & 1014.52 & 28.200(20) & 2.83(20) & 2.90(19)\\%2.829(202)&2.901(185)
              %&         &           &               &        & average & 3.0288(85) >2.87 & 2.986(100)>2.83\\
              &     & & && $\Delta I_\gamma^{abs}/I_\gamma^{abs}$=0.11\% & 2.92(8)(77)$^\dag$ & 2.91(10)(71)$^\dag$ & 2.91(6)(71)\footnotemark[5]\\%2.918(82)(767)&2.907(98)(712)&2.912(64)(713)
    $^{27}$Na & 5/2$^+$ & $\beta^-$ & 301(6) ms & 984.66 & 87.4(6) & $<$0.157 & $<$0.156\\
    $^{26}$Na & 3$^+$   & $\beta^-$ & 1.07128(25) s & 1808.71 & 99.08(50)\footnotemark[2] & $<$0.115 &  \\
              &         &           &               & 1128.89 & 5.88(3) &  & $<$1.89  \\
    $^{25}$Na & 5/2$^+$ & $\beta^-$ & 59.1(6) s & 389.710 & 12.68(22) & - & 0.652(130)\footnotemark[4]\\
              &         &           &           & 585.028 & 13.00(18) & 1.04(25) & 1.15(18)\\
              &         &           &           & 974.742 & 14.95(22) & 0.98(27) & 1.22(24)\\
              %&         &           &           &         & average   & 1.05(28) >0.88 &1.21(15)>1.04\\
              &     & & && $\Delta I_\gamma^{abs}/I_\gamma^{abs}$=5.4\% & 1.01(28)(27)$^\dag$ & 1.18(16)(29)$^\dag$& 1.11(14)(27)\footnotemark[6]\\
    $^{24}$Na & 4$^+$   & $\beta^-$ & 14.956(3) h  & 1368.625& 99.994(2)& 0.56(22) & -\\
              &     & & && $\Delta I_\gamma^{abs}/I_\gamma^{abs}$=incl. & 0.56(22)(15)$^\dag$$^\ddag$ & & 0.56(22)(15)\\
              & 1$^+$   & IT        & 20.18(10) ms  & 472.2023& 99.95(5)\footnotemark[3]   & 0.824(45)& 0.81(6)\\%0.808(56)
              &     & & && $\Delta I_\gamma^{abs}/I_\gamma^{abs}$=0.00\% & 0.825(45)(218)$^\dag$ &0.81(6)(20)$^\dag$ & 0.816(35)(200)\footnotemark[7]\\%0.809(56)(198)
    $^{25}$Ne & 1/2$^+$ & $\beta^-$ & 602(8) ms     & 89.53  & 95.4(8)   & $<$0.166 & $<$0.217\\
    $^{24}$Ne & 0$^+$   & $\beta^-$ & 3.38(2) min   & 472.2023 & 100.0(2)&  & $<$0.647 & \\%$<$0.667 \footnotemark[7]\\%0.846(46)
              %&     & & && $\Delta I_\gamma^{abs}/I_\gamma^{abs}$=0.00\% & & & $<$0.667 \footnotemark[7]\\
    $^{21}$Na & 3/2$^+$ & $\beta^-$ & 22.49(4) s  & 350.725 & 5.07(13) & 2.6(10) & -\\%2.60(102)
              &     & & && $\Delta I_\gamma^{abs}/I_\gamma^{abs}$=incl. & 2.6(10)(7)$^\dag$ & & $<$2.6(10)(7)\footnotemark[8]\\%2.62(103)(69)
    $^{21}$F  & 5/2$^+$ & $\beta^-$ & 4.158(20) s & 350.725 & 89.6(18)\footnotemark[2] & 0.16(6) & - \\%0.156(57)%Ig:89.55(180)
              %&         &           &             & 1395.13 & 15.4(3) & $<$ & $<$\\
              &     & & && $\Delta I_\gamma^{abs}/I_\gamma^{abs}$=3.32\% & 0.16(6)(4)$^\dag$ & & $<$0.16(6)(4)\footnotemark[8]\\%0.157(58)(41)
    %\hline
    \end{tabular}
    \end{ruledtabular}
    \footnotetext[0]{* The notation of the uncertainties is the same as that in Table~\ref{tab:Result_28Si}.}
    \footnotetext[0]{** $I_\gamma$ uncertainty includes both $\Delta I_\gamma^{rel}$ and $\Delta I_\gamma^{abs}$ for the nuclei denoted as $\Delta I_\gamma^{abs}/I_\gamma^{abs}$=incl.}
    %\footnotetext[0]{$\dag$: Contamination of the other isotope (Table.~\ref{tab:targets_info_measure}) in the enriched target was evaluated.}
    \footnotetext[0]{$\dag$: The impurity of the enriched target (Table~\ref{tab:targets_info_measure}) was taken into account with the decomposition process.}
    \footnotetext[0]{$\ddag$: The decay component from the mother nuclei was subtracted.}
    \footnotetext[1]{Absolute $I_\gamma$ was provided based on the estimation, and the uncertainty was not provided in the ENSDF. 5\% uncertainty was added~\cite{Alburger1969-ic}.}
    \footnotetext[2]{$\Delta I_\gamma^{rel}$ of these peaks are not provided in the ENSDF database and were estimated from other $\Delta I_\gamma$.}
    \footnotetext[3]{$\Delta I_\gamma^{rel}$ is not provided in the ENSDF database. The uncertainty originating from the probability of beta decay from $^{24m}$Na (0.05\%) was estimated as 0.05\%.}
    %The uncertainty of $I_\gamma$ is 0.05\% even if the probability of beta decay from $^{24m}$Na has 100\% uncertainty, which is negligible compared with the statistical uncertainty.}
    \footnotetext[4]{The BR determined with 389.7 keV was minimal compared with the result of the other peaks obtained at J-PARC. This result was excluded from the weighted average value.}
    \footnotetext[5]{The lower limit of $^{27}$Mg is 2.75\% calculated from the upper limit of $^{27}$Na.}
    \footnotetext[6]{The lower limit of $^{25}$Na is 1.01\% calculated from the upper limit of $^{25}$Ne.}
    \footnotetext[7]{A peak at 472.2 keV originated from $^{24\textit{m}}$Na and $^{24}$Ne. The upper limit of $^{24}$Ne was estimated from the statistics during the decay measurement after stopping the beam irradiation. The lower limit of the BR of $^{24\textit{m}}$Na is 0.160\% calculated from the upper limit of $^{24}$Ne.} 
    \footnotetext[8]{A peak at 350.7 keV could originate from $^{21}$Na and $^{21}$F. The background component was subtracted since the peak was also observed in the empty target measurement.}
    \label{tab:Result_29Si}
\end{table*}
%\begingroup
%\squeezetable
\begin{table*}[htbp]
    \centering \small
    \vspace*{-0.5cm}
    \caption{Absolute production BR of each isotope produced from muonic $^{30}$Si. $P_\mathrm{cap}=61.21(12)\%$ was utilized. The stopping rates were 0.349(26)\% and 0.261(19)\% in the RAL and J-PARC experiments, respectively. The same notations as those in Table~\ref{tab:Result_27Al} were utilized.}
    %\hspace*{-0cm}
    \begin{ruledtabular}
    \begin{tabular}{cccccclll}%\hline
    %Nucleus   & State & Decay & $T_{1/2}$ & $E_\gamma$ (keV) & $I_\gamma$ (\%) & $b_\gamma^{\tt{RAL}}$ (\%) & $b_\gamma^{\tt{J-PARC}}$ & $b$ (\%) \\\hline
    Nucleus   & State & Decay & $T_{1/2}$ & $E_\gamma$ (keV) & $I_\gamma$ (\%) & \multicolumn{1}{c}{$b_\gamma^{\tt{RAL}}$ (\%)} & \multicolumn{1}{c}{$b_\gamma^{\tt{J-PARC}}$ (\%)} & \multicolumn{1}{c}{$b$ (\%)} \\\hline
    $^{30}$Al & 3$^+$ & $\beta^-$ & 3.62(6) s   & 1263.13 & 40.1(15)\footnotemark[1] & 14.9(6) & 14.19(40) \\%14.85(58)
              &       &           &             & 1311.80 & 2.54(23)\footnotemark[1] & -         & 14.7(19)\\%14.67(193)
              &       &           &             & 1332.48 & 0.94(14)\footnotemark[1] & 12.5(48)&  -        \\%12.48(476)
              &       &           &             & 2235.23 & 65.1(22)\footnotemark[1] & 13.53(49) & 13.84(49) \\ 
              &       &           &             & 2595.39 & 5.77(22)\footnotemark[1] & -         & 13.4(10)\\%13.34(101)
              &       &           &             & 3498.33 & 32.6(14)\footnotemark[1] & -         & 13.9(7) \\ %13.89(67)
              %&    & & & &                                      average & 13.94(37) & 13.90(27)  \\
        & & & & & $\Delta I_\gamma^{abs}/I_\gamma^{abs}$=incl.  & 14.12(44)(115)$^\dag$ & 14.04(36)(113)$^\dag$ & 14.08(32)(113)\\
    $^{29}$Al & 5/2$^+$ & $\beta^-$ & 6.56(6) min & 1152.57  & 1.031(27)     & 68.(8) & 58.2(32) \\%67.55(777) & 58.16(321)
              &         &           &             & 1273.36  & 91.3(9)\footnotemark[2] & 65.8(7)  & 64.23(48)\\%65.71(69)%Ig:91.26(90)
              &         &           &             & 2028.09  & 3.514(27)     & 62.5(19) & 68.8(17) \\%62.47(194) & 68.75(170)
              &         &           &             & 2425.73  & 5.23(6)     & 68.3(16) & 67.3(15) \\%68.23(162) & 67.22(151)%Ig: 5.229(55)
              %&         &           &             &         &     average   & 65.17(59) & 64.26(44)  \\
              & & & & & $\Delta I_\gamma^{abs}/I_\gamma^{abs}$=0.07\% & 65.4(9)(53)$^\dag$$^\ddag$ & 64.3(7)(52)$^\dag$$^\ddag$ & 64.8(7)(52) \\%65.34(88)(533)& 64.25(75)(517) & 64.78(72)(519)
    $^{28}$Al & 3$^+$ & $\beta^-$ & 2.245(2) min & 1778.987 & 100(0)    & 15.39(20) & 14.13(10)  \\
              &     & & && $\Delta I_\gamma^{abs}/I_\gamma^{abs}$=0.0\% & 13.94(22)(114)$^\dag$$^\ddag$ & 13.94(10)(113)$^\dag$$^\ddag$ & 13.94(10)(112)\\
    $^{29}$Mg & 3/2$^+$ & $\beta^-$ & 1.30(12) s  & 1398.0 & 16.4(11) & - & 0.54(22) \\%0.537(218)
              &         &           &             & 1754.2 & 9.90(72)  & - & 0.82(26) \\%0.822(261)
              %&         &           &             &         & average  & & 0.650(217) &\\
              & & & & & $\Delta I_\gamma^{abs}/I_\gamma^{abs}$=13.89\%  & & 0.66(24)(5)$^\dag$ & 0.66(24)(5)\\%0.656(244)(53)
    $^{28}$Mg & 0$^+$   & $\beta^-$ & 20.915(9) h & 30.6383 & 89.0(45)\footnotemark[3]& 1.99(41) & - \\
              &         &           &             & 400.6   & 35.9(18)\footnotemark[3] & 1.97(15) & - \\
              &         &           &             & 941.7   & 36.3(18)\footnotemark[3] & 1.79(16) & - \\
              %&         &           &             &         & average  & 1.88(11) & >1.59\\
              &     & & & & $\Delta I_\gamma^{abs}/I_\gamma^{abs}$=incl. & 1.90(11)(15)$^\dag$ & & 1.90(11)(15)\footnotemark[7] \\
    $^{27}$Mg & 1/2$^+$ & $\beta^-$ & 9.458(12) min & 843.76 & 71.800(20) & 1.59(6)  & 1.70(12) \\%1.586(58)
              &         &           &               & 1014.52 & 28.200(20)& 1.91(18) & 1.51(15) \\%1.912(185)
              %&         &           &               &        & average & 1.601(55) >1.50 & 1.62(9) > 1.52\\
              &     & & & & $\Delta I_\gamma^{abs}/I_\gamma^{abs}$=0.11\% & 1.61(6)(13)$^\dag$ & 1.63(9)(13)$^\dag$ & 1.62(5)(13)\footnotemark[8] \\%1.610(56)(132)&1.624(92)(131)&1.617(48)(130)
    $^{28}$Na & 1$^+$ & $\beta^-$ & 30.5(4) ms  & 1473.5 & 37(5) & $<$0.43 & $<$0.29 \\
    $^{27}$Na & 5/2$^+$ & $\beta^-$ & 301(6) ms & 984.66 & 87.4(6) & $<$0.12 & $<$0.10 \\
    $^{26}$Na & 3$^+$ & $\beta^-$ & 1.07128(25) s & 1808.71 & 99.08(50)\footnotemark[2] & 0.290(44) & -\\
              &     & & && $\Delta I_\gamma^{abs}/I_\gamma^{abs}$=0.03\% & 0.291(44)(24)$^\dag$ & & 0.291(44)(24)\footnotemark[9] \\
    $^{25}$Na & 5/2$^+$ & $\beta^-$ & 59.1(6) s & 389.710 & 12.68(22) & 1.33(46) & 0.60(18)\footnotemark[6]\\%0.600(183)
              &         &           &           & 585.028 & 13.00(18) & 1.4(6) & 1.04(19)\\%1.40(57)
              &         &           &           & 974.742 & 14.95(22) & -        & 1.05(25)\\
              %&         &           &           &         & average  & 1.34(35) & 1.03(15) \\
              &     & & && $\Delta I_\gamma^{abs}/I_\gamma^{abs}$=5.4\% & 1.36(37)(11)$^\dag$ & 1.04(16)(8)$^\dag$ & 1.10(15)(9)\footnotemark[10]  \\
    $^{24}$Na & 4$^+$   & $\beta^-$ & 14.956(3) h  & 1368.625& 99.994(2)& 0.21(7) & - \\%0.209(69)
              & & && & $\Delta I_\gamma^{abs}/I_\gamma^{abs}$=incl. & 0.21(7)(2)$^\dag$$^\ddag$ &  & 0.21(7)(2)\\%0.208(69)(17)
              & 1$^+$   & IT        & 20.18(10) ms  & 472.2023& 99.95(5)\footnotemark[4]   & 0.520(37)& 0.541(36)\\
              & & && & $\Delta I_\gamma^{abs}/I_\gamma^{abs}$=0.0\% & 0.517(37)(43)$^\dag$ & 0.539(36)(44)$^\dag$& $<$0.528(26)(43)\footnotemark[11]  \\
    $^{25}$Ne & 1/2$^+$ & $\beta^-$ & 602(8) ms     & 89.53  & 95.4(8)   & $<$0.33 & $<$0.26\\
    $^{24}$Ne & 0$^+$   & $\beta^-$ & 3.38(2) min   & 472.2023 & 100.0(2)& 0.513(35) & 0.518(34)\\
              & & && & $\Delta I_\gamma^{abs}/I_\gamma^{abs}$=0.0\% & 0.513(35)(42)$^\dag$ & 0.517(35)(42)$^\dag$& $<$0.515(25)(41)\footnotemark[11]  \\
              %&         &           &               & 874.420  & 7.9(2)  & $<$1.18 & $<$ \\
    $^{22}$F  & (4$^+$) & $\beta^-$ & 4.230(40) s & 2082.6 & 81.9(20) & $<$0.113 & $<$0.104\\%0.104(38)
    %          & & && & $\Delta I_\gamma^{abs}/I_\gamma^{abs}$=incl. & & 0.104(38)(40)$^\dag$ & 0.104(40)\\
    $^{20}$Na & 2$^+$   & $\epsilon$& 447.9(23) ms  & 1633.602 & 79.3(11)& - & 0.13(6)\\%0.133(56)
              &     & & && $\Delta I_\gamma^{abs}/I_\gamma^{abs}$=incl. &  &0.13(6)(1)$^\dag$ & $<$0.13(6)(1)\footnotemark[12]\\%0.133(57)(11)
    $^{20}$F  & 2$^+$   & $\beta^-$ & 11.163(8) s   & 1633.602 & 99.9995(10)\footnotemark[5] & - & 0.105(45)\\
              &     & & && $\Delta I_\gamma^{abs}/I_\gamma^{abs}$=incl. &  & 0.105(45)(8)$^\dag$& $<$0.105(45)(8)\footnotemark[12]\\
    \end{tabular}
    \end{ruledtabular}
    \footnotetext[0]{* The notation of the uncertainties is the same as those in Table~\ref{tab:Result_28Si}.}
    \footnotetext[0]{** $I_\gamma$ uncertainty includes both $\Delta I_\gamma^{rel}$ and $\Delta I_\gamma^{abs}$ for the nuclei denoted as $\Delta I_\gamma^{abs}/I_\gamma^{abs}$=incl.}
    %\footnotetext[0]{$\dag$: Contamination of the other isotope (Table.~\ref{tab:targets_info_measure}) in the enriched target was evaluated.}
    \footnotetext[0]{$\dag$: The impurity of the enriched target (Table~\ref{tab:targets_info_measure}) was taken into account with the decomposition process.}
    \footnotetext[0]{$\ddag$: The decay component from the mother nuclei was subtracted.}% For example, the decay component from the $^{28}$Mg in the BR of $^{28}$Al was subtracted.}
    \footnotetext[1]{Absolute $I_\gamma$ was not provided in the ENSDF and was calculated from the intensity of beta decay to each energy level and the probability of $\gamma$-ray transition from each state~\cite{Endt1974-tz, Klotz1974-mi, Alburger1974-ie}. The uncertainties contained absolute uncertainty.}
    \footnotetext[2]{$\Delta I_\gamma^{rel}$ of these peaks are not provided in the ENSDF database and were estimated from other $\Delta I_\gamma$.}
    \footnotetext[3]{Absolute $I_\gamma$ was provided based on the estimation. The uncertainty was not provided in the ENSDF. 5\% uncertainty was added~\cite{Alburger1969-ic}.}
    \footnotetext[4]{$\Delta I_\gamma^{rel}$ is not provided in the ENSDF database. The uncertainty originating from the probability of beta decay from $^{24m}$Na (0.05\%) was estimated as 0.05\%.}
    %The uncertainty of $I_\gamma$ is 0.05\% even if the probability of beta decay from $^{24m}$Na has 100\% uncertainty, which is negligible compared with the statistical uncertainty.}
    \footnotetext[5]{$\Delta I_\gamma^{rel}$ of this peak is not given in the ENSDF database and was estimated from the probability of the beta decay to the ground state. }
    \footnotetext[6]{The BR determined with 389.7 keV was minimal compared with the result of the other peaks obtained at J-PARC. This result was excluded from the weighted average value.}
    \footnotetext[7]{The lower limit of $^{28}$Mg is 1.65\% calculated from the upper limit of $^{28}$Na.}
    \footnotetext[8]{The lower limit of $^{27}$Mg is 1.53\% calculated from the upper limit of $^{27}$Na.}
    \footnotetext[9]{The lower limit of $^{26}$Na could not be restricted from the upper limit of $^{26}$Ne. However, the 1n3p channel can be assumed as negligible inferring from the other results.}
    \footnotetext[10]{The lower limit of $^{25}$Na is 1.10\% calculated from the upper limit of $^{25}$Ne.}
    \footnotetext[11]{A peak at 472.2 keV originated from $^{24\textit{m}}$Na and $^{24}$Ne. The upper limit of these isotopes could not be restricted by the other energy peaks, and the production BR of these isotopes both represent the upper limit.} 
    \footnotetext[12]{A peak at 1633.6 keV could originate from $^{20}$Na and $^{20}$F. The upper limit of these isotopes could not be restricted by the other energy peaks, and the production BR of these isotopes both represent the upper limit.}
    \label{tab:Result_30Si}
\end{table*}
%\endgroup

The absolute BR results of produced nuclei following muon nuclear capture and subsequent particle emissions are listed in Tables~\ref{tab:Result_27Al}-\ref{tab:Result_30Si} for $^{27}$Al, $^\mathrm{nat}$Si, and $^{28,29,30}$Si, respectively.
The tables include information on observed $\gamma$-rays and BR derived from each $\gamma$-ray yield from two experiments, as well as compiled values.
The weighted average values of BR for each residual nucleus are provided below the results for each $\gamma$-ray, accounting for decay components from the mother nuclei and isomer, and the impurities in enriched targets.
The first parentheses in the uncertainty of weighted average and compiled value represent the relative uncertainty, including statistical uncertainty, uncertainties from $I_\gamma$ (both relative and absolute), and the half-life of the nuclei. The second parentheses account for the uncertainties associated with the efficiency of the germanium detector, muon capture rate, and the number of muons detected by the plastic scintillator while excluding the relative uncertainty. 
For enriched silicon targets, the uncertainty in the stopping rate is also included in the second set of parentheses.
%The uncertainties stemming from the lifetime and muon capture rate were negligible compared with other uncertainties.
The uncertainty in the first parenthesis corresponds to the relative uncertainty of each residue in the target, transitioning to absolute uncertainty when the uncertainty in the second set of parentheses is added.
%upper limitの説明
Furthermore, the upper limits of BRs of some reaction residues %channels 
were presented.
For undetected nuclei, the upper limit evaluated with the detection limit was provided.
In the case of two candidates for the origin of detected $\gamma$-ray, the upper limit was provided along with uncertainties in the final column.
%two potential sources for a detected

%0n0pとか表記の定義
The reaction channels were denoted based on the emitted number of neutrons and protons, hereinafter; for example, the production of $^{27}$Mg and $^{23}$Ne from muon nuclear capture of $^{27}$Al would be denoted as 0n0p and 2n2p channels, respectively.
Notably, the activation method utilized cannot differentiate between the types of emitted particles that result in the formation of identical residual nuclei, such as processes involving the emission of one neutron and one proton, as well as deuteron emissions processes.

%見えた核・見えなかった核
From muonic $^{27}$Al, no particle emission (0n0p, $^{27}$Mg), several neutrons and one proton emission ($x$n1p, $^{26, 25, 24}$Na) were detected.
The two-neutron two-proton emission (2n2p, $^{23}$Ne) channels may be detected as well, although the detected energy peak could potentially originating from the $^{23}$Mg (4n0p channel).
For $^{28,29,30}$Si, neutron emission channels ($^{30, 29, 28}$Al), one proton emission channels ($^{29, 28, 27}$Mg), two protons emission channels ($^{26, 25, 24}$Na) were detected. 
Reaction channels not listed in the table were not detected, either because they were stable nuclei or below the detection limit.
Notably, the detection limit for in-beam activation measurement varies depending on the produced nuclei because the detection limit depends not only on the production BR but also on the strength of $I_\gamma$ of the decay $\gamma$-ray, efficiency of the germanium detector at the energy, lifetime of the produced nuclei, and measurement time.
%Furthermore, the activation method cannot distinguish the kind of emitted particles that create the same residual nuclei, such as the one neutron and one proton emission and deuteron emissions processes. 
%Furthermore, the activation method utilized cannot differentiate between the types of emitted particles that result in the formation of identical residual nuclei, such as processes involving the emission of one neutron and one proton, as well as deuteron emissions processes.

%enrichedの組成の細かい混ざりの考慮、decay由来の生成量の考慮について
The decay components from the mother nuclei, which decayed into the same unstable nuclei as the produced nuclei from the muon nuclear capture, were evaluated if the BR of the mother nuclei was detected. For example, the decay component from the $^{28}$Mg in the measured BR of $^{28}$Al was subtracted in the analysis of silicon targets. In cases where the BRs of the mother nuclei were not detected, the lower limits of the daughter nuclei were calculated based on the upper limits of the mother nuclei and are detailed in footnotes. %28Al,29Al,24Na(b.s.)で評価、24mNa, 25Na, 27Mg,28Mgはlower見積もり
%For the isotopically enriched targets, corrections were made for the contamination of the other isotopes present in the target during the decomposition process.
For the isotopically enriched targets, the component originating from the contamination of the other isotopes present in the target was calculated and considered during the decomposition process.

%特筆するべき項目：uncertaintyが大きい部分、24Naの解析、MLFでlong lifeは評価できない、b.g.の考慮の仕方、25Naの389keVについてはコメントする、くらい？（複数でかぶる核に関してどっちが最もらしいとかはDiscussionで）
Large uncertainty was observed in the BR of $^{29,30}$Si owing to the large uncertainty of $\epsilon_\mathrm{stop}$, as listed in Table~\ref{tab:targets_info_measure}. Decomposition was conducted using natural abundance silicon, which had limited statistics for nuclei produced from $^{29,30}$Si in a $^\mathrm{nat}$Si target owing to the predominance of $^{28}$Si in natural abundance silicon. For example, the values of $b_{^{30}\mathrm{Al}}$ and $b_{^{29}\mathrm{Al}}$ for $^\mathrm{nat}$Si were 0.396(31)\% and 2.57(15)\%, respectively, with uncertainties of 8\% and 6\%, primarily attributed to the statistical uncertainty.
The determination of $\epsilon_\mathrm{stop}^{^{30}\mathrm{Si}}$ was primarily based on the absolute BR of $^{30}$Al, as it can only be produced from the muon nuclear capture of $^{30}$Si in $^\mathrm{nat}$Si. 
Therefore, the 8\% uncertainty of $b_{^{30}\mathrm{Al}}$ of $^\mathrm{nat}$Si propagated to the uncertainty of $\epsilon_\mathrm{stop}^{^{30}\mathrm{Si}}$.
The determination of $\epsilon_\mathrm{stop}^{^{29}\mathrm{Si}}$ relied significantly on the  absolute BR of $^{29}$Al.
Because $^{29}$Al was produced from both $^{29}$Si and $^{30}$Si, $b_{^{29}\mathrm{Al}}$ of $^\mathrm{nat}$Si was expressed as a combination of $b_{^{29}\mathrm{Al}}^{^{29}\mathrm{Si}}$ and $b_{^{29}\mathrm{Al}}^{^{30}\mathrm{Si}}$, 
where $b_i^j = b_{i~\mathrm{rel}}^j/\epsilon_\mathrm{stop}^j$.
Considering the %significant difference between the
large $b_{^{29}\mathrm{Al}}^{^{30}\mathrm{Si}}=64.8(52)$\%, and smaller $b_{^{29}\mathrm{Al}}^{^{29}\mathrm{Si}}=15.0(38)$\%, 
the uncertainty of $b_{^{30}\mathrm{Al}}^{^{30}\mathrm{Si}}$ strongly influenced the absolute value of $b_{^{29}\mathrm{Al}}^{^{29}\mathrm{Si}}$,resulting in a substantial uncertainty of approximately 25\% in $\epsilon_\mathrm{stop}^{^{29}\mathrm{Si}}$.

%MLFでlong lifeは評価できない
The BRs of $^{24}$Na ($T_{1/2}=15.0$ h) and $^{28}$Mg ($T_{1/2}=20.9$ h) were not evaluated in the J-PARC experiment for all targets owing to the background component originating from the existence of silicon in the beam duct.
The background component was evaluated and subtracted for short-lifetime isotopes; however, the evaluation of long-lifetime isotopes was hindered by a significant impact owing to the cumulative amount from muon irradiation prior to each measurement. %could not be evaluated due to the large effect.
In consequence, the uncertainties of the other BRs in the J-PARC experiment increased owing to the uncertainty resulting from the background subtraction process.

%24Na (isomer, g.s.)
In $^{24}$Na, an isomeric state with a lifetime of 20 ms was observed, decaying through the isomeric transition to the ground state by emitting the $\gamma$-ray with the energy of 472.2 keV.
The population of the isomeric state at 472.2 keV is not only a result of direct production following the particle emission from muon nuclear capture but also a result of the $\beta$ decay of $^{24}$Ne.
%Utilizing
%Since the $^{24}$Ne is decaying into $^{24}$Na by emitting $\gamma$-ray with the energy of 472.2 keV and 874.4 keV, the upper limit of the BR of $^{24}$Ne was deduced using the upper limit of the count at 874.4 keV.
By leveraging the decay of $^{24}$Ne into $^{24}$Na through the emission of $\gamma$-rays at 472.2 keV and 874.4 keV, the upper limit of the BR of $^{24}$Ne was estimated using the upper count limit at 874.4 keV.
The peak count of 472.2 keV after the beam irradiation was also utilized to the estimation of the upper limit of $^{24}$Ne, considering 20.2 ms lifetime of $^{24m}$Na, which implies that its decay $\gamma$-ray should not be detected in decay measurement after the beam irradiation. 
The lower limit of $^{24m}$Na was inferred based on the upper limit of $^{24}$Ne except for the $^{30}$Si target.
For the $^{30}$Si target, only the upper limits of the BR of $^{24m}$Na and $^{24}$Ne were determined.
The BR of $^{24}$Na (ground state) was inferred by considering the creation from the $^{24m}$Na, assuming that all the yield of 472.2 keV peaks were produced by $^{24m}$Na.
The longer lifetime of $^{24}$Na compared to both $^{24m}$Na and $^{24}$Ne, along with minimal differences in the upper limits of their BRs, ensured that the breakdown of these isotopes did not impact the determination of the ground state BR of $^{24}$Na.

%25Na, 389keV
%In contrast,
The production BR of $^{25}$Na determined by the yield of a 389.7-keV peak demonstrated a notably lower value when compared with results obtained from other energy peaks at 585.0 and 974.7 keV.
This discrepancy was observed solely in the analysis of silicon targets in the J-PARC experiment, as the RAL experiment and the analysis of the aluminum target yielded a result (389.7 keV) consistent with the estimation from other energy peaks.
The efficiency of the germanium detector does not strongly affect the yield of 389.7-keV, even considering the effect of self-absorption.
Background energy peaks at these energies were absent, and precautions were taken to avoid the potential contamination from single or double escape of higher energy $\gamma$-rays. The other possibilities of the contamination, such as overlap with the $\gamma$-ray emission from $^{25}$Al and $^{22}$Mg, were considered, but were insufficient to explain the discrepancy.
The cause of the significantly lower yield of 389.7 keV peak remains uncertain, leading to the exclusion of this data when determining the BR of $^{25}$Na.

%20Neの2+->0+ transitionに対応する1633keV peakが検出された
The transition $\gamma$-ray from $^{20}$Ne at an energy level of 1633.6 keV was detected with all the targets except for $^{29}$Si, with the transition $\gamma$-ray from $^{21}$Ne at 350.7 keV also observed in the $^{27}$Al and $^{29}$Si targets.
The detected $\gamma$-rays of $^{20}$Ne and $^{21}$Ne could originate from either $^{20}$F and $^{20}$Na, or $^{21}$F and $^{21}$Na, respectively.
The origin could not be determined based on available information, such as lifetime and the upper limit of the other non-detected $\gamma$-rays. Consequently, the BRs of $^{20,21}$F and $^{20,21}$Na were listed as upper limits.
In the $^{27}$Al target, the presence of $^{23}$Ne and $^{23}$Mg was also presented as an upper limit owing to the possibility that the detected transition $\gamma$-ray from 5/2$^+$ to 3/2$^+$ state in $^{23}$Na could originate from either isotope.

%comtamination from neutron & electron
Notably, the production BR may be influenced by electrons and neutrons present in the beam or the surroundings.
Electrons within the muon beam or generated during the muon decay process ($\mu^- \to \nu_\mu +\bar{\nu_e}+e^-$) can transmute target material through ($\gamma$, $xnyp$) reactions, facilitated by photons from bremsstrahlung radiation.
Environmental neutrons, as well as neutrons produced by muon nuclear capture reactions, can also react with the target material.
The potential for the background component to be generated by neutrons and electrons was assessed through a PHITS calculation, a Monte-Calro simulation code that will be explained in Sect.~\ref{sec:Discussion}.
The impact of electrons was deemed negligible, assuming the number of electrons was roughly twice as large as the stopping muon in the target for decay electrons and thirty times larger for beam-contaminated electrons.
Environmental neutrons were estimated as thermal neutrons, which did not influence the production BR owing to the prevalence of ($n$, $\gamma$) or ($n, n$).
However, the neutron emitted from the muon nuclear capture reaction could potentially impact the production BR results.
The maximum potential contamination resulting from neutron irradiation was estimated to be approximately 3\% for the 0n0p channel of $\mu^-+$$^{27}$Al and $\mu^-+$$^{28}$Si, assuming that the number of muons stopping in the surrounding material was equivalent to that of the target.
This contamination level was comparable to the uncertainty of the present measurement and was not corrected in the result.
Any potential contamination from these neutrons in channels other than the 0n0p channel was smaller than the uncertainty observed.
\section{Discussion}\label{sec:Discussion}
In this section, the present results were compared with previous measurements and discussed in terms of isotope dependence and the even-odd effect of neutron and proton numbers.
Additionally, the results were compared with theoretical model calculations.
The excitation energy distribution resulting from the muon nuclear capture reaction was estimated using the present results and model calculations.

\subsection{Comparison with previous measurements}\label{sec:previous}
%昔の測定がいくつか間違っている、今回のabs測定が大事、という話。
The production BR results of $^{27}$Al and $^{\mathrm{nat},28}$Si were compared with those of previous studies, utilizing the activation method~\cite{Bunatyan1970-mo, Vilgelmona1970-ec, Heusser1972-vs, Wyttenbach1978-ks, Heisinger2002-nm, Miller1972-wm} and prompt $\gamma$-ray measurements~\cite{Pratt1969-sk, Miller1972-wm, L_E_Temple_S_N_Kaplan_R_V_Pyle_and_G_F_Valby1971-mu, Gorringe1999-nv, Measday2007-os}, as listed in Table~\ref{tab:BR_previous}. 
%手法の紹介
Notably, the muon nuclear capture probabilities ($P_\mathrm{cap}$) utilized in the analysis of the previous studies differed from those in the present study~\cite{Mizuno2024} by approximately 1\%.
The discrepancy was not corrected for the comparison owing to the lack of description of values in some research, and the magnitude of the effect was smaller than the uncertainty of BRs.

%activatin測定：合ってたり合ってなかったり。信頼できるものとは矛盾しない。（否定の文章になってしまう。。なかったことにした方がいいか。）
Previous activation measurements were conducted for $^{27}$Al, $^{28}$Si, and $^\mathrm{nat}$Si.
Wyttenbach et al.~\cite{Wyttenbach1978-ks} measured the activation of $^{27}$Al targets, with the results of the production BRs of $^{25}$Na and $^{23}$Ne being comparable with the present results.
Miller et al.~measured enriched $^{28}$SiO$_2$, whreas Heisinger et al.~measured natural abundance Al$_2$O$_3$ and SiO$_2$ targets~\cite{Miller1972-wm, Heisinger2002-nm}, both showing systematically larger BRs.
The discrepancy between the present results and those of Miller et al.~and Heisinger et al.~may be attributed to the uncertainty of the stopping number of irradiated muons.
Muonic X-rays were employed to measure the muon attachment probability for each isotope, as the measurements were performed with oxide targets, potentially underestimating the stopping number of irradiated muons.
Discrepancies with the other previous studies~\cite{Bunatyan1970-mo, Vilgelmona1970-ec, Heusser1972-vs} could result from the measurement method of the stopping number of irradiated muons and the inferior energy resolution of the detector. 
Additionally, previous activation measurements did not consider the effect of production from mother nuclei, which are also generated by the muon nuclear capture reaction, such as the $\beta^-$ decay component from the $^{28}$Mg in the BR of $^{28}$Al.

%in-beamとはコンパラ。(Millerは大きめ)
%0n0pとgA,gPについて
The prompt $\gamma$-ray measurement technique determines BR by analyzing the $\gamma$-ray emissions that occur during the de-excitation of nuclei following particle emission. %to determine BR. %by measuring the de-excitation $\gamma$-ray emitted from the produced nuclei after the particle emission. 
By identifying the reaction residues based on the characteristic $\gamma$-ray energies, the BRs can be calculated by summing the total transition yield to the ground state.
It is important to note that prompt $\gamma$-ray measurements provide the lower limit of BR, as there may be missing yields with undefined or weak transitions, and the direct population to the ground state. 
Previous studies utilizing prompt $\gamma$-ray measurements have yielded results that are consistent with current findings, except for the study conducted by Miller et al. The overestimation of the results presented by Miller et al.~can be attributed to the incorrect assignment of certain $\gamma$-ray energies of $^{28}$Al, as highlighted by Measday et al.~\cite{Measday2007-os}.
Upon comparing the results of our current activation measurement with the previous prompt $\gamma$-ray measurement~\cite{Measday2007-os} for the 0n0p channel, 
the prompt $\gamma$-ray measurement covered nearly all (98(12)\%) the BR of 0n0p channel in $\mu^-+$$^{27}$Al, and 88(8)\% of that of the $^{28}$Si target. 
This indicates minimal missing yields and direct ground state populations in the 0n0p channel, contrary to the larger estimates of the missing yield in absolute BR proposed by Measday et~al.~of 3\% for $^{27}$Al and 9.4\% for $^{28}$Si~\cite{Measday2007-os}, which were based on the comparison of their results with previuos activation measurements as can be seen in Table~\ref{tab:BR_previous}.

%その他の先行研究との比較
The direct measurement of the emitted charged particles~\cite{Sobottka1968-px, Budyashov1971-wc, Krane1979-hc, Gaponenko2020-io, AlCap_Collaboration2022-ne, Manabe2023-lc}, neutron~\cite{Sundelin1973-gw, Kozlowski1985-tf}, and neutron multiplicity measurement~\cite{Macdonald1965-ab} have been previously reported.
However, direct comparison with most charged particle measurements is challenging because they usually focus on high-energy particle emission, whereas most components of the emitted particles typically have low energy.
For example, the charged particle emission probability above 1.5 MeV was reported as 15(2)\%~\cite{Sobottka1968-px} for a natural silicon target without particle identification. The measured BR of the charged particles was 5.2(2)\%, which is the sum of the charged particle emission channels in Table~\ref{tab:Result_natSi}. 
The discrepancy in the BR is primarily attributed to the production of $^{26}$Mg, with a BR exceeding 8.4(8)\%~\cite{Measday2007-os}, which could not be measured in the present experiment.
The distribution of neutron multiplicity was measured~\cite{Macdonald1965-ab} and evaluated~\cite{Measday2001-mw} as 
0n: 9(6)\%, 1n: 75(10)\%, 2n: 5(10)\%, and 3n: 9(6)\% for $^{27}$Al, and 
0n: 36(6)\%, 1n: 49(10)\%, 2n: 14(6)\%, and 3n: 1(1)\%  for a natural silicon target, 
assuming that higher neutron emissions exceeding three did not occur.
The no-neutron emission channel includes the 0n0p channel as well as charged particle emission channels that do not involve neutron emission, such as deuteron or alpha emission.
The present result of no neutron emission probabilities were 11.7(4)\% and $>$23.7\% for $^{27}$Al and $^{28}$Si, respectively. These probabilities were derived from a combination of channels, including 0n0p, 0n1p, a portion of 1n1p, 1n2p, and 2n2p channels, assuming that the $x$n2p channels represent helium isotope emission. The ratio of one neutron and one proton emission to deuteron emission in 1n1p channel 
was estimated from the number of proton and deuteron emissions from the direct particle measurement~\cite{AlCap_Collaboration2022-ne} to be 1:5, consistent with the estimation in Ref.~\cite{Wyttenbach1978-ks}.
Despite the significant uncertainty in multiplicity, this ratio remains consistent with previous research.
%It is comparable with the previous result for both targets, considering the large uncertainty in the multiplicity. 
The several neutron emission probabilities were difficult to compare owing to the complexity of neutron emission accompanied by charged particle emission and the substantial uncertainty in multiplicities.
%The higher neutron emission from aluminum than silicon was suggested. 
%However,
%Aluminum demonstrated a higher neutron emission rate compared with silicon.

Overall, the present results were comparable with the previous findings from prompt $\gamma$-ray measurements and direct particle detection. This measurement represents the first reliable absolute data of the production BR for muon nuclear capture of $^{27}$Al and silicon isotopes by utilizing the advantage of the in-beam activation method.

\begin{table}[htbp]
\centering
\caption{Comparison of the BR obtained in the present study with that from previous measurements. The absolute production BR obtained with the present study, previous activation measurement, and the lower limit of BR obtained from the prompt $\gamma$-ray measurement were presented for each reaction channel.} 
\begin{ruledtabular}
\begin{tabular}{D{,}{,}{12}llc}%\hline\hline D{.}{.}{-3}
   \multicolumn{1}{c}{Reaction}  & \multicolumn{1}{c}{Present} & Previous & Methods\\ \hline
   ^{27}\mathrm{Al}(\mu^-,~\nu_\mu)^{27}\mathrm{Mg}   & 9.90(33) & 12.0(7) & Activation~\cite{Heisinger2002-nm} \\%9.9(3)
                                         &        & $>$9.7(11)  & Prompt $\gamma$~\cite{Measday2007-os} \\
                                         &        & $>$6.8(7)  & Prompt $\gamma$~\cite{L_E_Temple_S_N_Kaplan_R_V_Pyle_and_G_F_Valby1971-mu} \\
    ^{27}\mathrm{Al}(\mu^-,~\nu_\mu\mathrm{np})^{25}\mathrm{Na} & 2.52(18) & 2.8(4) & Activation~\cite{Wyttenbach1978-ks} \\ %2.5(2)            
    ^{27}\mathrm{Al}(\mu^-,~\nu_\mu\mathrm{2np})^{24}\mathrm{Na} & 1.61(19) & 2.1(2) & Activation~\cite{Heisinger2002-nm} \\%1.6(2)
                                            &        & 3.5(8) & Activation~\cite{Heusser1972-vs} \\
    ^{27}\mathrm{Al}(\mu^-,~\nu_\mu\mathrm{2n2p})^{23}\mathrm{Ne} & 0.80(6) & 0.76(11) & Activation~\cite{Wyttenbach1978-ks} \\%0.80(7)
    ^\mathrm{nat}\mathrm{Si}(\mu^-,~\nu_\mu x)^{28}\mathrm{Al} & 20.1(7) & 22.8(25) & Activation~\cite{Heisinger2002-nm} \\%20.1(7)
                               &         & 31(3) & Activation~\cite{Bunatyan1970-mo} \\
                               %&         & $>$16.6(12)  & Prompt $\gamma$~\cite{Measday2007-os} \\
                               &         & $>$4.2(6)  & Prompt $\gamma$~\cite{L_E_Temple_S_N_Kaplan_R_V_Pyle_and_G_F_Valby1971-mu} \\
    ^\mathrm{nat}\mathrm{Si}(\mu^-,~\nu_\mu x)^{27}\mathrm{Mg} & 2.94(15) & 5.6(8) & Activation~\cite{Vilgelmona1970-ec} \\%2.9(2)
                               %&        & $>$2.5(4)  & Prompt $\gamma$~\cite{Measday2007-os} \\
    ^\mathrm{nat}\mathrm{Si}(\mu^-,~\nu_\mu x)^{24}\mathrm{Na} & 1.67(12) & 3.4(2)     & Activation~\cite{Heisinger2002-nm}\\%1.7(1)
                                           %&        & $>$0.5(5)  & Prompt $\gamma$~\cite{Measday2007-os} \\
    ^{28}\mathrm{Si}(\mu^-,~\nu_\mu)^{28}\mathrm{Al}  & 18.9(9) %& 28(4)$^e$ & Activation \\
                                                       & 26(3) & Activation~\cite{Miller1972-wm} \\%18.9(10)
                                &         & $>$16.6(12)\footnotemark[1]  & Prompt $\gamma$~\cite{Measday2007-os} \\
                                            &         & $>$16.9(15)  & Prompt $\gamma$~\cite{Gorringe1999-nv} \\
                                            &         & $>$20.3    & Prompt $\gamma$~\cite{Miller1972-wm} \\
    ^{28}\mathrm{Si}(\mu^-,~\nu_\mu\mathrm{p})^{27}\mathrm{Mg}  & 2.87(14) & $>$2.5(4)\footnotemark[1]  & Prompt $\gamma$~\cite{Measday2007-os} \\%2.9(1)
                                &   & $>$1.9(2) & Prompt $\gamma$~\cite{Miller1972-wm} \\
    ^{28}\mathrm{Si}(\mu^-,~\nu_\mu\mathrm{2n2p})^{24}\mathrm{Na} & 1.71(13)& $>$0.5(5)\footnotemark[1]  & Prompt $\gamma$~\cite{Measday2007-os} \\%1.70(13)
    %$^{27}$Al($\mu^-$, $\nu_\mu$)$^{27}$Mg   & 9.9(3) & 12.0(7) & Activation~\cite{Heisinger2002-nm} \\
    %                                     &        & $>$9.7(11)  & Prompt $\gamma$~\cite{Measday2007-os} \\
    %                                     &        & $>$6.8(7)  & Prompt $\gamma$~\cite{L_E_Temple_S_N_Kaplan_R_V_Pyle_and_G_F_Valby1971-mu} \\
    %$^{27}$Al($\mu^-$, $\nu_\mu$np)$^{25}$Na & 2.5(2) & 2.8(4) & Activation~\cite{Wyttenbach1978-ks} \\                        
    %$^{27}$Al($\mu^-$, $\nu_\mu$2np)$^{24}$Na & 1.6(2) & 2.1(2) & Activation~\cite{Heisinger2002-nm} \\
    %                                        &        & 3.5(8) & Activation~\cite{Heusser1972-vs} \\
    %$^{27}$Al($\mu^-$, $\nu_\mu$2n2p)$^{23}$Ne & 0.80(7) & 0.76(11) & Activation~\cite{Wyttenbach1978-ks} \\
   %\hline
\end{tabular}
\end{ruledtabular}
%\footnotetext[1]{These values were obtained using natural abundance silicon and modified to the BR of $^{28}$Si in Ref.~\cite{Measday2007-os} by the authors.}
\footnotetext[1]{These values were obtained using natural abundance silicon and estimated to the BR of $^{28}$Si in the paper~\cite{Measday2007-os}.}
\label{tab:BR_previous}
\end{table}

\subsection{Trend of the results}
\label{sec:general_trends}
In this section, %the current trend will be discussed regarding 
the structure of the excitation function and the particle emission mechanism following the muon nuclear capture reaction is discussed.
%This section delves into the variance in particle emission probability between aluminum and various silicon isotopes, particularly focusing on the emission process properties and the structure of the excitation function.

\begin{figure}[htbp]
    \centering
    \includegraphics[width=0.98\linewidth]{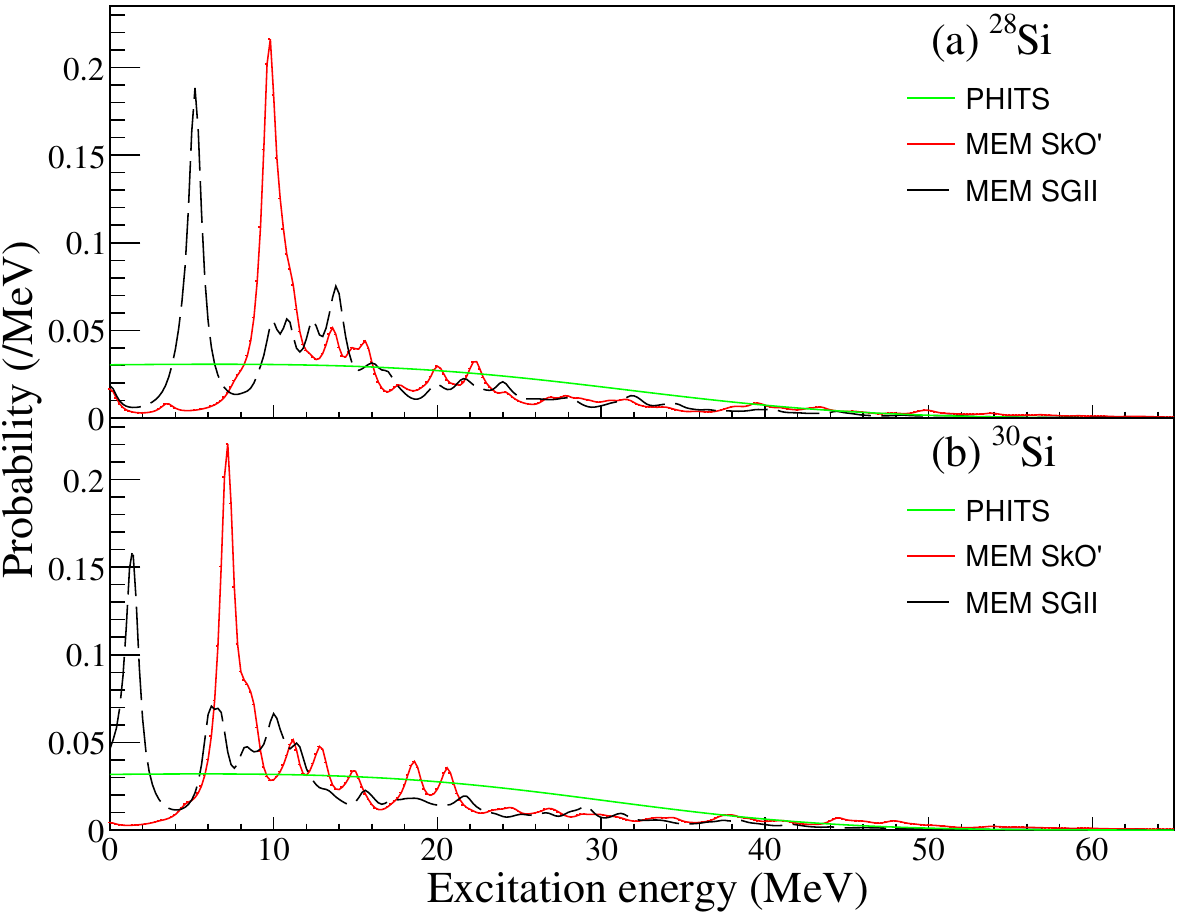}
    \caption{Excitation function of the muon nuclear capture reaction of (a) $^{28}$Si and (b) $^{30}$Si, calculated using the Singer model~\cite{Singer1962-lj} used in PHITS (green solid line), as well as the microscopic and evaporation model (MEM)~\cite{Minato2023-wu} with the effective interaction of SkO' (red solid line) and SGII (black dotted line).}
    \label{fig:excitation_function}
\end{figure}

\begin{figure*}[htbp]
    \centering
    \includegraphics[width=0.98\linewidth]{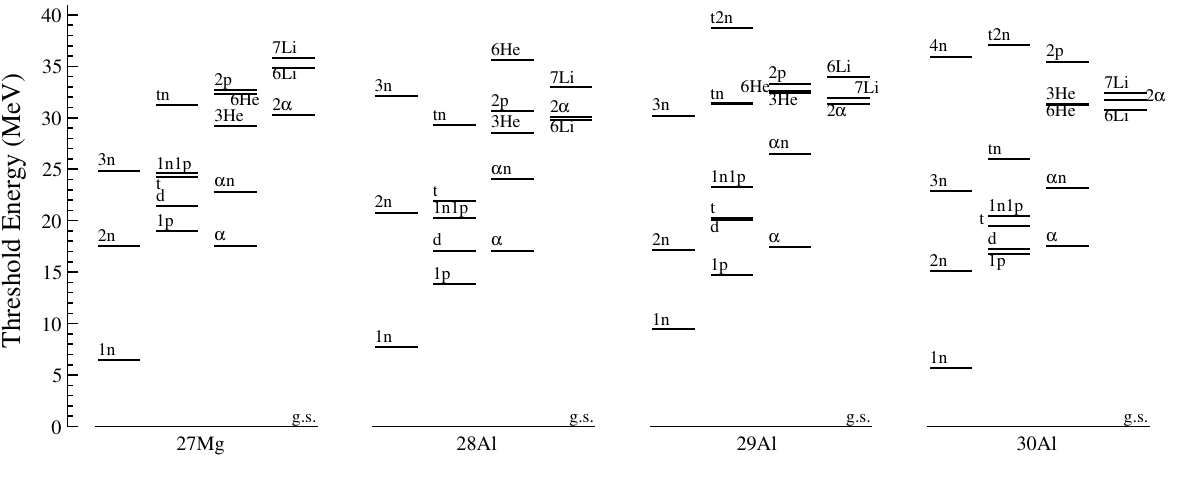}%
    \caption{Threshold energy of each particle emission for $^{27}$Mg and $^{28,29,30}$Al, representing reaction products of muon nuclear capture on $^{27}$Al and $^{28,29,30}$Si. For the charged particle emission, the contribution of the Coulomb barrier was added to the threshold energy, utilizing the classical model (See text). The thresholds of both one neutron and one proton emission, as well as deuteron emission, are presented for 1n1p emission. The threshold energy for other multiple particle emissions was calculated assuming that they would be emitted as light complex particles, such as helium isotopes, for the $x$n2p emission channels.} %1pは軽い粒子を作ってSxが一番小さいものを採用
    \label{fig:thresholdE}
\end{figure*}

%The probability of particle emission after the muon nuclear capture process depends on the excitation energy of the nucleus, particle emission processes, and the threshold energy required for each particle emission.
The probability of particle emission after muon nuclear capture depends on the particle emission processes, and the excitation energy of the nucleus, and the threshold energy required for each emissions.
Three particle emission processes from the muon nuclear capture reaction have been proposed: direct, pre-equilibrium, and evaporation processes.
Notably, particles emitted through the direct and pre-equilibrium processes carry higher energy compared with those produced by the evaporation process.
The excitation function, which represents the energy distribution of the excitation, has been discussed phenomenologically~\cite{Singer1962-lj, Singer1974-zv, Lifshitz1978-zz, Lifshitz1980-bu, Balashov1967-ml, Foldy1964-ea, Uberall1974-tk, Mukhopadhyay1977-qf, Kozlowski1978-di} and using some microscopic calculations~\cite{Auerbach1984-gf, Kolbe1994-zz, Minato2023-wu}.%Kolbe2001, Jokiniemi2019はcapture rateだけ、RPA計算で励起関数はあんまりない？ あっても低励起だけ？
%Typical shape of the excitation function, with an energy peak at approximately 5--20 MeV, extending up to 100 MeV, is shown in Fig.~\ref{fig:excitation_function}, with the details will be discussed in Sect.~\ref{sec:comparison_with_theory}.
The excitation functions calculated using the Singer model~\cite{Singer1962-lj}, as well as the microscopic and evaporation model~\cite{Minato2023-wu} with the effective interaction of SkO' and SGII are shown in Fig.~\ref{fig:excitation_function}, details will be discussed in Sect.~\ref{sec:comparison_with_theory}. 
The excitation function has an energy peak at approximately 5--20 MeV, extending up to 100 MeV. However, the detailed structure of the excitation function remains elusive as direct measurement is hindered by the difficulty associated with detecting emitted muon neutrinos.
%Three particle emission processes from the muon nuclear capture reaction have been proposed: direct, pre-equilibrium, and evaporation processes. Notably, particles emitted through the direct and pre-equilibrium processes carry higher energy compared with those produced by the evaporation process.
The probability of particle emission is also influenced by the threshold energy required for each particle emission, particularly for the evaporation process. Figure~\ref{fig:thresholdE} shows the threshold energy below 40 MeV calculated from the atomic mass table (AME2020)~\cite{Wang2021-uw} for $^{27}$Mg and $^{28,29,30}$Al. These isotopes were initially formed through muon nuclear capture of $^{27}$Al and $^{28,29,30}$Si.
For the charged particle emission, the Coulomb barrier is added to the threshold energy in the figure. The Coulomb barrier was calculated 
%As the threshold energy of charged particle emission is influenced by the Coulomb barrier of the nucleus, the effect was considered in Fig.~\ref{fig:thresholdE} 
using Eq.~(6) of Ref.~\cite{Wyttenbach1978-ks}, 
\begin{equation}
    V_c = \frac{e^2}{4\pi\epsilon_0}\frac{zZ}{r_0A^{1/3}+\rho_0},
\end{equation}
which expresses the systematic Coulomb barrier with mass and atomic numbers. 
The threshold energy for the emission of multiple particles was estimated based on the assumption that they are emitted as light particles, such as helium isotopes for 2p emission channels.


%\begin{comment}
%activationで測定できない核に対しては-をつける、検出されなかった部分は空白
\begin{table}[htbp]
\centering
\caption{Summary of the production BR: neutron emission}
\begin{ruledtabular}
\begin{tabular}{clllll}%\hline\hline
   %target nucleus  & ($\mu^-$,$\nu_\mu$) & ($\mu^-$,$\nu_\mu$ n) & ($\mu^-$,$\nu_\mu$ 2n) & ($\mu^-$,$\nu_\mu$ 3n) \\\hline
   Target & \multicolumn{1}{c}{0n0p} & \multicolumn{1}{c}{1n0p} & \multicolumn{1}{c}{2n0p} & \multicolumn{1}{c}{3n0p} & \multicolumn{1}{c}{Sum}\\\hline
   $^{27}$Al & 9.90(33) & $>$70(6)\footnotemark[2]& $>$7(1)\footnotemark[3]&$>$2\footnotemark[3] & $>$89(6)\\%$>$53(5)~\cite{Measday2007}
   $^{28}$Si & 18.9(9) & $>$28(2)\footnotemark[1] & $>$2.8(5)\footnotemark[1] & &$>$49.7(23)\\
    %2.2(2):g.s.~\cite{Heisinger2002} & -\\%18.87(95)
   $^{29}$Si & 15.0(38)  & 49(12) & - & - & 64(16)\\ %50.5(142)
   $^{30}$Si & 14.1(12)  & 65(5)  & 13.9(11) & - & 93(7)\\ %64.2(52)%\hline
   %$^{27}$Al & 9.90(33) & $>$70(6)\footnotemark[2]& $>$7(1)\footnotemark[3]&$>$2\footnotemark[3] & $>$89(6)\\%$>$53(5)~\cite{Measday2007}
   %$^{28}$Si & 18.9(10) & $>$28(2)\footnotemark[1] & $>$2.8(5)\footnotemark[1] & &$>$49.7(23)\\
    %2.2(2):g.s.~\cite{Heisinger2002} & -\\%18.87(95)
   %$^{29}$Si & 15.6(46)  & 51(14) & - & - & 66(15)\\ %50.5(142)
   %$^{30}$Si & 14.0(11)  & 64(5)  & 13.8(11) & - & 92(5)\\ %64.2(52)%\hline
\end{tabular}
\end{ruledtabular}
%\item{a: The lower limit of the sum of $^{26}$Mg and $^{26}$Na was obtained previously~\cite{Heisinger2002}. The lower limit of the $^{26}$Mg was calculated using the result of $^{26}$Na obtained in the present study. Same as the result of $^{25}$Mg and $^{25}$Na.: 36.9, 3.86}
\footnotetext[1]{These values were obtained with natural silicon and modified to the BR of $^{28}$Si with some assumptions in the reference~\cite{Measday2007-os}.}
\footnotetext[2]{Sourced from Ref.~\cite{L_E_Temple_S_N_Kaplan_R_V_Pyle_and_G_F_Valby1971-mu}}
\footnotetext[3]{Sourced from Ref.~\cite{Measday2007-os}}
%Temple Pcapの記載なし、Measdayは引用していない
\label{tab:neutronBR}
%\end{table}

%\begin{table}[htbp]
\centering
\caption{Summary of the production BR: one proton (including deuteron and triton) emission}
\begin{ruledtabular}
\begin{tabular}{cllllll}%\hline\hline
   %target nucleus  & ($\mu^-$,$\nu_\mu$ p) & ($\mu^-$,$\nu_\mu$ np) & ($\mu^-$,$\nu_\mu$ 2np) & ($\mu^-$,$\nu_\mu$ 3np) & ($\mu^-$,$\nu_\mu$4n1p)\\\hline
   Target  & \multicolumn{1}{c}{0n1p} & \multicolumn{1}{c}{1n1p} & \multicolumn{1}{c}{2n1p} & \multicolumn{1}{c}{3n1p} & \multicolumn{1}{c}{4n1p} & \multicolumn{1}{c}{Sum}\\\hline
   $^{27}$Al & 0.78(6) & 2.52(18) & 1.61(19) & - & 0.05(1)\footnotemark[2] & 4.97(30)\\
   $^{28}$Si & 2.87(14) & $>$8.4(8)\footnotemark[1] & $>$1.5(1)\footnotemark[1] & - & & $>$12.8(8)\\
   $^{29}$Si & 1.6(6) & 2.9(7) & - & - & - & 4.5(12) \\
   $^{30}$Si & 0.66(25) & 1.90(19) & 1.62(14) & - & - & 4.17(43)\\ %\hline
   %$^{27}$Al & 0.78(8) & 2.54(18) & 1.61(23) & - & 0.05(1)\footnotemark[2] & 4.98(9)\\
   %$^{28}$Si & 2.9(1) & $>$8.4(8)\footnotemark[1] & $>$1.5(1)\footnotemark[1] & - & & $>$12.8(8)\\
   %$^{29}$Si & 1.6(7) & 3.0(9) & - & - & - & 4.7(11) \\
   %$^{30}$Si & 0.65(24) & 1.9(2) & 1.6(1) & - & - & 4.1(3)\\ %\hline
\end{tabular}
\end{ruledtabular}
\footnotetext[1]{These values were obtained with natural silicon and modified to the BR of $^{28}$Si with some assumptions in the reference~\cite{Measday2007-os}.}
%\footnotetext[1]{These values are the result of natural silicon~\cite{Measday2007-os}, but 92\% components of natural silicon are $^{28}$Si, and the lower limit of $^{28}$Si is not different from that of natural silicon.}
\footnotetext[2]{Sourced from Ref.~\cite{Heisinger2002-nm}}
%27Al(4n1p): Heisingerのofflilne測定(22Na)、どの程度信頼できるか不明だけど他に情報もないので使う
\label{tab:chargedBR-1p}
%\end{table}

%\begin{table}[htbp]
\centering
\caption{Summary of the production BR: two protons (helium isotopes) emission}
\begin{ruledtabular}
\begin{tabular}{cllllll}%\hline\hline
   %target nucleus  & ($\mu^-$,$\nu_\mu$ 2p) & ($\mu^-$,$\nu_\mu$ n2p) & ($\mu^-$,$\nu_\mu$ 2n2p) & ($\mu^-$,$\nu_\mu$ 3n2p) & ($\mu^-$,$\nu_\mu$ 4n2p)\\\hline
   Target & \multicolumn{1}{c}{0n2p} & \multicolumn{1}{c}{1n2p} & \multicolumn{1}{c}{2n2p} & \multicolumn{1}{c}{3n2p} & \multicolumn{1}{c}{4n2p} & \multicolumn{1}{c}{Sum}\\\hline
   $^{27}$Al &  & 0.038(7)\footnotemark[4] & 0.80(6)\footnotemark[5] & - & - & 0.84(6)\\
   $^{28}$Si &  & 0.24(17) & 1.71(13) & $>$0.5(5)\footnotemark[1] & 0.15(3)\footnotemark[2] & $>$2.6(5)\\
   $^{29}$Si &  &  & 1.11(31) & 1.39(42) & - & 2.5(7)\\
   $^{30}$Si &  &  & 0.29(5) & 1.10(18) & 0.73(8)\footnotemark[3] & 2.11(24)\\ %\hline
   %$^{27}$Al &  & 0.038(7)\footnotemark[4] & 0.80(7)\footnotemark[5] & - & - & 0.83(7)\\
   %$^{28}$Si &  & 0.24(17) & 1.7(1) & $>$0.5(5)\footnotemark[1] & 0.15(3)\footnotemark[2] & $>$2.6(5)\\
   %$^{29}$Si &  &  & 1.2(4) & 1.4(5) & - & 2.6(6)\\
   %$^{30}$Si &  &  & 0.29(5) & 1.1(2) & 0.72(8)\footnotemark[3] & 2.1(2)\\ %\hline
\end{tabular}
\end{ruledtabular}
\footnotetext[1]{This value was obtained with natural silicon and modified to the BR of $^{28}$Si with some assumptions in the reference~\cite{Measday2007-os}.}
\footnotetext[2]{This value resulted from natural silicon~\cite{Heisinger2002-nm}; however 92\% of the components of natural silicon are $^{28}$Si, which was not corrected.}%the lower limit of $^{28}$Si is not different from that of natural silicon.}
%\footnotetext[2]{cite from Ref.~\cite{Heisinger2002-nm}}
\footnotetext[3]{A BR of $^{24}$Ne was deemed negligible.}
\footnotetext[4]{Sourced from Ref.~\cite{Wyttenbach1978-ks}}
\footnotetext[5]{A BR of $^{23}$Mg was deemed negligible.}
%28Si(4n2p): Heisingerのofflilne測定(22Na)、どの程度信頼できるか不明だけど他に情報もないので使う
\label{tab:chargedBR-2p}
%\end{table}

%\begin{table}[htbp]
\centering
\caption{Summary of the production BR: other charged particle emissions}
\begin{ruledtabular}
\begin{tabular}{cllll}%\hline\hline
   %nucleus  & ($\mu^-$,$\nu_\mu$ 3n3p) & ($\mu^-$,$\nu_\mu$ 4n3p) & ($\mu^-$,$\nu_\mu$ 4n4p) & ($\mu^-$,$\nu_\mu$ 6n4p)\\\hline
   Target  & \multicolumn{1}{c}{3n3p} & \multicolumn{1}{c}{4n3p} & \multicolumn{1}{c}{4n4p} & \multicolumn{1}{c}{6n4p}\\\hline
   $^{27}$Al & 0.096(37)\footnotemark[2] & 0.102(21)\footnotemark[1] &  & - \\
   $^{28}$Si & - & - & 0.116(21)\footnotemark[1] & \\
   $^{29}$Si &  & - & 0.16(7)\footnotemark[2] & -\\
   $^{30}$Si &  &  & - & 0.105(46)\footnotemark[1] \\ %\hline
   %$^{27}$Al & 0.09(4)\footnotemark[2] & 0.10(2)\footnotemark[1] &  & - \\
   %$^{28}$Si & - & - & 0.12(2)\footnotemark[1] & \\
   %$^{29}$Si &  & - & 0.16(8)\footnotemark[2] & -\\
   %$^{30}$Si &  &  & - & 0.10(3)\footnotemark[1] \\ %\hline
\end{tabular}
\end{ruledtabular}
\footnotetext[1]{A BR of $^{20}$Na was deemed negligible.}
\footnotetext[2]{A BR of $^{21}$Na was deemed negligible.}
%30Siの2n2pはSEと被って評価できない
\label{tab:chargedBR-rare}
\end{table}

%%%%%%% General Trend of the result %%%%%%%%%%%%
The results of the absolute BR obtained through the present measurement are listed with the absolute uncertainty in Tables~\ref{tab:neutronBR}--\ref{tab:chargedBR-rare}. 
In the case where channels could not be distinguished based on detected $\beta$-delayed $\gamma$-ray energy, for example, such as the 4n0p channel ($^{23}$Mg) and 2n2p channel ($^{23}$Na) of $^{27}$Al, the higher threshold energy channel was excluded.
%the existence of the channel having higher threshold energy were omitted.
Previous values for the long-life isotopes and stable nuclei that were not measured in this experiment are also included.
The lower limits from the prompt $\gamma$-ray measurement and reliable values from the activation measurement, which were consistent with the present study, were selected as the previous results.
The sum of experimentally measured BRs for each isotope was also provided.
Tables~\ref{tab:neutronBR}--\ref{tab:chargedBR-rare} reveal that neutron emission without charged particle emission is predominant for all target isotopes, with the 1n0p channel demonstrating the maximum BR. 
By considering the threshold energy of each particle emission and the present result, the average excitation function was estimated at approximately 10--20 MeV, with the excitation energy exceeding 30 MeV being observed.
The discussion below focuses on the variance in particle emission probability between aluminum and various silicon isotopes. %, especially from the point of view of the emission process properties and the structure of the excitation function.
%The discussion below delves into the variance in particle emission probability between aluminum and various silicon isotopes. %, particularly focusing on the emission process properties and the structure of the excitation function.

%\subsubsection{No particle emission}
%\subsubsection{Systematics of BR for no particle emission channel}
\subsubsection{Even-odd isotope dependence of no particle emission channel}
%%% Alの方が中性子はきやすい %%%
Comparing the neutron emission probabilities of $^{27}$Al and $^{28,29,30}$Si in Table~\ref{tab:neutronBR}, the 0n0p channel of $\mu^-+$$^{27}$Al, 9.9(3)\%, is obviously smaller than that of silicon isotopes.
Although the 1n0p channel of $^{27}$Al and $^{28}$Si were not measured in the current measurement owing to their stable residuals, the 1n0p channel appeared to be larger in aluminum compared with silicon isotopes, drawn from the lower limit observed in previous measurements and estimation based on measured charged particle emission probabilities.
The smaller 0n0p channel in $\mu^-+$$^{27}$Al cannot be explained by the difference of separation energy between $^{27}$Mg and $^{28,29,30}$Al. %$^{27}$Al and $^{28,29,30}$Si.
The mass differences of the initial and final nuclei of the muon nuclear capture reaction were 2.6, 4.6, 3.7, and 8.6 MeV for $^{27}$Al and $^{28,29,30}$Si.
Although the mass difference of the reaction of $^{27}$Al was lower than those of the other silicon isotopes, it did not directly influence the average excitation energy. For example, in the Singer model~\cite{Singer1962-lj} that describes a phenomenological excitation function, the amount of the mass difference influences the available energy for the reaction, with the difference in mean excitation energy estimated to be less than 1 MeV. This difference does not result in a twofold-change in the BR of the 0n0p channel. %neutron emission probability. %($E_0$)
Furthermore, the even-odd and shell effects do not impact the direct and pre-equilibrium emission processes~\cite{Watanabe1987-jr} when assuming that the excitation energy remains constant across isotopes.
Therefore, the discrepancy observed in the 0n0p channel cannot be explained by the characteristics of particle emission processes, but rather may result from the difference in the excitation function. As the average excitation energy does not differ drastically with the mass and atomic number~\cite{Lifshitz1980-bu}, the difference could lie in the strength of the transition peak below the threshold energy, influenced by the shell structure. 

%\subsubsection{Gamov-Teller resonance peak in the excitation distribution}

\begin{figure}[htbp]
    \centering
    \includegraphics[width=0.98\linewidth]{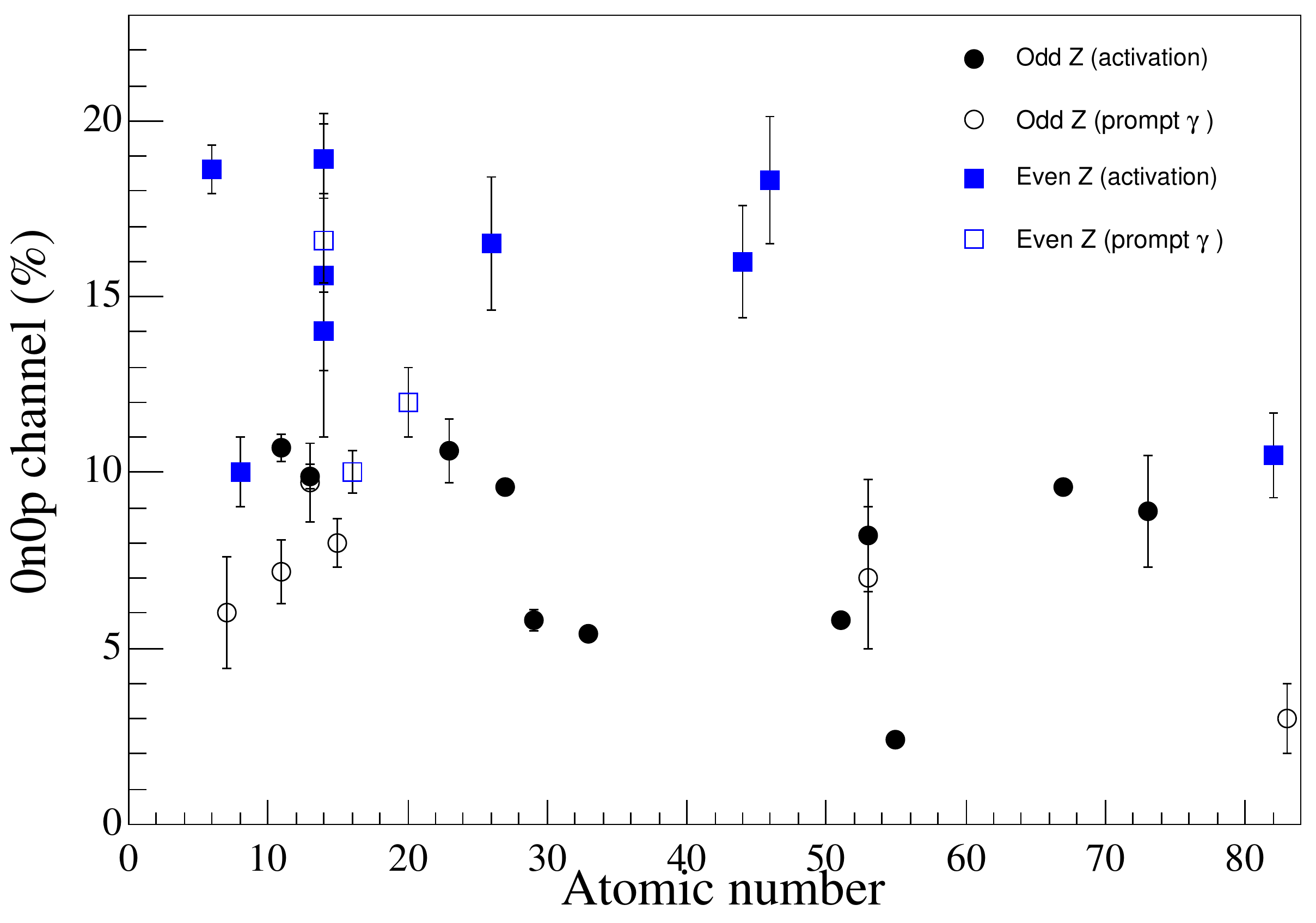}
    \caption{Summary of the BR of the 0n0p channel obtained previously together with the present result. The values are summarized focusing on the even-odd distribution of the proton number. %Previous results of activation were utilized if available; otherwise, the total yield of prompt $\gamma$-ray measurements was adopted.
    Previous results of the activation method are indicated using closed square and circle symbols for even-Z and odd-Z nuclei, respectively. The data obtained using prompt $\gamma$-ray measurement are represented with open symbols.
    %\footnote{The present results of $^{27}$Al and $^{28,29,30}$Si were utilized. The data for the isotopes $^{23}$Na, $^{51}$V, $^{56}$Fe, $^{59}$Co, $^{65}$Cu, $^{75}$As, $^{104}$Ru, $^{123}$Sb, $^{133}$Cs, $^{165}$Ho, $^{181}$Ta, and $^{208}$Pb were obtained through activation measurements conducted by Wyttenbach et al.,~as referenced in note 9 added in proof of Ref.~\cite{Mukhopadhyay1977-qf}. 
    %The results for $^{56}$Fe, $^{65}$Cu, $^{123}$Sb and $^{208}$Pb were likely obtained using natural abundance targets, and be corrected to the value of enriched isotopes owing to their presence as the heaviest isotopes in natural abundance.
    %The isotopes $^{12}$C and $^{16}$O were evaluated by Measday~\cite{Measday2001-mw}.
    %$^{14}$N was sourced from Ref.~\cite{Stocki2002-bw} (prompt $\gamma$), $^{31}$P and $^{32}$S were sourced from Ref.~\cite{Gorringe1999-nv} (prompt $\gamma$), $^{40}$Ca was sourced from Ref.~\cite{Measday2006-kb} (prompt $\gamma$), $^{104}$Pd was sourced from Ref.~\cite{Niikura2024-ck} (activation), $^{127}$I was sourced from Ref.~\cite{Winsberg1954-lj} (activation), and $^{209}$Bi was sourced from Ref.~\cite{Measday2007-zh} (prompt $\gamma$).
    %The natural abundance targets with high enrichment were not adjusted for the BR of enriched isotopes, including $_6$C$_6$:98.9\%, $_7$N$_7$:99.6\%, $_8$O$_8$:99.8\%, $_{16}$S$_{16}$:95.0\%, $_{20}$Ca$_{20}$:96.9\%, and $_{23}$V$_{28}$:99.8\%.}
    \footnote{The present results of $^{27}$Al and $^{28,29,30}$Si were utilized in activation results, and the results of prompt $\gamma$-ray measurement of $^{27}$Al and $^\mathrm{nat}$Si were sourced from Ref.~\cite{Measday2007-os}.
    The activation data for the isotopes $^{23}$Na, $^{51}$V, $^{56}$Fe, $^{59}$Co, $^{65}$Cu, $^{75}$As, $^{104}$Ru, $^{123}$Sb, $^{133}$Cs, $^{165}$Ho, $^{181}$Ta, and $^{208}$Pb were obtained through activation measurements conducted by Wyttenbach et al.,~as referenced in note 9 added in proof of Ref.~\cite{Mukhopadhyay1977-qf}. The results for $^{56}$Fe, $^{65}$Cu, $^{123}$Sb and $^{208}$Pb were likely obtained using natural abundance targets, and be corrected to the value of enriched isotopes owing to their presence as the heaviest isotopes in natural abundance.
    The isotopes $^{12}$C and $^{16}$O were evaluated by Measday~\cite{Measday2001-mw}.
    $^{14}$N was sourced from Ref.~\cite{Stocki2002-bw} (prompt $\gamma$),  
    the prompt $\gamma$ ray result of $^{23}$Na, $^{31}$P, and $^{32}$S were sourced from Ref.~\cite{Gorringe1999-nv}, %(prompt $\gamma$), 
    $^{40}$Ca was sourced from Ref.~\cite{Measday2006-kb} (prompt $\gamma$),
    $^{104}$Pd was sourced from Ref.~\cite{Niikura2024-ck} (activation),
    $^{127}$I was sourced from Ref.~\cite{Winsberg1954-lj} for activation and Ref.~\cite{Measday2007-zh} for prompt $\gamma$ measurement, and
    $^{209}$Bi was sourced from Ref.~\cite{Measday2007-zh} (prompt $\gamma$).
    The natural abundance targets with high enrichment were not adjusted for the BR of enriched isotopes, including $_6$C$_6$:98.9\%, $_7$N$_7$:99.6\%, $_8$O$_8$:99.8\%, $_{16}$S$_{16}$:95.0\%, $_{20}$Ca$_{20}$:96.9\%, and $_{23}$V$_{28}$:99.8\%.}
    }
    \label{fig:even-odd_Z}
\end{figure}

The increased neutron emission probabilities in odd-$Z$ nuclei have been consistent with previous measurements, particularly in medium-mass nuclei.
%Figure~\ref{fig:even-odd_Z} shows a systematical summary of the BRs of 0n0p channel focusing on the even-odd of atomic number.
A comprehensive overview of the BRs of the 0n0p channel, focusing on the even-odd nature of atomic numbers, is shown in Fig.~\ref{fig:even-odd_Z}.
%The BRs were estimated assuming that the prompt $\gamma$-ray measurements covered most BR in the 0n0p channel, as discussed in Sect.~\ref{sec:previous}. 
The BRs obtained using the activation method were indicated using closed square and circle symbols for even-$Z$ and odd-$Z$ nuclei, respectively. 
The data obtained using prompt $\gamma$-ray measurement are represented with open symbols, assuming that the prompt $\gamma$-ray measurements covered most BR in the 0n0p channel, as discussed in Sect.~\ref{sec:previous}. 
%Looking at Fig.~\ref{fig:even-odd_Z}, the BR of the 0n0p channel in other nuclei follows a similar trend to that of aluminum and silicon. Specifically, muon nuclear capture in odd-$Z$ nuclei has a smaller BR of 0n0p, whereas even-$Z$ nuclei demonstrate a larger BR.
Looking at Fig.~\ref{fig:even-odd_Z}, the BR of the 0n0p channel in muon nuclear capture of odd-$Z$ nuclei has a smaller BR of 0n0p, whereas even-$Z$ nuclei demonstrate a larger BR. 
This trend is consistent across medium-mass even-$Z$-even-$N$ and odd-$Z$-even-$N$ nuclei in light and medium mass nuclei.

In the muon nuclear capture reaction, a strong transition to the 1$^+$ state has been observed, as evidenced by the excited levels obtained through prompt $\gamma$-ray measurements~\cite{Miller1972-wm, Cannata1977-vm, Measday2001-mw}.
This knowledge suggests that the Gamow-Teller (GT) transition plays a crucial role in muon capture, analogous to its importance in pion capture and $\beta^+$ decay. %play an important role, is expected to be crucial
%and the analogy with the pion capture and $\beta^+$ decay,
%A possible explanation for 
Thus, the even-odd $Z$ dependence observed in the 0n0p channel may be attributed to the even-odd effect in GT transition, induced by the proton-neutron (pn) pair creation, and known to influence the GT strength in the low-energy regions~\cite{Yoshida2013-vv, Bai2014-au, Sagawa2016-zx, Yuksel2020-se}.
When the configuration of the GT state comprises a pn pair on a closed-shell nucleus with particle-particle properties, specifically when the pn pair is separated from the core and demonstrates deuteron-like behavior in the final states, a strong low-energy peak appears as a result of strong $T=0$ pairing correlations.
%So far, 
This phenomenon has thus far been investigated through the examination of GT strength in closed-shell and two nucleons with $N=Z$ nuclei~\cite{Fujita2014-qz, Sagawa2016-zx, Matsubara2021-hl}.
Additionally, a similar effect owing to the pn pair in spin-dipole excitation has been investigated~\cite{Yoshida2021-fp}.
In the transition from $^{28}$Si to $^{28}$Al with the muon nuclear capture reaction, the primary shell configuration is the initial (0$^+$, $T=1$) state of two protons in $d_{5/2}$ to the final (1$^+$, $T=0$) state of one proton in $d_{5/2}$ and one neutron in either $d_{5/2}$ or $d_{3/2}$, in which the pn pairing lower the energy of GT strength peak and enhance its intensity.
Conversely, the transition from $^{27}$Al to $^{27}$Mg remains unaffected by the pn pair owing to the even-$Z$ of the final state.

\begin{figure}
    \centering
    \includegraphics[width=0.98\linewidth]{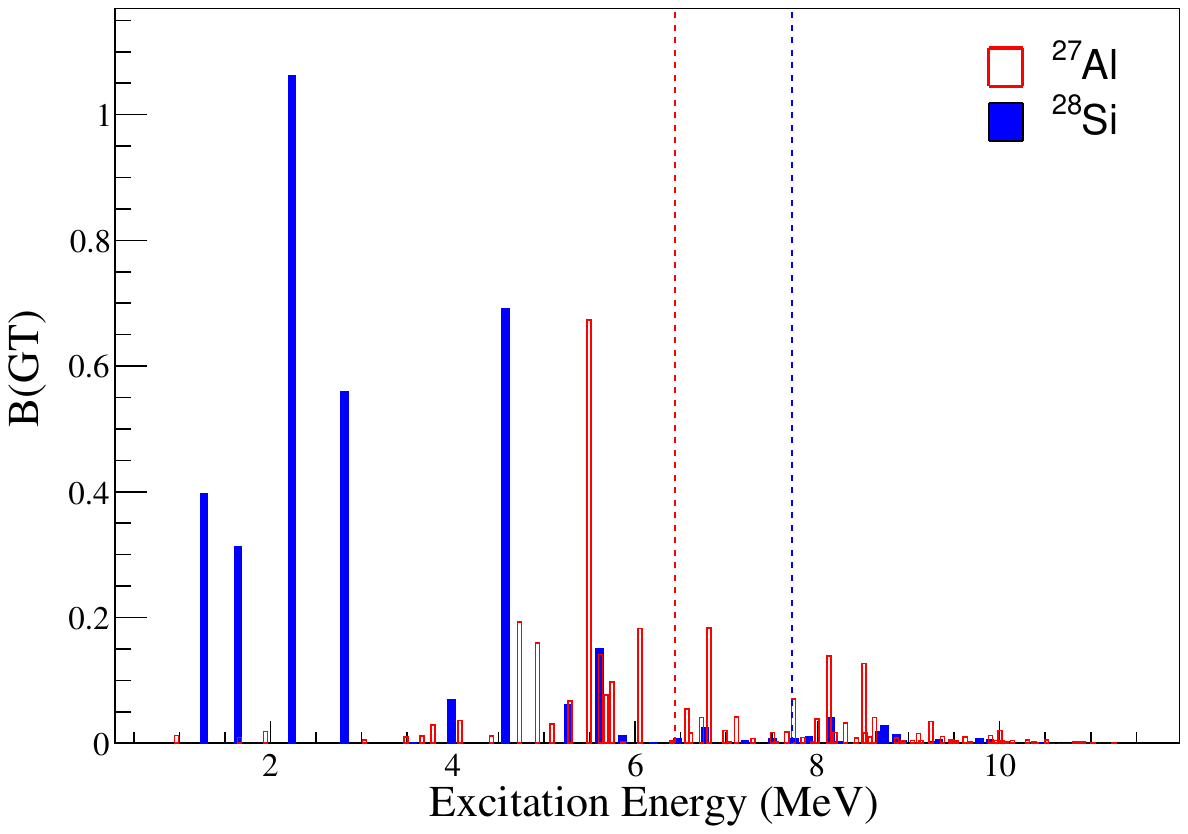}
    \caption{Gamow-Teller strength of the transition from $^{27}$Al to $^{27}$Mg (red) and from $^{28}$Si to $^{28}$Al (blue) calculated by utilizing KSHELL. The one neutron separation energies of $^{27}$Mg (6.44 MeV) and $^{28}$Al (7.72 MeV) are also represented using vertical dotted lines.}
    \label{fig:GT_strength}
\end{figure}

%kshellを使って、AlとSiで確かめてみると、確かにSn以下のB(GT)のtotalが倍くらい違う
To discuss the relationship between the enhancement of GT strength at low energy and the even-odd $Z$ dependence of the 0n0p channel, the GT strength, B(GT), of $^{27}$Al and $^{28}$Si was calculated using the KSHELL code~\cite{Shimizu2019-va} with USD-A interaction.
The calculated B(GT) values, along with the one neutron separation energy are shown in Fig.~\ref{fig:GT_strength}.
The cumulative GT strength below the one neutron separation energy of the final state post muon nuclear capture reaction amounts to 1.77 and 3.36 for $^{27}$Al and $^{28}$Si, respectively. These ratio is comparable with that of the BR of 0n0p channels of 9.9(3)\% and 18.9(9)\% for $^{27}$Al and $^{28}$Si, respectively.
Therefore, the GT transition strength at low energy is enhanced by the pn pairing effect, and %based on the systematic analysis and evaluation of B(GT) below the one neutron separation energy,
%This enhancement influences the even-odd effect observed in the peak structure of the excitation function resulting from muon nuclear capture and particle emission probabilities.
influences the peak structure of the excitation function resulting from muon nuclear capture. 
This variation in the peak structure at low energy leads to the even-odd effect observed in particle emission probabilities.
The enhancement in B(GT) does not significantly impact the excitation function in the high-energy region, which is consistent with the experimental findings indicating minimal differences in multiple charged particle emissions or pre-equilibrium components~\cite{AlCap_Collaboration2022-ne, Manabe2023-lc}. 
Heavy nuclei are unaffected by the pn pair owing to an excess of neutrons, leading to a particle-hole-type excitation. 
This observation is further consistent with the small BR of lead, reported at 10.5(12)\%~\cite{Mukhopadhyay1977-qf}, and the previous result of the temperature of compound nuclei~\cite{Schroder1974-us}, suggesting a higher neutron emission probability for lead compared with bismuth. 
The limited increase in the 0n0p in closed-shell nuclei, such as $^{32}$S (0n0p: 9.9(6)\%~\cite{Gorringe1999-nv}) and $^{40}$Ca (0n0p: 12(1)\%~\cite{Measday2006-kb}), can be attributed to the filled shells, making it challenging to generate pairs through GT transition. %Gorringe modified
Notably, the impact of the enhanced low-energy transition strength owing to the pn pair on the 0n0p channel may involve not only GT but also spin-dipole resonance components.
Furthermore, to eliminate the possibility of atomic number dependence on pre-equilibrium emission and significant shifts in excitation energy, direct energy distribution measurements of emitted particles are required.

%%% Siの同位体内での傾向 %%%
%N増えるほど0n0p減る、1n0p, 2n0pは多分増えている
%When the results of $^{28,29,30}$Si are compared, 
In $^{28,29,30}$Si, the BRs of the 0n0p channel decreases along the neutron number, as shown in Table~\ref{tab:neutronBR}. 
The discrepancy in the 0n0p channel between $^{28}$Si and $^{30}$Si can be understood by the neutron separation energy.
When considering the mean excitation energy ranging from 10 to 20 MeV and assuming it does not differ much in $^{28}$Si and $^{30}$Si, a higher neutron separation energy results in a higher 0n0p channel for $^{28}$Si. 
%However, the ratio of the 0n0p channels for $^{28}$Si and $^{29}$Si deviated from the expected trend based on the variance in neutron separation energy.
However, the BR of 0n0p channels of $\mu^-+$$^{28}$Si is larger than that of $^{29}$Si, although the neutron separation energy of $^{28}$Al is smaller than that of $^{29}$Al.
%The even-odd and shell effects have been found to have no impact on the direct and pre-equilibrium emission~\cite{Watanabe1987-jr}.
The direct and pre-equilibrium emissions have been known not to be influenced by the even-odd and shell effects~\cite{Watanabe1987-jr}.
The mass differences of the initial and final nuclei of the muon nuclear capture reaction may influence the average excitation energy up to 1 MeV and may reduce the BR of the 0n0p channel in $\mu^-+$$^{29}$Si. 
The difference in the detailed structure of the excitation function below the threshold energy, as discussed, may also affect the lower 0n0p channel of $^{29}$Si than $^{28}$Si. 
The impact of strong GT strength on the low energy region resulting from the pn pair is believed to be decreased for $^{29}$Si owing to the even neutron number in the final state, leading to a decrease in BR of the 0n0p channel.
These cumulative effects may lead to an increased neutron emission from the compound nuclei of odd $N$ nuclei.

\subsubsection{charged particle emission}
%見えた傾向をつらつらと。
The BRs of one-proton and two-proton emission channels are listed in Tables~\ref{tab:chargedBR-1p} and \ref{tab:chargedBR-2p}.
When comparing the BRs of charged particle emission from isotopically enriched silicon targets, the BRs of the 0n1p and 2n2p channels decrease with increasing neutron excess.
This trend indicates that the probability of charged particle emission without neutron emission depends on the neutron excess, assuming most of the 2n2p channel corresponds to alpha particle emission.
Focusing on the $x$n2p ($x>2$) channels of $^{28}$Si and $^{30}$Si, the neutron emission accompanied by charged particle emissions becomes more prominent with increasing neutron excess.
The presence of large 3n2p and 4n2p channels in $^{30}$Si compared with the 2n2p channel cannot be explained by the separation energy of these channels, considering the decrease in excitation energy distribution above approximately 10--20 MeV.
The total charged particle emission probabilities also appear to be influenced by the neutron excess, with a decrease in total charged particle emission as neutron excess increases. %, although not all the BRs were detected.
This trend is generally attributed to differences in threshold energy. Still, the probabilities of charged particle emission are affected by the Coulomb barrier, competing channels having similar threshold energies, and emission from direct and pre-equilibrium processes, as well.

%Siの1n1p(d)放出数について考察。(10%くらいはありそう、他と比較してとても大きい、Wyttenbachとかと比べる？）先行との比較で書いてもいいかも。
The 1n1p channel of $\mu^-+$$^{28}$Si (production of $^{26}$Mg) exhibits a significantly large BR compared with other isotopes, although the channel could not be measured in the current experiment. 
Previous measurements indicated that more than 15(2)\% charged particle emission for natural abundance silicon~\cite{Sobottka1968-px}, with prompt $\gamma$-ray measurements showing a lower limit of 8.4\%~\cite{Measday2007-os}.
This high BR is also supported by PHITS and MEM calculations, as detailed in Sect.~\ref{sec:comparison_with_theory} and Fig.~\ref{fig:compare_calc_charged}, %and may be due to the low threshold energy of one proton emission from $^{28}$Al.
potentially resulting from the low threshold energy required for one-proton emission from $^{28}$Al.

%Wyttenbachの系統の議論についてコメント（D論だけでいいかも？）
For the charged particle emission probabilities, Wyttenbach et al.~\cite{Wyttenbach1978-ks} reported systematics that the probability of each charged particle emission channel decreased exponentially with increasing Coulomb barrier.
Based on these findings, BRs of 0n1p, 1n1p, 2n1p, 3n1p, and alpha channels of silicon isotopes were estimated as approximately 0.7, 4.1, 2.8, 2.8, and 0.9\%, respectively.
The results obtained for silicon isotopes in this study did not align with the systematics, although the results of $^{27}$Al were comparable.
This discrepancy may be attributed to the fact that most of the reported targets in the previous research were odd $Z$ and even $N$ nuclei, and the particle emission probability depends on the proton number and isotopes, as discussed in this section.

%minor channelについてコメント(これはsummaryでもいいのかもしれない)
The BRs corresponding to minor particle emission channels have been observed and summarized in Table~\ref{tab:chargedBR-rare}.
These BRs are too large to be understood from the high-energy component of the excitation function when considering the other particle BRs.
In particular, the 4n4p and 6n4p channels observed with silicon isotopes, which can be assumed as two alpha and two-alpha two-neutron emission, cannot be explained without a mechanism that enhances the emission of alpha particles, such as a cluster structure in the excited state.
Notably, the observed nuclei are $^{20,21}$F, which were successfully measured owing to their high sensitivity to the in-beam activation method.
Systematic measurements, including minor channels, are essential for further discussions.

\subsection{Comparison with theoretical calculations}\label{sec:comparison_with_theory}
\begin{figure}[htbp]
    \centering
    \includegraphics[width=0.98\linewidth]{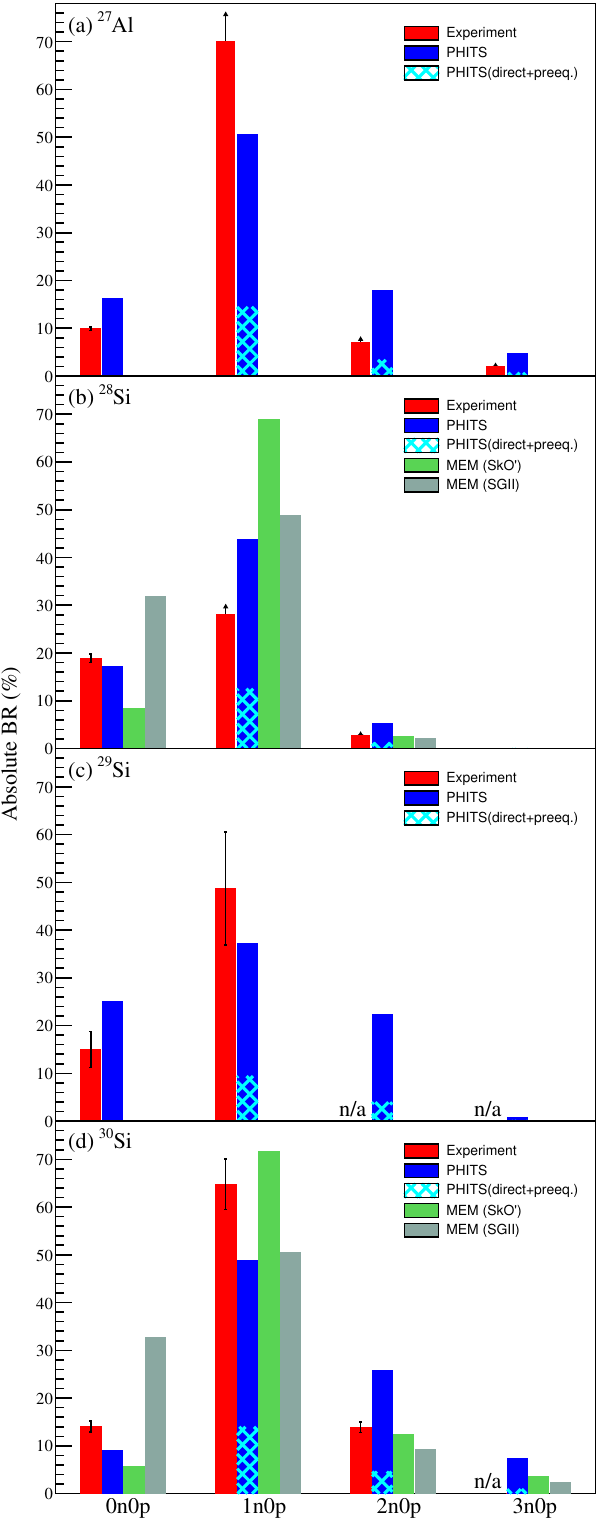}
    \caption{Comparison of the BR of neutron emission channels between the experimental results and theoretical calculations. The experimental results (in red), the PHITS calculation (in blue), and the MEM calculation using SkO' (in green) and SGII (in grey-green) are presented from left to right for each channel. The experimental results are values listed in Table~\ref{tab:neutronBR}, including the present and previous results. The experimental values with upper arrow error bars represent lower limits.
    The components of the direct and pre-equilibrium are represented with shading for PHITS.}
    \label{fig:compare_calc_neutron}
\end{figure}

\begin{figure}[htbp]
    \centering
    \includegraphics[width=0.98\linewidth]{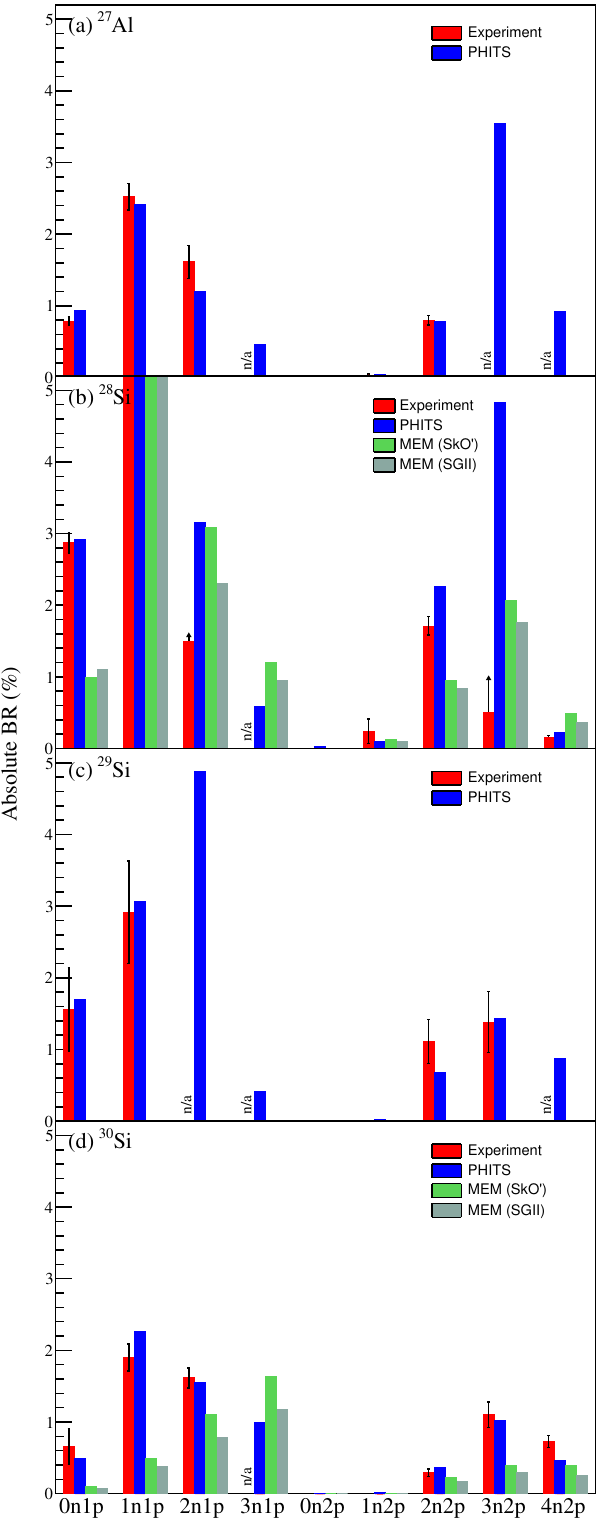}
    \caption{Same as Fig.~\ref{fig:compare_calc_neutron} for charged particle emission channels. The experimental results are taken from the values in Tables~\ref{tab:chargedBR-1p}--\ref{tab:chargedBR-2p}. The BR of the 1n1p channel of $\mu^-+$$^{28}$Si exceeds the range, $>$8.4(8) for the experimental value, 17\% for PHITS, and 9.91\% and 8.50\% for MEM calculation with SkO' and SGII, respectively.}
    \label{fig:compare_calc_charged}
\end{figure}

%今回PHITSと湊さん計算と比較した（それぞれのEex, 粒子放出のモデルの紹介）
The experimental results were compared with two model calculations: a Monte-Carlo simulation using the particle and heavy ions transport code system ver.~3.33 (PHITS)~\cite{Sato2013-ym} and the Microscopic and Evaporation Model~\cite{Minato2023-wu} referred to as the MEM calculation hereinafter. 
In Figs.~\ref{fig:compare_calc_neutron} and \ref{fig:compare_calc_charged}, the experimental values, including the present result of the absolute BRs and the previous result of the absolute and lower limit of the BRs summarized in Tables~\ref{tab:neutronBR}--\ref{tab:chargedBR-2p}, were compared with the model calculations.

%%%%%%% PHITS %%%%%%%%%
%手法の説明
%PHITS: Singerモデルで励起関数を作って前平衡がJQMD, evapolationがGEM, SCMのパラメータは＊
%MECはBRにほぼ効いていないのでないものを使う、コメントは入れてもいいけど、公式には出ていないので書かないことにするか。
PHITS is a general-purpose Monte-Carlo particle transport simulation code that utilizes various nuclear reaction models and nuclear data libraries. 
Muon interaction model was also incorporated in PHITS~\cite{Abe2017-ol}.
The excitation function following the muon nuclear capture process is described using a phenomenological function proposed by Singer~\cite{Singer1962-lj}, with the momentum distribution of nucleon in the nucleus derived by Amado~\cite{Amado1976-nd}.
After the neutron acquires energy, the energy dissipation in the nucleus is calculated using the Jaeri Quantum Molecular Dynamics (JQMD)~\cite{Niita1995-yt, Ogawa2015-qq}. Subsequently, the evaporation process was determined utilizing the Generalized Evaporation Model (GEM)~\cite{Furihata2000-da}. To account for the surface effect, which enhances the emission probabilities of light complex particles such as deuteron, triton, and alpha particles, the Surface Coalescence Model (SCM)~\cite{Watanabe2007-cq} was employed. The SCM model was included to reproduce the energy distribution of emitted charged particles, particularly in high-energy regions~\cite{Manabe2023-lc}.
Since the default parameters in the PHITS code were different from the originally proposed values, modifications were made in this study, with the SCM parameters $h_0=0.26$ GeV $\mathrm{fm}/c$ and $D=2.3$ fm~\cite{Watanabe2007-cq}. These parameters define the emission probabilities of light complex particles in the SCM model.
The root-mean-square charge radius parameters were set as $R_0=3.0610$, 3.1224, 3.1176, and 3.1336 fm for $^{27}$Al and $^{28,29,30}$Si\cite{Angeli2013-to}, respectively. 
The components of the direct and pre-equilibrium emissions are also represented with shade for PHITS calculation in Fig.~\ref{fig:compare_calc_neutron}.
%Figs.~\ref{fig:compare_calc_neutron} and \ref{fig:compare_calc_charged}.
%For the 2n0p and 3n0p channels, the direct and pre-equilibrium component was calculated with weighted by the ratio of emitted number in each process, for example, one neutron emitted with

%中性子放出が＊%, 荷電粒子が＊%、という大まかな傾向は実験と一致。
%細かくみると、中性子は合わなくて、荷電粒子は大体合う。
%level densityの違い(paring効果)や、separation energyの影響、親と子のmassの違いはPHITSには入っている、がどこかで書けると良いが。。
The PHITS calculation reproduced the overall trend of the BR: predominantly emitting neutrons and emitting smaller charged particles, and especially accurately reproduced the BR of charged particle emissions. 
%
%Alを再現しないことについては、励起peakの影響がありそうなので(前述)PHITSで再現ができないことは比較的自明。
%これらの状況を合わせると、
%平均励起energyはもう少し高くて、evenZだとpeakが低いところに出てくる、みたいな形が推定される(これは書くか微妙)
When focusing on the BR of the 0n0p channel, we observed that the 0n0p channels of $^{28,30}$Si were slightly underestimated, whereas those of $^{27}$Al and $^{29}$Si were overestimated.
In the 0n0p channel, the energy peaks in the excitation function are crucial in the result, as discussed in Sect.~\ref{sec:general_trends}. %The excitation function in the PHITS calculation lacks detailed structures, potentially leading to discrepancies in the 0p0n channels. 
The excitation function in the PHITS calculation lacks detailed structures and does not differ among isotopes. Therefore, the ratio of BRs of the 0n0p channels of PHITS results was aligned with the threshold energy levels of particle emission for each isotope, inconsistently with the experimental result.

\begin{figure}
    \centering
    \includegraphics[width=0.98\linewidth]{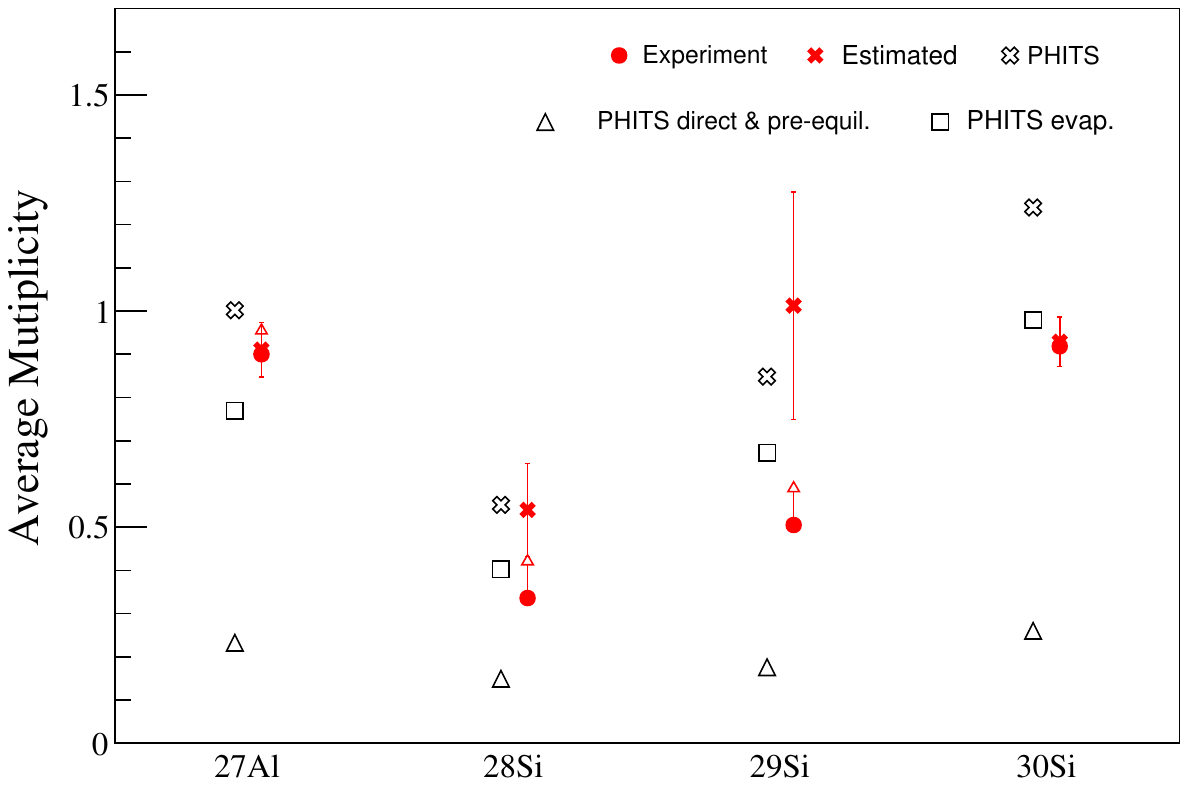}
    \caption{Average multiplicity of neutron emission, calculated using the $x$n0p ($x=1$, 2, 3) channels. 
    The multiplicities, calculated using PHITS, for direct and pre-equilibrium emissions were represented by open triangle symbols, whereas those for the evaporation process were denoted by open square symbols. 
    The sum of these processes is indicated using open cross symbols.
    Experimental results, listed in Table~\ref{tab:neutronBR}, are represented by filled symbols.
    If the absolute BR of certain channels was not measured, the multiplicities are denoted by an upper arrow, indicating a lower limit.
    The estimated average multiplicities, evaluated using the experimental values, are represented with red-cross-filled symbols. }
    \label{fig:PHITS_multiplicity}
\end{figure}

With a closer examination of the BR of neutron emission channels,
1n0p channel were underestimated for all isotopes (though $^{28}$Si was measured as the lower limit), and 2n0p channel was overestimated for $^{27}$Al and $^{30}$Si.
The discrepancy in neutron emission channels can be attributed to the distribution of the excitation function and characteristics of the particle emission process.
%To investigate the cause of the discrepancies in neutron emission channels, the particle emission from each process in the PHITS calculation was examined in terms of neutron multiplicities. %, direct and pre-equilibrium, and evaporation
To investigate the effect of the latter, the particle emission from each process in the PHITS calculation was examined in terms of neutron multiplicities.
%The direct and pre-equilibrium emission were calculated during the JQMD calculation, whereas evaporation was calculated using the GEM model.
%Figure~\ref{fig:PHITS_multiplicity} shows the average multiplicity of neutrons emitted based on the JQMD and GEM calculations, respectively, calculated with the weighted sum of $x$n0p ($x=1$,2,3) channels.
Figure~\ref{fig:PHITS_multiplicity} shows the average multiplicity of neutrons for experimental results, estimation based on the experimental results, and PHITS, calculated with the sum of $x$n0p ($x=1$,2,3) channels weighted by the absolute BR and the emitted neutron numbers. 
%The particle emission, including charged particles, was eliminated here to compare with the present result. 
Only the $x$n0p channels were adopted here for comparing with the present result. 
%The filled symbols represent the average multiplicities of experimental results, calculated using the values listed in Table~\ref{tab:neutronBR}. 
If the absolute BR of certain channels was not measured, the multiplicities are denoted by an upper arrow, indicating the lower limit. 
``The estimated average multiplicities'' derived from experimental values are also denoted by red-cross-filled symbols. 
The estimated average multiplicity of $^{28}$Si was determined by utilizing the measured neutron multiplicity~\cite{Macdonald1965-ab}, calculating the contribution from the total $x$n0p channel of the PHITS calculation. For other isotopes, the average multiplicity was calculated by distributing the BR among unmeasured channels until the total BR of the $x$n0p channel matched that of PHITS. For example, the 2n0p and 3n0p channels of $^{29}$Si were estimated to have approximately 10\% BR, resulting in a total BR of 0n0p--3n0p channels of 86.4\%, consistent with the total BR calculated by PHITS. The ambiguity of the distributed BR was included in the uncertainty.
The average multiplicity from the direct and pre-equilibrium emissions, calculated during the JQMD, was represented with open square symbols, whereas that of the evaporation process, calculated during the GEM, was represented with open triangle symbols. The summed multiplicity of these processes is shown with open cross symbols.

Focusing on the neutron multiplicity of PHITS, 
the direct and pre-equilibrium emission were relatively constant across all isotopes, whereas the average multiplicity from the compound nuclei increased with the neutron excess, $(A-Z)/A$. 
%The trend is roughly comparable with the estimated values, considering the significant uncertainty associated with $^{29}$Si.
%Given that the evaporation process is dominant for multiple neutron emission channels, as shown in Fig.~\ref{fig:compare_calc_neutron}, the trend suggests a higher probability of multiple neutron emission with increasing neutron excess.
Since the evaporation process is dominant for multiple neutron emission channels, neutron excess dependence on compound emission leads to a higher probability of multiple neutron emission with increasing neutron excess.
This neutron multiplicity dependence on the neutron excess is roughly comparable with both experimental and estimated values.
%, considering the significant uncertainty associated with $^{29}$Si. 
%
When comparing the estimated value with the summed multiplicity of PHITS results for each isotope, PHITS overestimated the average multiplicity for $^{27}$Al and $^{30}$Si. 
%
%Regarding the direct and pre-equilibrium emission, previous measurements obtained 19(3)\% of emissions had energies higher than 10 MeV for $^\mathrm{nat}$Si~\cite{Kozlowski1985-tf}, whereas PHITS reported a lower percentage of 11\%, indicagting that PHITS underestimated the pre-equilibrium emission. 
Regarding the direct and pre-equilibrium emission, the emitted neutron energy distribution has been measured for $^\mathrm{nat}$Si in the previous research~\cite{Kozlowski1985-tf}, reporting 19(3)\% of emissions in energies higher than 10 MeV, in contrast, PHITS obtained a lower percentage of 11\%.
The underestimation of high-energy neutron emission was also reported in the previous measurement of neutron energy spectrum on palladium isotopes~\cite{Saito2022-vc}.
This underestimation in high-energy emission indicates that PHITS underestimated the number of direct and pre-equilibrium emissions, considering the predominance of those processes in the high-energy neutron emission.
%Given that pre-equilibrium emission carries a significant amount of excitation energy, a smaller pre-equilibrium component results in high excitation energy in compound nuclei. 
%Assuming a constant probability of direct and pre-equilibrium emission probability for $^{28,29,30}$Si and $^{27}$Al, PHITS consistently underestimated the pre-equilibrium state for all isotopes, therefore, potentially leading to an overestimation of the number of neutrons emitted from the compound nucleus.
Assuming the excitation energy remains constant across isotopes, the direct and pre-equilibrium emission probabilities do not differ among $^{28,29,30}$Si and $^{27}$Al. %~\cite{Watanabe1987-jr}. 
Upon this assumption, the underestimation of direct and pre-equilibrium emission for $^\mathrm{nat}$Si suggests that PHITS consistently underestimated the pre-equilibrium emission for $^{28,29,30}$Si and $^{27}$Al. 
%Given that pre-equilibrium emission carries a significant amount of excitation energy, the lower pre-equilibrium emission potentially leads to an overestimation of the number of neutrons emitted from the compound nucleus, since the energy distribution at the beginning of the evaporation calculation remains high.
Given that pre-equilibrium emission carries a significant amount of excitation energy, a lower pre-equilibrium emission could potentially keep the energy distribution at the beginning of the evaporation calculation high. This makes an overestimation of the number of neutrons emitted from the compound nucleus, consistent with the observed overestimation in $^{27}$Al and $^{30}$Si.
The impact of the overestimating multiplicity owing to the lack of the pre-equilibrium emission became more apparent when the 2n and 3n channels are more prominent. 
This could explain why overestimation of multiplicity was observed in $^{27}$Al and $^{30}$Si, with larger neutron excess, but not in $^{28}$Si.
When the direct and pre-equilibrium emission increased, the 1n0p channel increased owing to a significant portion of direct and pre-equilibrium emission belonging to the 1n0p channel, as shown in Fig.~\ref{fig:compare_calc_neutron}. Additionally, the reduction of higher multiple neutron emission from the evaporation process further contributes to enhancing the 1n0p channel.
The underestimation of pre-equilibrium emission may be attributed to low excitation energy or the property of the JQMD model.

%Alを再現しないことについては、励起peakの影響がありそうなので(前述)PHITSで再現ができないことは比較的自明。
%これらの状況を合わせると、
%平均励起energyはもう少し高くて、evenZだとpeakが低いところに出てくる、みたいな形が推定される(これは書くか微妙)
%When focusing on the BR of the 0n0p channel, the energy peaks in the excitation function are crucial in the result, as discussed in Sect.~\ref{sec:general_trends}. %The excitation function in the PHITS calculation lacks detailed structures, potentially leading to discrepancies in the 0p0n channels. 
%The excitation function in the PHITS calculation lacks detailed structures and does not differ among isotopes. Therefore, the ratio of BRs of the 0n0p channels was aligned with the threshold energy levels of particle emission for each isotope in the PHITS calculation, inconsistently with the experimental result.

For charged particle emissions, the calculated BRs are consistent with those of the present results, although there are a few exceptions.
%Additionally, the calculated summed BRs of the proton emission channels ($x$n1p, $x=0$--4) were 4.99, 23.9, 10.2, and 5.48\%, whreas that of alpha emission channels ($x$n2p, $x=2$, 3, 4) were 5.54, 7.59, 3.20, and 2.08\% for $^{27}$Al and $^{28,29,30}$Si, respectively. 
%次の一文は必要か。。
%Similar to the neutron emission, the probabilities of direct and pre-equilibrium emission remained relatively constant across all isotopes, at approximately 2--3\% for proton emission and 0.2\% for alpha emission. 
The summed BRs of the proton emission channels ($x$n1p, $x=0$--4) and that of alpha emission channels ($x$n2p, $x=2$, 3, 4) are listed in Table~\ref{tab:sumChargedBR}.
These summed BRs of PHITS were comparable with the experimental results. %listed in Tables~\ref{tab:chargedBR-1p} and~\ref{tab:chargedBR-2p}. %, both in terms of absolute BR values and the isotope-dependent trends, showing a decrease in total BR with neutron excess.
However, by considering an overestimation of the charged particle emission on palladium isotopes~\cite{Niikura2024-ck} and the discrepancy in neutron emission probabilities, improvements are necessary for PHITS, despite showing consistent values on the BR of charged particle emission in the mass region of aluminum and silicon.

\begin{table}[htbp]
\centering
\caption{Sum of the BRs of charged particle emission channels.}
\begin{ruledtabular}
\begin{tabular}{ccllll}%\multicolumn{1}{c}{0n0p}
Channel         & Target & Expt. & PHITS & \multicolumn{2}{c}{MEM} \\
& & & & SkO' & SGII\\ \hline
%\multirow{4}{*}{$x$n1p\\($x=0-4$)}  
           & $^{27}$Al & 4.97(7)    & 4.99 & - & - \\
$x$n1p     & $^{28}$Si & $>$12.8(8) & 23.9 & 15.2 & 12.9 \\
($x=0$--4)\footnotemark[1] & $^{29}$Si & 4.5(12)     & 10.2 & - & - \\
           & $^{30}$Si & 4.17(43)   & 5.48 & 3.32 & 2.40 \\\hline
%\multirow{4}{*}{$x$n2p} 
           & $^{27}$Al & 0.80(6)    & 5.54 & - & - \\
$x$n2p     & $^{28}$Si & $>$2.4(5)  & 7.59 & 3.51 & 2.97 \\
($x=2$--4)\footnotemark[2] & $^{29}$Si & 2.5(7)     & 3.20 & - & - \\
           & $^{30}$Si & 2.11(24)   & 2.08 & 1.01 & 0.72 \\ %1.02 & 0.73 \\
\end{tabular}
\label{tab:sumChargedBR}
\end{ruledtabular}
\footnotetext[1]{The experimental values cover $x=0$--2, 4 for $^{27}$Al, $x=0$--2 for $^{28}$Si, $x=0$, 1 for $^{29}$Si, and $x=0$--2 for $^{30}$Si.}
\footnotetext[2]{The experimental values cover $x=2$ for $^{27}$Al, $x=2$--4 for $^{28}$Si, $x=2$, 3 for $^{29}$Si, and $x=2$--4 for $^{30}$Si.}
\end{table}

%%%%%%% microscopic calculation %%%%%%%%
%手法の説明
%Minato: 説明は湊さん共著に入ってもらえるなら書いてもらう？
%2p2hをdoorwayに入れたのが特徴的な近年の計算：capture rate, particle emission(E分布)で実験を再現
%偶々核でしか計算ができない(どこに起因？)ので28,30Siのみで比較を行う
%荷電粒子のenergy分布を再現する計算、ということは書く
Recently, MEM has been developed to provide a comprehensive description of the muon nuclear capture process. This model calculates nuclear excitation resulting from the capture process using the second Tamm-Dankoff approximation~\cite{Da_Providencia1965-cb, Minato2016-qm}, with the treatment of 2p-2h model space.
In this study, two effective interactions, SkO'~\cite{sko2} and SGII~\cite{sg2}, which are often used to describe charge exchange reactions, were employed. 
The initial state of the nucleus and the muon were calculated using the Skyrm-Hartree-Fock method~\cite{Vautherin1972Phys.Rev.C5_626}, assuming spherical symmetry and the density functional theory, respectively.
Furthermore, the two-body meson-exchange current was included phenomenologically to introduce a high-energy component to the excitation function.
Particle emissions from the pre-equilibrium states were estimated using a two-component exciton model~\cite{Koning2004-bl, Iwamoto2016-pl}, whereas emissions from the compound states were computed using a Hauser-Feshbach statistical model~\cite{Hauser1952-cs}.
The MEM calculation well reproduced the energy distribution of emitted particles following muon nuclear capture reaction in high-energy regions above 5--10 MeV for $^{28}$Si and $^{40}$Ca~\cite{Minato2023-wu}.
This calculation can be applied to even-even nuclei, allowing for a comparison of experimental results and calculations for $^{28,30}$Si in this study.

%中性子放出が＊%, 荷電粒子が＊%、という大まかな傾向は実験と一致。絶対値のBRは実験を再現しない
%モデルによって、中性子放出の分岐が大きく違う、荷電粒子はモデルごとの違いはない
The MEM calculation reproduced the general trend of predominantly emitting neutrons, particularly the 1n0p channel, and emitting a smaller amount of charged particles. The order of the BR for both the total neutron emission and charged particle emission channels is also consistent. However, the absolute BRs of each channel showed discrepancies with the experimental results.
Regarding the neutron emission channel, the 0n0p channel was underestimated with SkO' and overestimated with SGII for both $^{28,30}$Si. The 0n0p channel is sensitive to the energy peak in the excitation energy and the neutron separation energy. The calculated peak energy is depends on effective forces, SkO’ and SGII as shown in Fig.~\ref{fig:excitation_function}, therefore, the BR significantly depends on the model utilized in the calculation.
Additionally, the MEM calculation did not consider the pn correlation in the final state, which could impact the BR of the 0n0p channel, as discussed in Sect.~\ref{sec:general_trends}. 
%荷電粒子放出に関しては、totalのorderは合っているけど、測定されたBRでは測定値の半分くらいになっている
%荷電粒子＋中性子放出が多すぎているようにも見える
The measured absolute BRs of the charged particle emission channels exceeded the calculated values, as listed in Table~\ref{tab:sumChargedBR}. %For example, the sum of the BR of the 1p channel of $^{30}$Si is 4.2(3)\% for the experiment, and 3.32\% and 2.40\% for the MEM calculation with SkO' and SGII, respectively. The sum of the 2p channel of $^{30}$Si is 2.1(2)\%, 1.02\%, and 0.73\%, respectively. 
When comparing the BRs of the 2p channel of $\mu^-+$$^{28}$Si and the 1p channel of $\mu^-+$$^{30}$Si, the MEM calculation predicts a higher BR for channels involving higher neutron-accompanied charged particle emissions, such as the 3n1p channel. However, this fact is inconsistent with the experimental results.
The calculation seems to overestimate the neutron emission probability accompanied by charged particle emission.
%MECの効果の入れ方が効いている可能性：MECは現象論的に加えられている＋高energy励起の粒子放出モデルによる原因が考えられる。:次の章に繋げにくいから書かなくていいか。
%変形が入っていないことも原因の一つだろう。
Given that the high-energy component of the excitation function influenced the emission of multiple particles, the inclusion of meson-exchange current %, which was added phenomenologically in high-energy regions, 
could impact these emission channels. 
Additionally, the neglected nuclear deformation effect could contribute to discrepancies in production BRs.
While the MEM calculation accurately reproduces the energy distribution of emitted charged particles in high-energy regions, a significant discrepancy was observed in the production BRs.
Further model refinement is necessary to reprocude the experimental results.

%MEMとの比較と分けるかはEexの結果次第。
%1: MEM計算が合わない原因はEexにあると考え、この実験の結果から評価をした
%2: 全く別立ての文章にする(現在)：

\subsection{Evaluating excitation energy}
\label{sec:Eex_estimation}
The absolute BR provides information regarding the excitation energy distribution following the muon nuclear capture reaction.
As the primary particle emission results from the evaporation process, the excitation function can be estimated from the production BR, using the statistical evaporation model.
The structures of the excitation function of $^{28,30}$Si were estimated based on the results of the current measurement.

%計算手法
The production BR of residual nuclei $i$, %following muon nuclear capture
$b_{i}$, can be expressed with the excitation function populated by muon nuclear capture, $\omega(E^{*})$, as 
\begin{equation}
%Q_{i}=\int P_{i}(E^{*})\omega(E^{*}) dE^{*},
b_{i}=\int_0^\infty P_{i}(E^{*})\omega(E^{*}) dE^{*},
\label{eq:third}
\end{equation}
where $P_{i}(E^{*})$ represents the probability of the creation of a residual nuclei $i$ with a certain excitation energy at $E^{*}$.
$P_{i}(E^{*})$ was calculated using the Hauser-Feshbach statistical model implemented in CCONE~\cite{Iwamoto2007-xf, Iwamoto2016-pl}. 
The initial angular momentum distribution of the excitation function was estimated based on the known average angular momentum distribution, which was calculated from the spin-parity-dependent level density. 
The excitation function $\omega(E^{*})$ was assumed to be expressed as a linear combination of the Gaussian function $G(E^{*}; E_{j},\sigma_{j})$ as 
\begin{equation}\label{eq:Eex_evaluate}
\omega(E^{*})=\sum_{j}^{H}c_{j}G(E^{*}; E_{j},\sigma_{j}),
\end{equation}
where $c_j$ represents the coefficient of each Gaussian function. $E_{j}$, $\sigma_{j}$, and $H$ represent the mean energy, standard deviation, and number of Gaussian functions with different mean energies, respectively. 
Substituting Eq.~(\ref{eq:Eex_evaluate}) into Eq.~\eqref{eq:third} results in a chi-square $\chi^{2}$ of the production BR ($b_{i}$) and experimental production BR ($b_{\mathrm{exp},i}$), expressed as
\begin{equation}
\chi^{2}
=\sum_{i}^{N} \left(
\frac{\displaystyle b_{\mathrm{exp},i}-\sum_{j}^{H} c_{j} \int P_{i}(E^{*}) G(E^{*}; E_{j},\sigma_{j}) dE^{*}}{\Delta b_{\mathrm{exp},i}}
\right)^{2},
\end{equation}
where $N$ represents the number of experimental data and $\Delta b_{\mathrm{exp},i}$ represents an uncertainty of the experiment. 
The optimum excitation function is determined by deriving $c_{j}$ to minimize $\chi^{2}$.
Bayesian optimization with the Gaussian process was employed to deduce the optimum excitation function while constraining the range of $\omega(E^{*})$ to avoid unphysical outcomes, namely $\omega(E^{*})<0$. 
The Radius basis function (RBF) was utilized as the kernel function, with hyperparameters set at 3 and 1 for length and variance, respectively. 
%ガウス過程のカーネルにはRBFを採用し、ハイパーパラメータlengthを3, varianceを1とした。
%これらのパラメータは、系統的に調べた結果、もっとも$\chi^{2}$を小さくするものを選んでいる。
$H=20, E_{j}=2.5~\mathrm{MeV} \times j, \sigma_{j}=5.0~\mathrm{MeV}$ were adopted.
These parameters were selected to minimize $\chi^{2}$.
The total area of the excitation function was normalized to the sum of experimental BRs.

%使用するBR（実験データのevaluation）
The experimental data utilized in this study included the measured production BR of $^{28,30}$Si, supplemented by the evaluated production BR for $^{28}$Si.
The decision to incorporate by the evaluated production BR for $^{28}$Si resulted from the limitations in the measured BR in our study, particularly in the dominant component (1n0p and 2n0p channels) which could not be measured with the activation method.
Previous studies involving prompt $\gamma$-ray measurement~\cite{Measday2007-os}, total charged particle emission probability~\cite{Sobottka1968-px}, and neutron multiplicity measurement~\cite{Macdonald1965-ab} were incorporated into the evaluation process, combined with the present result.
The estimated values of BRs were as follows: 50(10)\% of 1n0p, 13.5(60)\% of 2n0p, 10.8(30)\% of 1n1p, 1.5(5)\% of 2n1p, 0.5(5)\% of 3n2p, and 0.15(3)\% of 4n2p channels.

\begin{figure}
    \centering
    \includegraphics[width=0.95\linewidth]{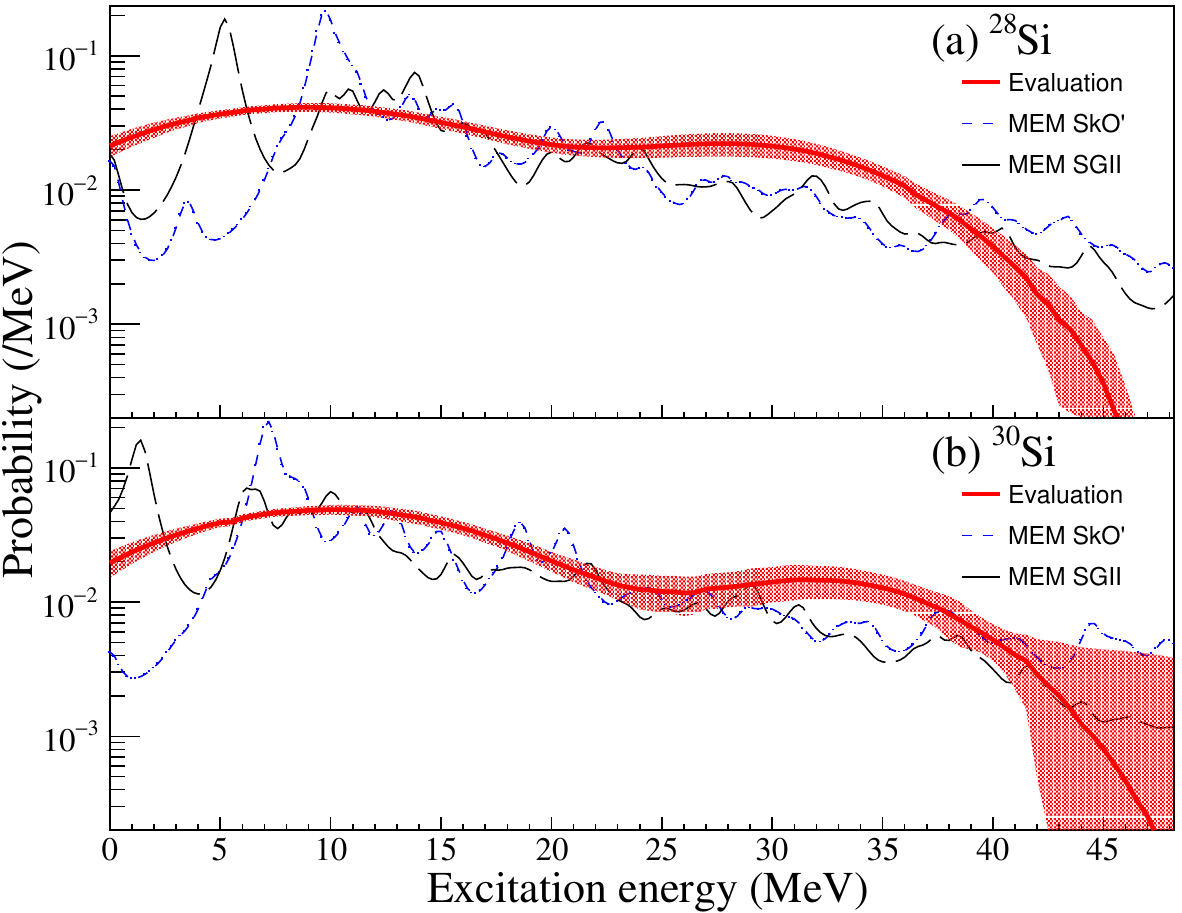}
    \caption[Estimated excitation function from the experimentally measured BR]{Estimated excitation function from the experimentally measured BR. The theoretically calculated excitation function using MEM is represented by black and blue dotted lines.}
    \label{fig:estimated_Eex}
\end{figure}

%結果
Figure~\ref{fig:estimated_Eex} (a) and (b) show the evaluated excitation functions of $^{28}$Si and $^{30}$Si, respectively, with the theoretically calculated excitation function using MEM represented by black and blue dotted lines.
The evaluated excitation functions based on the experimental data provided insights into the general shape of the excitation function, particularly in the region below approximately 40 MeV, where the estimation by BR is effective. 
The average excitation energies of $^{28}$Si and $^{30}$Si are 16.4(3) and 15.4(3) MeV, respectively, and the distributions are exponentially decreasing above 35--40 MeV.
%The obtained average excitation energies of $^{28}$Si and $^{30}$Si are 13.8 and 16.9 MeV, respectively, and the distributions are exponentially decreasing above 30--40 MeV. The peaks around 30 MeV in the excitation function of $^{30}$Si may correspond to the 3n2p (alpha and one neutron) and 4n2p (alpha and two neutrons) emission BRs by considering threshold. %, but it is not possible to say for sure without comparing with the peaks excluding these.
At higher energies, direct and pre-equilibrium particle emission mechanisms come into play, so this discussion alone cannot provide conclusive evidence.
The present proposed excitation functions do not have detailed structures of transition peaks because only the experimental results of production BR were included, which were expressed with the width of intervals of each threshold energy.
%
%Notably, the initial angular momentum distribution of the excitation function was estimated based on the known average angular momentum distribution, which was calculated from the spin-parity-dependent level density.  
Further discussion is required to evaluate %the reliability of the presented excitation function and 
the initial angular momentum distribution, and to consider other measurements, such as emitted particle energies and prompt $\gamma$-ray measurements. However, the possibility and method of deriving the excitation function from measured production BR were demonstrated.

\section{Summary and Conclusions}\label{sec:summary}
The absolute production BRs of residual nuclei following muon nuclear capture of $^{27}$Al, $^\mathrm{nat}$Si, and $^{28,29,30}$Si were measured at two pulsed muon beam facilities: RIKEN-RAL and MLF, J-PARC. 
The experiments employed the in-beam activation method, utilizing a plastic scintillator for muon counting and high-purity germanium detectors for measuring $\beta$-delayed $\gamma$-rays emitted by the produced nuclei.
Absolute BRs were measured using a countable intensity muon beam at RAL, and a high-statistic BR measurement using a high-intensity pulsed muon beam was conducted at J-PARC.
The irradiated muon number was calibrated for the J-PARC experiment to determine the absolute BRs.
The results for $^{28,29,30}$Si were derived through a decomposition analysis utilizing data from the $^\mathrm{nat}$Si target.

The measured BRs were discussed, focusing on the excitation function, particle emission mechanism following the muon nuclear capture reaction, and nuclear properties.
%中性子放出(0n0p）
The results revealed that neutron emission without charged particle emission was the predominant process for all target isotopes, with the 1n0p channel demonstrating the highest BR. The production BRs also suggested an average excitation function peaking between 10--20 MeV, with evidence of excitation energies exceeding 30 MeV.
Comparson of the 0n0p channel between $^{27}$Al and $^{28,29,30}$Si revealed a significantly smaller 0n0p channel for $^{27}$Al. This trend of higher neutron emission probabilities in odd-$Z$ nuclei %, particularly in medium-mass nuclei, 
was also found in previous measurements.
%Based on this systematics and evaluation of the Gamow-Teller (GT) strength below the one neutron separation energy, the importance of considering the final interaction of proton-neutron pairs was emphasized for the muon nuclear capture reaction. 
Small BRs in the 0n0p channel of muon nuclear capture of odd-$Z$ nuclei were discussed with the evaluation of the Gamow-Teller (GT) strength below the one neutron separation energy, emphasizing the importance of considering the final interaction of proton-neutron pairs for the muon nuclear capture reaction. 
This means the potential of muon nuclear capture for investigating isovector transitions and nucleon pairing effects.
%荷電粒子放出
For the charged particle emission channels, the measured BRs for the 0n1p and 2n2p channels decreased as neutron excess increased. 
Additionally, the neutron emission, coupled with charged particle emissions, demonstrated an increase with neutron excess, indicating a correlation between neutron excess and the probability of charged particle emission.
The observation of BRs for minor particle emission channels, such as the 4n4p and 6n4p channels, indicated a potential mechanism that enhances the emission of alpha particles, such as cluster structures in the excited state.
These results of the BR of neutron and charged particle emission channels indicate that the even-odd effect of neutron and proton numbers, as well as neutron excess, influence particle emission following muon nuclear capture.
%To delve deeper into the origin and characteristics of these observed features of particle emissions, systematic measurements considering the even-odd effect and neutron excess, and additional experiments, such as direct charged particle measurement coinciding with neutron measurement, are imperative.
%%The production BR can be measured with relatively easy methods, which is an advantage for carrying out the measurements systematically.
%The relative ease of measuring production BRs presents an advantage for conducting systematic measurements.

%先行と理論との比較
The experimental results were compared with previous measurements and theoretical model calculations. The present results were generally consistent with previous prompt $\gamma$-ray measurements and direct particle detection experiments. 
This study provided the most accurate absolute BR for $^{27}$Al, $^\mathrm{nat}$Si, and $^{28,29,30}$Si.
By comparing the 0n0p channel obtained through prompt $\gamma$-ray measurements with the results of this study, over 88\% of the 0n0p channel was covered by the prompt $\gamma$-ray measurement either in $^{27}$Al and $^{28}$Si, suggesting a few ground state transitions in the 0n0p channels.
Comparisons with theoretical model calculations using PHITS and MEM revealed both agreements and discrepancies. PHITS reproduced overall trends but overestimated neutron multiplicities and the 0n0p channel for some isotopes. 
MEM reproduced the general order of BRs for total neutron and charged particle emission channels, but underestimated charged particle emission BRs.
Notably, a significant model dependence was observed, particularly in neutron emission channels.
Even models that accurately reproduced the energy distribution of neutrons and charged particles struggled to accurately represent the production BRs. %excitation functions in the intermediate energy region below approximately 30 MeV. 
This indicates the necessity of investigating muon nuclear capture, including BR measurements.
Additionally, a method for estimating the excitation function from the measured production BR was proposed utilizing the theoretical model of statistical evaporation emission. 

To delve deeper into the origin and characteristics of observed features of particle emissions, systematic measurements considering the even-odd effect and neutron excess, and additional experiments, such as direct charged particle measurement coinciding with neutron measurement, are imperative.
The relative ease of measuring production BRs presents an advantage for conducting systematic measurements.
Further theoretical refinements are necessary to accurately predict the observed BRs, which would enhance our comprehensive understanding of weak response in nuclei and the particle emission mechanism from the highly excited states.

%まとめ
In conclusion, this study highlights the significance of considering nuclear structure effects, even-odd effects of neutron and proton numbers, neutron excess, and different particle emission mechanisms to effectively model this complex reaction. 
This study also demonstrated the potential of production BR measurement in extracting properties of excitation transitions and reaction dynamics of muon nuclear capture.
%the excitation energy distribution following muon nuclear capture, as well as in investigating subsequent particle emission. 
%Further theoretical refinements are necessary to accurately predict the observed BRs, which would enhance our comprehensive understanding of weak response in nuclei and the particle emission mechanism from the highly excited states.

\begin{acknowledgments}
We are grateful to Dr.~I.~Umegaki for the support on the experiment.
I would like to thank Dr. H.~Sagawa and Dr.~M.~Sasano for valuable discussions. 
The muon experiment at the Materials and Life Science Experimental Facility of the J-PARC was performed under a user program (Proposal No.2021B0537).
The RAL experiment was conducted under a program No.2070005 at the ISIS Neutron and Muon Source.
% JSPS Grant No.
This research is partially supported by the JSPS Grant-in-Aid under Grant Nos.JP19H05664 and JP24H00073. %~JP19H05664, JP22K20372, JP23H04526, JP23H01845, JP23K01845, JP23K03426, JP24H00073, JP24K00647, and JP24K17057**.
% Mizuno
R.M. is supported by the Force-front Physics and Mathematics Program in Drive Transformation (FoPM), a World-leading Innovative Graduate Study (WINGS) program, and the JSR Fellowship from the University of Tokyo.
% Naito
%T.N. acknowledges the RIKEN Special Postdoctoral Researcher Program.
% the JSPS Grant-in-Aid for Research Activity Start-up under Grant No.~JP22K20372,
% the JSPS Grant-in-Aid for Transformative Research Areas (A) under Grant No.~JP23H04526,
% the JSPS Grant-in-Aid for Scientific Research (B) under Grant Nos.~JP23H01845 and JP23K01845,
% the JSPS Grant-in-Aid for Scientific Research (C) under Grant No.~JP23K03426,
% and 
% the JSPS Grant-in-Aid for Early-Career Scientists under Grant No.~JP24K17057.
%The numerical calculations were pertially performed on cluster computers at the RIKEN iTHEMS program.
\end{acknowledgments}

\bibliography{References,References_manual,References_add}

\begin{thebibliography}{93}%
\makeatletter
\providecommand \@ifxundefined [1]{%
 \@ifx{#1\undefined}
}%
\providecommand \@ifnum [1]{%
 \ifnum #1\expandafter \@firstoftwo
 \else \expandafter \@secondoftwo
 \fi
}%
\providecommand \@ifx [1]{%
 \ifx #1\expandafter \@firstoftwo
 \else \expandafter \@secondoftwo
 \fi
}%
\providecommand \natexlab [1]{#1}%
\providecommand \enquote  [1]{``#1''}%
\providecommand \bibnamefont  [1]{#1}%
\providecommand \bibfnamefont [1]{#1}%
\providecommand \citenamefont [1]{#1}%
\providecommand \href@noop [0]{\@secondoftwo}%
\providecommand \href [0]{\begingroup \@sanitize@url \@href}%
\providecommand \@href[1]{\@@startlink{#1}\@@href}%
\providecommand \@@href[1]{\endgroup#1\@@endlink}%
\providecommand \@sanitize@url [0]{\catcode `\\12\catcode `\$12\catcode `\&12\catcode `\#12\catcode `\^12\catcode `\_12\catcode `\%12\relax}%
\providecommand \@@startlink[1]{}%
\providecommand \@@endlink[0]{}%
\providecommand \url  [0]{\begingroup\@sanitize@url \@url }%
\providecommand \@url [1]{\endgroup\@href {#1}{\urlprefix }}%
\providecommand \urlprefix  [0]{URL }%
\providecommand \Eprint [0]{\href }%
\providecommand \doibase [0]{https://doi.org/}%
\providecommand \selectlanguage [0]{\@gobble}%
\providecommand \bibinfo  [0]{\@secondoftwo}%
\providecommand \bibfield  [0]{\@secondoftwo}%
\providecommand \translation [1]{[#1]}%
\providecommand \BibitemOpen [0]{}%
\providecommand \bibitemStop [0]{}%
\providecommand \bibitemNoStop [0]{.\EOS\space}%
\providecommand \EOS [0]{\spacefactor3000\relax}%
\providecommand \BibitemShut  [1]{\csname bibitem#1\endcsname}%
\let\auto@bib@innerbib\@empty
%</preamble>
\bibitem [{\citenamefont {Niikura}\ \emph {et~al.}(2024)\citenamefont {Niikura}, \citenamefont {Saito}, \citenamefont {Matsuzaki}, \citenamefont {Ishida},\ and\ \citenamefont {Hillier}}]{Niikura2024-ck}%
  \BibitemOpen
  \bibfield  {author} {\bibinfo {author} {\bibfnamefont {M.}~\bibnamefont {Niikura}}, \bibinfo {author} {\bibfnamefont {T.~Y.}\ \bibnamefont {Saito}}, \bibinfo {author} {\bibfnamefont {T.}~\bibnamefont {Matsuzaki}}, \bibinfo {author} {\bibfnamefont {K.}~\bibnamefont {Ishida}},\ and\ \bibinfo {author} {\bibfnamefont {A.}~\bibnamefont {Hillier}},\ }\href@noop {} {\bibfield  {journal} {\bibinfo  {journal} {Phys. Rev. C}\ }\textbf {\bibinfo {volume} {109}},\ \bibinfo {pages} {014328} (\bibinfo {year} {2024})}\BibitemShut {NoStop}%
\bibitem [{\citenamefont {Measday}(2001)}]{Measday2001-mw}%
  \BibitemOpen
  \bibfield  {author} {\bibinfo {author} {\bibfnamefont {D.~F.}\ \bibnamefont {Measday}},\ }\href@noop {} {\bibfield  {journal} {\bibinfo  {journal} {Phys. Rep.}\ }\textbf {\bibinfo {volume} {354}},\ \bibinfo {pages} {243} (\bibinfo {year} {2001})}\BibitemShut {NoStop}%
\bibitem [{\citenamefont {{AlCap Collaboration}}\ \emph {et~al.}(2022)\citenamefont {{AlCap Collaboration}}, \citenamefont {Edmonds}, \citenamefont {Quirk}, \citenamefont {Wong}, \citenamefont {Alexander}, \citenamefont {Bernstein}, \citenamefont {Daniel}, \citenamefont {Diociaiuti}, \citenamefont {Donghia}, \citenamefont {Gillies}, \citenamefont {Hungerford}, \citenamefont {Kammel}, \citenamefont {Krikler}, \citenamefont {Kuno}, \citenamefont {Lancaster}, \citenamefont {Litchfield}, \citenamefont {Miller}, \citenamefont {Palladino}, \citenamefont {Repond}, \citenamefont {Sato}, \citenamefont {Sarra}, \citenamefont {Soleti}, \citenamefont {Tishchenko}, \citenamefont {Tran}, \citenamefont {Uchida}, \citenamefont {Winter},\ and\ \citenamefont {Wu}}]{AlCap_Collaboration2022-ne}%
  \BibitemOpen
  \bibfield  {author} {\bibinfo {author} {\bibnamefont {{AlCap Collaboration}}}, \bibinfo {author} {\bibfnamefont {A.}~\bibnamefont {Edmonds}}, \bibinfo {author} {\bibfnamefont {J.}~\bibnamefont {Quirk}}, \bibinfo {author} {\bibfnamefont {M.-L.}\ \bibnamefont {Wong}}, \bibinfo {author} {\bibfnamefont {D.}~\bibnamefont {Alexander}}, \bibinfo {author} {\bibfnamefont {R.~H.}\ \bibnamefont {Bernstein}}, \bibinfo {author} {\bibfnamefont {A.}~\bibnamefont {Daniel}}, \bibinfo {author} {\bibfnamefont {E.}~\bibnamefont {Diociaiuti}}, \bibinfo {author} {\bibfnamefont {R.}~\bibnamefont {Donghia}}, \bibinfo {author} {\bibfnamefont {E.~L.}\ \bibnamefont {Gillies}}, \bibinfo {author} {\bibfnamefont {E.~V.}\ \bibnamefont {Hungerford}}, \bibinfo {author} {\bibfnamefont {P.}~\bibnamefont {Kammel}}, \bibinfo {author} {\bibfnamefont {B.~E.}\ \bibnamefont {Krikler}}, \bibinfo {author} {\bibfnamefont {Y.}~\bibnamefont {Kuno}}, \bibinfo {author} {\bibfnamefont {M.}~\bibnamefont {Lancaster}}, \bibinfo {author} {\bibfnamefont
  {R.~P.}\ \bibnamefont {Litchfield}}, \bibinfo {author} {\bibfnamefont {J.~P.}\ \bibnamefont {Miller}}, \bibinfo {author} {\bibfnamefont {A.}~\bibnamefont {Palladino}}, \bibinfo {author} {\bibfnamefont {J.}~\bibnamefont {Repond}}, \bibinfo {author} {\bibfnamefont {A.}~\bibnamefont {Sato}}, \bibinfo {author} {\bibfnamefont {I.}~\bibnamefont {Sarra}}, \bibinfo {author} {\bibfnamefont {S.~R.}\ \bibnamefont {Soleti}}, \bibinfo {author} {\bibfnamefont {V.}~\bibnamefont {Tishchenko}}, \bibinfo {author} {\bibfnamefont {N.~H.}\ \bibnamefont {Tran}}, \bibinfo {author} {\bibfnamefont {Y.}~\bibnamefont {Uchida}}, \bibinfo {author} {\bibfnamefont {P.}~\bibnamefont {Winter}},\ and\ \bibinfo {author} {\bibfnamefont {C.}~\bibnamefont {Wu}},\ }\href@noop {} {\bibfield  {journal} {\bibinfo  {journal} {Phys. Rev. C}\ }\textbf {\bibinfo {volume} {105}},\ \bibinfo {pages} {035501} (\bibinfo {year} {2022})}\BibitemShut {NoStop}%
\bibitem [{\citenamefont {Manabe}\ \emph {et~al.}(2023)\citenamefont {Manabe}, \citenamefont {Watanabe}, \citenamefont {Niikura}, \citenamefont {Nakano}, \citenamefont {Saito}, \citenamefont {Suzuki}, \citenamefont {Kawashima}, \citenamefont {Tomono}, \citenamefont {Sato},\ and\ \citenamefont {Harano}}]{Manabe2023-lc}%
  \BibitemOpen
  \bibfield  {author} {\bibinfo {author} {\bibfnamefont {S.}~\bibnamefont {Manabe}}, \bibinfo {author} {\bibfnamefont {Y.}~\bibnamefont {Watanabe}}, \bibinfo {author} {\bibfnamefont {M.}~\bibnamefont {Niikura}}, \bibinfo {author} {\bibfnamefont {K.}~\bibnamefont {Nakano}}, \bibinfo {author} {\bibfnamefont {T.~Y.}\ \bibnamefont {Saito}}, \bibinfo {author} {\bibfnamefont {D.}~\bibnamefont {Suzuki}}, \bibinfo {author} {\bibfnamefont {Y.}~\bibnamefont {Kawashima}}, \bibinfo {author} {\bibfnamefont {D.}~\bibnamefont {Tomono}}, \bibinfo {author} {\bibfnamefont {A.}~\bibnamefont {Sato}},\ and\ \bibinfo {author} {\bibfnamefont {H.}~\bibnamefont {Harano}},\ }\href@noop {} {\bibfield  {journal} {\bibinfo  {journal} {EPJ Web of Conferences}\ }\textbf {\bibinfo {volume} {284}} (\bibinfo {year} {2023})}\BibitemShut {NoStop}%
\bibitem [{\citenamefont {Autran}\ and\ \citenamefont {Munteanu}(2024)}]{Autran2024-mp}%
  \BibitemOpen
  \bibfield  {author} {\bibinfo {author} {\bibfnamefont {J.-L.}\ \bibnamefont {Autran}}\ and\ \bibinfo {author} {\bibfnamefont {D.}~\bibnamefont {Munteanu}},\ }\href@noop {} {\bibfield  {journal} {\bibinfo  {journal} {Journal of Nuclear Engineering}\ }\textbf {\bibinfo {volume} {5}},\ \bibinfo {pages} {91} (\bibinfo {year} {2024})}\BibitemShut {NoStop}%
\bibitem [{\citenamefont {Minato}\ \emph {et~al.}(2023)\citenamefont {Minato}, \citenamefont {Naito},\ and\ \citenamefont {Iwamoto}}]{Minato2023-wu}%
  \BibitemOpen
  \bibfield  {author} {\bibinfo {author} {\bibfnamefont {F.}~\bibnamefont {Minato}}, \bibinfo {author} {\bibfnamefont {T.}~\bibnamefont {Naito}},\ and\ \bibinfo {author} {\bibfnamefont {O.}~\bibnamefont {Iwamoto}},\ }\href@noop {} {\bibfield  {journal} {\bibinfo  {journal} {Phys. Rev. C}\ }\textbf {\bibinfo {volume} {107}},\ \bibinfo {pages} {054314} (\bibinfo {year} {2023})}\BibitemShut {NoStop}%
\bibitem [{\citenamefont {Mizuno}\ \emph {et~al.}(2025)\citenamefont {Mizuno}, \citenamefont {Niikura}, \citenamefont {Akamatsu}, \citenamefont {Fujiie}, \citenamefont {Ito}, \citenamefont {Ishida}, \citenamefont {Kikuchi}, \citenamefont {Matsuzaki}, \citenamefont {Minato}, \citenamefont {Murata}, \citenamefont {Shimomura}, \citenamefont {Takeshita}, \citenamefont {Umegaki},\ and\ \citenamefont {Yamaguchi}}]{Mizuno2024}%
  \BibitemOpen
  \bibfield  {author} {\bibinfo {author} {\bibfnamefont {R.}~\bibnamefont {Mizuno}}, \bibinfo {author} {\bibfnamefont {M.}~\bibnamefont {Niikura}}, \bibinfo {author} {\bibfnamefont {S.}~\bibnamefont {Akamatsu}}, \bibinfo {author} {\bibfnamefont {T.}~\bibnamefont {Fujiie}}, \bibinfo {author} {\bibfnamefont {T.}~\bibnamefont {Ito}}, \bibinfo {author} {\bibfnamefont {K.}~\bibnamefont {Ishida}}, \bibinfo {author} {\bibfnamefont {T.}~\bibnamefont {Kikuchi}}, \bibinfo {author} {\bibfnamefont {T.}~\bibnamefont {Matsuzaki}}, \bibinfo {author} {\bibfnamefont {F.}~\bibnamefont {Minato}}, \bibinfo {author} {\bibfnamefont {J.}~\bibnamefont {Murata}}, \bibinfo {author} {\bibfnamefont {K.}~\bibnamefont {Shimomura}}, \bibinfo {author} {\bibfnamefont {S.}~\bibnamefont {Takeshita}}, \bibinfo {author} {\bibfnamefont {I.}~\bibnamefont {Umegaki}},\ and\ \bibinfo {author} {\bibfnamefont {Y.}~\bibnamefont {Yamaguchi}},\ }\href@noop {} {\bibfield  {journal} {\bibinfo  {journal} {Phys. Rev. C Nucl. Phys., Accepted}\ } (\bibinfo
  {year} {2025})}\BibitemShut {NoStop}%
\bibitem [{\citenamefont {Matsuzaki}\ \emph {et~al.}(2001)\citenamefont {Matsuzaki}, \citenamefont {Ishida}, \citenamefont {Nagamine}, \citenamefont {Watanabe}, \citenamefont {Eaton},\ and\ \citenamefont {Williams}}]{Matsuzaki2001-pg}%
  \BibitemOpen
  \bibfield  {author} {\bibinfo {author} {\bibfnamefont {T.}~\bibnamefont {Matsuzaki}}, \bibinfo {author} {\bibfnamefont {K.}~\bibnamefont {Ishida}}, \bibinfo {author} {\bibfnamefont {K.}~\bibnamefont {Nagamine}}, \bibinfo {author} {\bibfnamefont {I.}~\bibnamefont {Watanabe}}, \bibinfo {author} {\bibfnamefont {G.~H.}\ \bibnamefont {Eaton}},\ and\ \bibinfo {author} {\bibfnamefont {W.~G.}\ \bibnamefont {Williams}},\ }\href@noop {} {\bibfield  {journal} {\bibinfo  {journal} {Nucl. Instrum. Methods Phys. Res. A}\ }\textbf {\bibinfo {volume} {465}},\ \bibinfo {pages} {365} (\bibinfo {year} {2001})}\BibitemShut {NoStop}%
\bibitem [{\citenamefont {Hillier}\ \emph {et~al.}(2018)\citenamefont {Hillier}, \citenamefont {Lord}, \citenamefont {Ishida},\ and\ \citenamefont {Rogers}}]{Hillier2018-vd}%
  \BibitemOpen
  \bibfield  {author} {\bibinfo {author} {\bibfnamefont {A.~D.}\ \bibnamefont {Hillier}}, \bibinfo {author} {\bibfnamefont {J.~S.}\ \bibnamefont {Lord}}, \bibinfo {author} {\bibfnamefont {K.}~\bibnamefont {Ishida}},\ and\ \bibinfo {author} {\bibfnamefont {C.}~\bibnamefont {Rogers}},\ }\href@noop {} {\bibfield  {journal} {\bibinfo  {journal} {Philos. Trans. A Math. Phys. Eng. Sci.}\ }\textbf {\bibinfo {volume} {377}},\ \bibinfo {pages} {20180064} (\bibinfo {year} {2018})}\BibitemShut {NoStop}%
\bibitem [{\citenamefont {Higemoto}\ \emph {et~al.}(2017)\citenamefont {Higemoto}, \citenamefont {Kadono}, \citenamefont {Kawamura}, \citenamefont {Koda}, \citenamefont {Kojima}, \citenamefont {Makimura}, \citenamefont {Matoba}, \citenamefont {Miyake}, \citenamefont {Shimomura},\ and\ \citenamefont {Strasser}}]{Higemoto2017-qv}%
  \BibitemOpen
  \bibfield  {author} {\bibinfo {author} {\bibfnamefont {W.}~\bibnamefont {Higemoto}}, \bibinfo {author} {\bibfnamefont {R.}~\bibnamefont {Kadono}}, \bibinfo {author} {\bibfnamefont {N.}~\bibnamefont {Kawamura}}, \bibinfo {author} {\bibfnamefont {A.}~\bibnamefont {Koda}}, \bibinfo {author} {\bibfnamefont {K.}~\bibnamefont {Kojima}}, \bibinfo {author} {\bibfnamefont {S.}~\bibnamefont {Makimura}}, \bibinfo {author} {\bibfnamefont {S.}~\bibnamefont {Matoba}}, \bibinfo {author} {\bibfnamefont {Y.}~\bibnamefont {Miyake}}, \bibinfo {author} {\bibfnamefont {K.}~\bibnamefont {Shimomura}},\ and\ \bibinfo {author} {\bibfnamefont {P.}~\bibnamefont {Strasser}},\ }\href@noop {} {\bibfield  {journal} {\bibinfo  {journal} {Quantum Beam Sci.}\ }\textbf {\bibinfo {volume} {1}},\ \bibinfo {pages} {11} (\bibinfo {year} {2017})}\BibitemShut {NoStop}%
\bibitem [{\citenamefont {Mizuno}\ \emph {et~al.}(2024)\citenamefont {Mizuno}, \citenamefont {Niikura}, \citenamefont {Saito}, \citenamefont {Matsuzaki}, \citenamefont {Sakurai}, \citenamefont {Amato}, \citenamefont {Asari}, \citenamefont {Biswas}, \citenamefont {Chiu}, \citenamefont {Gerchow}, \citenamefont {Guguchia}, \citenamefont {Janka}, \citenamefont {Ninomiya}, \citenamefont {Ritjoho}, \citenamefont {Sato}, \citenamefont {von Schoeler}, \citenamefont {Tomono}, \citenamefont {Terada},\ and\ \citenamefont {Wang}}]{Mizuno2024-br}%
  \BibitemOpen
  \bibfield  {author} {\bibinfo {author} {\bibfnamefont {R.}~\bibnamefont {Mizuno}}, \bibinfo {author} {\bibfnamefont {M.}~\bibnamefont {Niikura}}, \bibinfo {author} {\bibfnamefont {T.~Y.}\ \bibnamefont {Saito}}, \bibinfo {author} {\bibfnamefont {T.}~\bibnamefont {Matsuzaki}}, \bibinfo {author} {\bibfnamefont {H.}~\bibnamefont {Sakurai}}, \bibinfo {author} {\bibfnamefont {A.}~\bibnamefont {Amato}}, \bibinfo {author} {\bibfnamefont {S.}~\bibnamefont {Asari}}, \bibinfo {author} {\bibfnamefont {S.}~\bibnamefont {Biswas}}, \bibinfo {author} {\bibfnamefont {I.}~\bibnamefont {Chiu}}, \bibinfo {author} {\bibfnamefont {L.}~\bibnamefont {Gerchow}}, \bibinfo {author} {\bibfnamefont {Z.}~\bibnamefont {Guguchia}}, \bibinfo {author} {\bibfnamefont {G.}~\bibnamefont {Janka}}, \bibinfo {author} {\bibfnamefont {K.}~\bibnamefont {Ninomiya}}, \bibinfo {author} {\bibfnamefont {N.}~\bibnamefont {Ritjoho}}, \bibinfo {author} {\bibfnamefont {A.}~\bibnamefont {Sato}}, \bibinfo {author} {\bibfnamefont {K.}~\bibnamefont {von
  Schoeler}}, \bibinfo {author} {\bibfnamefont {D.}~\bibnamefont {Tomono}}, \bibinfo {author} {\bibfnamefont {K.}~\bibnamefont {Terada}},\ and\ \bibinfo {author} {\bibfnamefont {C.}~\bibnamefont {Wang}},\ }\href@noop {} {\bibfield  {journal} {\bibinfo  {journal} {Nucl. Instrum. Methods Phys. Res. A}\ }\textbf {\bibinfo {volume} {1060}},\ \bibinfo {pages} {169029} (\bibinfo {year} {2024})}\BibitemShut {NoStop}%
\bibitem [{CAE()}]{CAENV1730}%
  \BibitemOpen
  \href@noop {} {\emph {\bibinfo {title} {V1730 \& V1725 Digitizer User Manual, UM2792}}},\ \bibinfo {organization} {CAEN S.p.A.}\BibitemShut {Stop}%
\bibitem [{\citenamefont {Agostinelli}\ \emph {et~al.}(2003)\citenamefont {Agostinelli}, \citenamefont {Allison}, \citenamefont {Amako}, \citenamefont {Apostolakis}, \citenamefont {Araujo}, \citenamefont {Arce}, \citenamefont {Asai}, \citenamefont {Axen}, \citenamefont {Banerjee}, \citenamefont {Barrand}, \citenamefont {Behner}, \citenamefont {Bellagamba}, \citenamefont {Boudreau}, \citenamefont {Broglia}, \citenamefont {Brunengo}, \citenamefont {Burkhardt}, \citenamefont {Chauvie}, \citenamefont {Chuma}, \citenamefont {Chytracek}, \citenamefont {Cooperman}, \citenamefont {Cosmo}, \citenamefont {Degtyarenko}, \citenamefont {Dell'Acqua}, \citenamefont {Depaola}, \citenamefont {Dietrich}, \citenamefont {Enami}, \citenamefont {Feliciello}, \citenamefont {Ferguson}, \citenamefont {Fesefeldt}, \citenamefont {Folger}, \citenamefont {Foppiano}, \citenamefont {Forti}, \citenamefont {Garelli}, \citenamefont {Giani}, \citenamefont {Giannitrapani}, \citenamefont {Gibin}, \citenamefont {Gómez~Cadenas}, \citenamefont
  {González}, \citenamefont {Gracia~Abril}, \citenamefont {Greeniaus}, \citenamefont {Greiner}, \citenamefont {Grichine}, \citenamefont {Grossheim}, \citenamefont {Guatelli}, \citenamefont {Gumplinger}, \citenamefont {Hamatsu}, \citenamefont {Hashimoto}, \citenamefont {Hasui}, \citenamefont {Heikkinen}, \citenamefont {Howard}, \citenamefont {Ivanchenko}, \citenamefont {Johnson}, \citenamefont {Jones}, \citenamefont {Kallenbach}, \citenamefont {Kanaya}, \citenamefont {Kawabata}, \citenamefont {Kawabata}, \citenamefont {Kawaguti}, \citenamefont {Kelner}, \citenamefont {Kent}, \citenamefont {Kimura}, \citenamefont {Kodama}, \citenamefont {Kokoulin}, \citenamefont {Kossov}, \citenamefont {Kurashige}, \citenamefont {Lamanna}, \citenamefont {Lampén}, \citenamefont {Lara}, \citenamefont {Lefebure}, \citenamefont {Lei}, \citenamefont {Liendl}, \citenamefont {Lockman}, \citenamefont {Longo}, \citenamefont {Magni}, \citenamefont {Maire}, \citenamefont {Medernach}, \citenamefont {Minamimoto}, \citenamefont {Mora~de
  Freitas}, \citenamefont {Morita}, \citenamefont {Murakami}, \citenamefont {Nagamatu}, \citenamefont {Nartallo}, \citenamefont {Nieminen}, \citenamefont {Nishimura}, \citenamefont {Ohtsubo}, \citenamefont {Okamura}, \citenamefont {O'Neale}, \citenamefont {Oohata}, \citenamefont {Paech}, \citenamefont {Perl}, \citenamefont {Pfeiffer}, \citenamefont {Pia}, \citenamefont {Ranjard}, \citenamefont {Rybin}, \citenamefont {Sadilov}, \citenamefont {Di~Salvo}, \citenamefont {Santin}, \citenamefont {Sasaki}, \citenamefont {Savvas}, \citenamefont {Sawada}, \citenamefont {Scherer}, \citenamefont {Sei}, \citenamefont {Sirotenko}, \citenamefont {Smith}, \citenamefont {Starkov}, \citenamefont {Stoecker}, \citenamefont {Sulkimo}, \citenamefont {Takahata}, \citenamefont {Tanaka}, \citenamefont {Tcherniaev}, \citenamefont {Safai~Tehrani}, \citenamefont {Tropeano}, \citenamefont {Truscott}, \citenamefont {Uno}, \citenamefont {Urban}, \citenamefont {Urban}, \citenamefont {Verderi}, \citenamefont {Walkden}, \citenamefont
  {Wander}, \citenamefont {Weber}, \citenamefont {Wellisch}, \citenamefont {Wenaus}, \citenamefont {Williams}, \citenamefont {Wright}, \citenamefont {Yamada}, \citenamefont {Yoshida},\ and\ \citenamefont {Zschiesche}}]{Agostinelli2003-eh}%
  \BibitemOpen
  \bibfield  {author} {\bibinfo {author} {\bibfnamefont {S.}~\bibnamefont {Agostinelli}}, \bibinfo {author} {\bibfnamefont {J.}~\bibnamefont {Allison}}, \bibinfo {author} {\bibfnamefont {K.}~\bibnamefont {Amako}}, \bibinfo {author} {\bibfnamefont {J.}~\bibnamefont {Apostolakis}}, \bibinfo {author} {\bibfnamefont {H.}~\bibnamefont {Araujo}}, \bibinfo {author} {\bibfnamefont {P.}~\bibnamefont {Arce}}, \bibinfo {author} {\bibfnamefont {M.}~\bibnamefont {Asai}}, \bibinfo {author} {\bibfnamefont {D.}~\bibnamefont {Axen}}, \bibinfo {author} {\bibfnamefont {S.}~\bibnamefont {Banerjee}}, \bibinfo {author} {\bibfnamefont {G.}~\bibnamefont {Barrand}}, \bibinfo {author} {\bibfnamefont {F.}~\bibnamefont {Behner}}, \bibinfo {author} {\bibfnamefont {L.}~\bibnamefont {Bellagamba}}, \bibinfo {author} {\bibfnamefont {J.}~\bibnamefont {Boudreau}}, \bibinfo {author} {\bibfnamefont {L.}~\bibnamefont {Broglia}}, \bibinfo {author} {\bibfnamefont {A.}~\bibnamefont {Brunengo}}, \bibinfo {author} {\bibfnamefont {H.}~\bibnamefont
  {Burkhardt}}, \bibinfo {author} {\bibfnamefont {S.}~\bibnamefont {Chauvie}}, \bibinfo {author} {\bibfnamefont {J.}~\bibnamefont {Chuma}}, \bibinfo {author} {\bibfnamefont {R.}~\bibnamefont {Chytracek}}, \bibinfo {author} {\bibfnamefont {G.}~\bibnamefont {Cooperman}}, \bibinfo {author} {\bibfnamefont {G.}~\bibnamefont {Cosmo}}, \bibinfo {author} {\bibfnamefont {P.}~\bibnamefont {Degtyarenko}}, \bibinfo {author} {\bibfnamefont {A.}~\bibnamefont {Dell'Acqua}}, \bibinfo {author} {\bibfnamefont {G.}~\bibnamefont {Depaola}}, \bibinfo {author} {\bibfnamefont {D.}~\bibnamefont {Dietrich}}, \bibinfo {author} {\bibfnamefont {R.}~\bibnamefont {Enami}}, \bibinfo {author} {\bibfnamefont {A.}~\bibnamefont {Feliciello}}, \bibinfo {author} {\bibfnamefont {C.}~\bibnamefont {Ferguson}}, \bibinfo {author} {\bibfnamefont {H.}~\bibnamefont {Fesefeldt}}, \bibinfo {author} {\bibfnamefont {G.}~\bibnamefont {Folger}}, \bibinfo {author} {\bibfnamefont {F.}~\bibnamefont {Foppiano}}, \bibinfo {author} {\bibfnamefont {A.}~\bibnamefont
  {Forti}}, \bibinfo {author} {\bibfnamefont {S.}~\bibnamefont {Garelli}}, \bibinfo {author} {\bibfnamefont {S.}~\bibnamefont {Giani}}, \bibinfo {author} {\bibfnamefont {R.}~\bibnamefont {Giannitrapani}}, \bibinfo {author} {\bibfnamefont {D.}~\bibnamefont {Gibin}}, \bibinfo {author} {\bibfnamefont {J.~J.}\ \bibnamefont {Gómez~Cadenas}}, \bibinfo {author} {\bibfnamefont {I.}~\bibnamefont {González}}, \bibinfo {author} {\bibfnamefont {G.}~\bibnamefont {Gracia~Abril}}, \bibinfo {author} {\bibfnamefont {G.}~\bibnamefont {Greeniaus}}, \bibinfo {author} {\bibfnamefont {W.}~\bibnamefont {Greiner}}, \bibinfo {author} {\bibfnamefont {V.}~\bibnamefont {Grichine}}, \bibinfo {author} {\bibfnamefont {A.}~\bibnamefont {Grossheim}}, \bibinfo {author} {\bibfnamefont {S.}~\bibnamefont {Guatelli}}, \bibinfo {author} {\bibfnamefont {P.}~\bibnamefont {Gumplinger}}, \bibinfo {author} {\bibfnamefont {R.}~\bibnamefont {Hamatsu}}, \bibinfo {author} {\bibfnamefont {K.}~\bibnamefont {Hashimoto}}, \bibinfo {author} {\bibfnamefont
  {H.}~\bibnamefont {Hasui}}, \bibinfo {author} {\bibfnamefont {A.}~\bibnamefont {Heikkinen}}, \bibinfo {author} {\bibfnamefont {A.}~\bibnamefont {Howard}}, \bibinfo {author} {\bibfnamefont {V.}~\bibnamefont {Ivanchenko}}, \bibinfo {author} {\bibfnamefont {A.}~\bibnamefont {Johnson}}, \bibinfo {author} {\bibfnamefont {F.~W.}\ \bibnamefont {Jones}}, \bibinfo {author} {\bibfnamefont {J.}~\bibnamefont {Kallenbach}}, \bibinfo {author} {\bibfnamefont {N.}~\bibnamefont {Kanaya}}, \bibinfo {author} {\bibfnamefont {M.}~\bibnamefont {Kawabata}}, \bibinfo {author} {\bibfnamefont {Y.}~\bibnamefont {Kawabata}}, \bibinfo {author} {\bibfnamefont {M.}~\bibnamefont {Kawaguti}}, \bibinfo {author} {\bibfnamefont {S.}~\bibnamefont {Kelner}}, \bibinfo {author} {\bibfnamefont {P.}~\bibnamefont {Kent}}, \bibinfo {author} {\bibfnamefont {A.}~\bibnamefont {Kimura}}, \bibinfo {author} {\bibfnamefont {T.}~\bibnamefont {Kodama}}, \bibinfo {author} {\bibfnamefont {R.}~\bibnamefont {Kokoulin}}, \bibinfo {author} {\bibfnamefont
  {M.}~\bibnamefont {Kossov}}, \bibinfo {author} {\bibfnamefont {H.}~\bibnamefont {Kurashige}}, \bibinfo {author} {\bibfnamefont {E.}~\bibnamefont {Lamanna}}, \bibinfo {author} {\bibfnamefont {T.}~\bibnamefont {Lampén}}, \bibinfo {author} {\bibfnamefont {V.}~\bibnamefont {Lara}}, \bibinfo {author} {\bibfnamefont {V.}~\bibnamefont {Lefebure}}, \bibinfo {author} {\bibfnamefont {F.}~\bibnamefont {Lei}}, \bibinfo {author} {\bibfnamefont {M.}~\bibnamefont {Liendl}}, \bibinfo {author} {\bibfnamefont {W.}~\bibnamefont {Lockman}}, \bibinfo {author} {\bibfnamefont {F.}~\bibnamefont {Longo}}, \bibinfo {author} {\bibfnamefont {S.}~\bibnamefont {Magni}}, \bibinfo {author} {\bibfnamefont {M.}~\bibnamefont {Maire}}, \bibinfo {author} {\bibfnamefont {E.}~\bibnamefont {Medernach}}, \bibinfo {author} {\bibfnamefont {K.}~\bibnamefont {Minamimoto}}, \bibinfo {author} {\bibfnamefont {P.}~\bibnamefont {Mora~de Freitas}}, \bibinfo {author} {\bibfnamefont {Y.}~\bibnamefont {Morita}}, \bibinfo {author} {\bibfnamefont
  {K.}~\bibnamefont {Murakami}}, \bibinfo {author} {\bibfnamefont {M.}~\bibnamefont {Nagamatu}}, \bibinfo {author} {\bibfnamefont {R.}~\bibnamefont {Nartallo}}, \bibinfo {author} {\bibfnamefont {P.}~\bibnamefont {Nieminen}}, \bibinfo {author} {\bibfnamefont {T.}~\bibnamefont {Nishimura}}, \bibinfo {author} {\bibfnamefont {K.}~\bibnamefont {Ohtsubo}}, \bibinfo {author} {\bibfnamefont {M.}~\bibnamefont {Okamura}}, \bibinfo {author} {\bibfnamefont {S.}~\bibnamefont {O'Neale}}, \bibinfo {author} {\bibfnamefont {Y.}~\bibnamefont {Oohata}}, \bibinfo {author} {\bibfnamefont {K.}~\bibnamefont {Paech}}, \bibinfo {author} {\bibfnamefont {J.}~\bibnamefont {Perl}}, \bibinfo {author} {\bibfnamefont {A.}~\bibnamefont {Pfeiffer}}, \bibinfo {author} {\bibfnamefont {M.~G.}\ \bibnamefont {Pia}}, \bibinfo {author} {\bibfnamefont {F.}~\bibnamefont {Ranjard}}, \bibinfo {author} {\bibfnamefont {A.}~\bibnamefont {Rybin}}, \bibinfo {author} {\bibfnamefont {S.}~\bibnamefont {Sadilov}}, \bibinfo {author} {\bibfnamefont
  {E.}~\bibnamefont {Di~Salvo}}, \bibinfo {author} {\bibfnamefont {G.}~\bibnamefont {Santin}}, \bibinfo {author} {\bibfnamefont {T.}~\bibnamefont {Sasaki}}, \bibinfo {author} {\bibfnamefont {N.}~\bibnamefont {Savvas}}, \bibinfo {author} {\bibfnamefont {Y.}~\bibnamefont {Sawada}}, \bibinfo {author} {\bibfnamefont {S.}~\bibnamefont {Scherer}}, \bibinfo {author} {\bibfnamefont {S.}~\bibnamefont {Sei}}, \bibinfo {author} {\bibfnamefont {V.}~\bibnamefont {Sirotenko}}, \bibinfo {author} {\bibfnamefont {D.}~\bibnamefont {Smith}}, \bibinfo {author} {\bibfnamefont {N.}~\bibnamefont {Starkov}}, \bibinfo {author} {\bibfnamefont {H.}~\bibnamefont {Stoecker}}, \bibinfo {author} {\bibfnamefont {J.}~\bibnamefont {Sulkimo}}, \bibinfo {author} {\bibfnamefont {M.}~\bibnamefont {Takahata}}, \bibinfo {author} {\bibfnamefont {S.}~\bibnamefont {Tanaka}}, \bibinfo {author} {\bibfnamefont {E.}~\bibnamefont {Tcherniaev}}, \bibinfo {author} {\bibfnamefont {E.}~\bibnamefont {Safai~Tehrani}}, \bibinfo {author} {\bibfnamefont
  {M.}~\bibnamefont {Tropeano}}, \bibinfo {author} {\bibfnamefont {P.}~\bibnamefont {Truscott}}, \bibinfo {author} {\bibfnamefont {H.}~\bibnamefont {Uno}}, \bibinfo {author} {\bibfnamefont {L.}~\bibnamefont {Urban}}, \bibinfo {author} {\bibfnamefont {P.}~\bibnamefont {Urban}}, \bibinfo {author} {\bibfnamefont {M.}~\bibnamefont {Verderi}}, \bibinfo {author} {\bibfnamefont {A.}~\bibnamefont {Walkden}}, \bibinfo {author} {\bibfnamefont {W.}~\bibnamefont {Wander}}, \bibinfo {author} {\bibfnamefont {H.}~\bibnamefont {Weber}}, \bibinfo {author} {\bibfnamefont {J.~P.}\ \bibnamefont {Wellisch}}, \bibinfo {author} {\bibfnamefont {T.}~\bibnamefont {Wenaus}}, \bibinfo {author} {\bibfnamefont {D.~C.}\ \bibnamefont {Williams}}, \bibinfo {author} {\bibfnamefont {D.}~\bibnamefont {Wright}}, \bibinfo {author} {\bibfnamefont {T.}~\bibnamefont {Yamada}}, \bibinfo {author} {\bibfnamefont {H.}~\bibnamefont {Yoshida}},\ and\ \bibinfo {author} {\bibfnamefont {D.}~\bibnamefont {Zschiesche}},\ }\href@noop {} {\bibfield  {journal}
  {\bibinfo  {journal} {Nucl. Instrum. Methods Phys. Res. A}\ }\textbf {\bibinfo {volume} {506}},\ \bibinfo {pages} {250} (\bibinfo {year} {2003})}\BibitemShut {NoStop}%
\bibitem [{\citenamefont {Allison}\ \emph {et~al.}(2006)\citenamefont {Allison}, \citenamefont {Amako}, \citenamefont {Apostolakis}, \citenamefont {Araujo}, \citenamefont {Arce~Dubois}, \citenamefont {Asai}, \citenamefont {Barrand}, \citenamefont {Capra}, \citenamefont {Chauvie}, \citenamefont {Chytracek}, \citenamefont {Cirrone}, \citenamefont {Cooperman}, \citenamefont {Cosmo}, \citenamefont {Cuttone}, \citenamefont {Daquino}, \citenamefont {Donszelmann}, \citenamefont {Dressel}, \citenamefont {Folger}, \citenamefont {Foppiano}, \citenamefont {Generowicz}, \citenamefont {Grichine}, \citenamefont {Guatelli}, \citenamefont {Gumplinger}, \citenamefont {Heikkinen}, \citenamefont {Hrivnacova}, \citenamefont {Howard}, \citenamefont {Incerti}, \citenamefont {Ivanchenko}, \citenamefont {Johnson}, \citenamefont {Jones}, \citenamefont {Koi}, \citenamefont {Kokoulin}, \citenamefont {Kossov}, \citenamefont {Kurashige}, \citenamefont {Lara}, \citenamefont {Larsson}, \citenamefont {Lei}, \citenamefont {Link},
  \citenamefont {Longo}, \citenamefont {Maire}, \citenamefont {Mantero}, \citenamefont {Mascialino}, \citenamefont {McLaren}, \citenamefont {Mendez~Lorenzo}, \citenamefont {Minamimoto}, \citenamefont {Murakami}, \citenamefont {Nieminen}, \citenamefont {Pandola}, \citenamefont {Parlati}, \citenamefont {Peralta}, \citenamefont {Perl}, \citenamefont {Pfeiffer}, \citenamefont {Pia}, \citenamefont {Ribon}, \citenamefont {Rodrigues}, \citenamefont {Russo}, \citenamefont {Sadilov}, \citenamefont {Santin}, \citenamefont {Sasaki}, \citenamefont {Smith}, \citenamefont {Starkov}, \citenamefont {Tanaka}, \citenamefont {Tcherniaev}, \citenamefont {Tome}, \citenamefont {Trindade}, \citenamefont {Truscott}, \citenamefont {Urban}, \citenamefont {Verderi}, \citenamefont {Walkden}, \citenamefont {Wellisch}, \citenamefont {Williams}, \citenamefont {Wright},\ and\ \citenamefont {Yoshida}}]{Allison2006-ho}%
  \BibitemOpen
  \bibfield  {author} {\bibinfo {author} {\bibfnamefont {J.}~\bibnamefont {Allison}}, \bibinfo {author} {\bibfnamefont {K.}~\bibnamefont {Amako}}, \bibinfo {author} {\bibfnamefont {J.}~\bibnamefont {Apostolakis}}, \bibinfo {author} {\bibfnamefont {H.}~\bibnamefont {Araujo}}, \bibinfo {author} {\bibfnamefont {P.}~\bibnamefont {Arce~Dubois}}, \bibinfo {author} {\bibfnamefont {M.}~\bibnamefont {Asai}}, \bibinfo {author} {\bibfnamefont {G.}~\bibnamefont {Barrand}}, \bibinfo {author} {\bibfnamefont {R.}~\bibnamefont {Capra}}, \bibinfo {author} {\bibfnamefont {S.}~\bibnamefont {Chauvie}}, \bibinfo {author} {\bibfnamefont {R.}~\bibnamefont {Chytracek}}, \bibinfo {author} {\bibfnamefont {G.~A.~P.}\ \bibnamefont {Cirrone}}, \bibinfo {author} {\bibfnamefont {G.}~\bibnamefont {Cooperman}}, \bibinfo {author} {\bibfnamefont {G.}~\bibnamefont {Cosmo}}, \bibinfo {author} {\bibfnamefont {G.}~\bibnamefont {Cuttone}}, \bibinfo {author} {\bibfnamefont {G.~G.}\ \bibnamefont {Daquino}}, \bibinfo {author} {\bibfnamefont
  {M.}~\bibnamefont {Donszelmann}}, \bibinfo {author} {\bibfnamefont {M.}~\bibnamefont {Dressel}}, \bibinfo {author} {\bibfnamefont {G.}~\bibnamefont {Folger}}, \bibinfo {author} {\bibfnamefont {F.}~\bibnamefont {Foppiano}}, \bibinfo {author} {\bibfnamefont {J.}~\bibnamefont {Generowicz}}, \bibinfo {author} {\bibfnamefont {V.}~\bibnamefont {Grichine}}, \bibinfo {author} {\bibfnamefont {S.}~\bibnamefont {Guatelli}}, \bibinfo {author} {\bibfnamefont {P.}~\bibnamefont {Gumplinger}}, \bibinfo {author} {\bibfnamefont {A.}~\bibnamefont {Heikkinen}}, \bibinfo {author} {\bibfnamefont {I.}~\bibnamefont {Hrivnacova}}, \bibinfo {author} {\bibfnamefont {A.}~\bibnamefont {Howard}}, \bibinfo {author} {\bibfnamefont {S.}~\bibnamefont {Incerti}}, \bibinfo {author} {\bibfnamefont {V.}~\bibnamefont {Ivanchenko}}, \bibinfo {author} {\bibfnamefont {T.}~\bibnamefont {Johnson}}, \bibinfo {author} {\bibfnamefont {F.}~\bibnamefont {Jones}}, \bibinfo {author} {\bibfnamefont {T.}~\bibnamefont {Koi}}, \bibinfo {author} {\bibfnamefont
  {R.}~\bibnamefont {Kokoulin}}, \bibinfo {author} {\bibfnamefont {M.}~\bibnamefont {Kossov}}, \bibinfo {author} {\bibfnamefont {H.}~\bibnamefont {Kurashige}}, \bibinfo {author} {\bibfnamefont {V.}~\bibnamefont {Lara}}, \bibinfo {author} {\bibfnamefont {S.}~\bibnamefont {Larsson}}, \bibinfo {author} {\bibfnamefont {F.}~\bibnamefont {Lei}}, \bibinfo {author} {\bibfnamefont {O.}~\bibnamefont {Link}}, \bibinfo {author} {\bibfnamefont {F.}~\bibnamefont {Longo}}, \bibinfo {author} {\bibfnamefont {M.}~\bibnamefont {Maire}}, \bibinfo {author} {\bibfnamefont {A.}~\bibnamefont {Mantero}}, \bibinfo {author} {\bibfnamefont {B.}~\bibnamefont {Mascialino}}, \bibinfo {author} {\bibfnamefont {I.}~\bibnamefont {McLaren}}, \bibinfo {author} {\bibfnamefont {P.}~\bibnamefont {Mendez~Lorenzo}}, \bibinfo {author} {\bibfnamefont {K.}~\bibnamefont {Minamimoto}}, \bibinfo {author} {\bibfnamefont {K.}~\bibnamefont {Murakami}}, \bibinfo {author} {\bibfnamefont {P.}~\bibnamefont {Nieminen}}, \bibinfo {author} {\bibfnamefont
  {L.}~\bibnamefont {Pandola}}, \bibinfo {author} {\bibfnamefont {S.}~\bibnamefont {Parlati}}, \bibinfo {author} {\bibfnamefont {L.}~\bibnamefont {Peralta}}, \bibinfo {author} {\bibfnamefont {J.}~\bibnamefont {Perl}}, \bibinfo {author} {\bibfnamefont {A.}~\bibnamefont {Pfeiffer}}, \bibinfo {author} {\bibfnamefont {M.~G.}\ \bibnamefont {Pia}}, \bibinfo {author} {\bibfnamefont {A.}~\bibnamefont {Ribon}}, \bibinfo {author} {\bibfnamefont {P.}~\bibnamefont {Rodrigues}}, \bibinfo {author} {\bibfnamefont {G.}~\bibnamefont {Russo}}, \bibinfo {author} {\bibfnamefont {S.}~\bibnamefont {Sadilov}}, \bibinfo {author} {\bibfnamefont {G.}~\bibnamefont {Santin}}, \bibinfo {author} {\bibfnamefont {T.}~\bibnamefont {Sasaki}}, \bibinfo {author} {\bibfnamefont {D.}~\bibnamefont {Smith}}, \bibinfo {author} {\bibfnamefont {N.}~\bibnamefont {Starkov}}, \bibinfo {author} {\bibfnamefont {S.}~\bibnamefont {Tanaka}}, \bibinfo {author} {\bibfnamefont {E.}~\bibnamefont {Tcherniaev}}, \bibinfo {author} {\bibfnamefont {B.}~\bibnamefont
  {Tome}}, \bibinfo {author} {\bibfnamefont {A.}~\bibnamefont {Trindade}}, \bibinfo {author} {\bibfnamefont {P.}~\bibnamefont {Truscott}}, \bibinfo {author} {\bibfnamefont {L.}~\bibnamefont {Urban}}, \bibinfo {author} {\bibfnamefont {M.}~\bibnamefont {Verderi}}, \bibinfo {author} {\bibfnamefont {A.}~\bibnamefont {Walkden}}, \bibinfo {author} {\bibfnamefont {J.~P.}\ \bibnamefont {Wellisch}}, \bibinfo {author} {\bibfnamefont {D.~C.}\ \bibnamefont {Williams}}, \bibinfo {author} {\bibfnamefont {D.}~\bibnamefont {Wright}},\ and\ \bibinfo {author} {\bibfnamefont {H.}~\bibnamefont {Yoshida}},\ }\href@noop {} {\bibfield  {journal} {\bibinfo  {journal} {IEEE Trans. Nucl. Sci.}\ }\textbf {\bibinfo {volume} {53}},\ \bibinfo {pages} {270} (\bibinfo {year} {2006})}\BibitemShut {NoStop}%
\bibitem [{\citenamefont {Allison}\ \emph {et~al.}(2016)\citenamefont {Allison}, \citenamefont {Amako}, \citenamefont {Apostolakis}, \citenamefont {Arce}, \citenamefont {Asai}, \citenamefont {Aso}, \citenamefont {Bagli}, \citenamefont {Bagulya}, \citenamefont {Banerjee}, \citenamefont {Barrand}, \citenamefont {Beck}, \citenamefont {Bogdanov}, \citenamefont {Brandt}, \citenamefont {Brown}, \citenamefont {Burkhardt}, \citenamefont {Canal}, \citenamefont {Cano-Ott}, \citenamefont {Chauvie}, \citenamefont {Cho}, \citenamefont {Cirrone}, \citenamefont {Cooperman}, \citenamefont {Cortés-Giraldo}, \citenamefont {Cosmo}, \citenamefont {Cuttone}, \citenamefont {Depaola}, \citenamefont {Desorgher}, \citenamefont {Dong}, \citenamefont {Dotti}, \citenamefont {Elvira}, \citenamefont {Folger}, \citenamefont {Francis}, \citenamefont {Galoyan}, \citenamefont {Garnier}, \citenamefont {Gayer}, \citenamefont {Genser}, \citenamefont {Grichine}, \citenamefont {Guatelli}, \citenamefont {Guèye}, \citenamefont {Gumplinger},
  \citenamefont {Howard}, \citenamefont {Hřivnáčová}, \citenamefont {Hwang}, \citenamefont {Incerti}, \citenamefont {Ivanchenko}, \citenamefont {Ivanchenko}, \citenamefont {Jones}, \citenamefont {Jun}, \citenamefont {Kaitaniemi}, \citenamefont {Karakatsanis}, \citenamefont {Karamitros}, \citenamefont {Kelsey}, \citenamefont {Kimura}, \citenamefont {Koi}, \citenamefont {Kurashige}, \citenamefont {Lechner}, \citenamefont {Lee}, \citenamefont {Longo}, \citenamefont {Maire}, \citenamefont {Mancusi}, \citenamefont {Mantero}, \citenamefont {Mendoza}, \citenamefont {Morgan}, \citenamefont {Murakami}, \citenamefont {Nikitina}, \citenamefont {Pandola}, \citenamefont {Paprocki}, \citenamefont {Perl}, \citenamefont {Petrović}, \citenamefont {Pia}, \citenamefont {Pokorski}, \citenamefont {Quesada}, \citenamefont {Raine}, \citenamefont {Reis}, \citenamefont {Ribon}, \citenamefont {Ristić~Fira}, \citenamefont {Romano}, \citenamefont {Russo}, \citenamefont {Santin}, \citenamefont {Sasaki}, \citenamefont {Sawkey},
  \citenamefont {Shin}, \citenamefont {Strakovsky}, \citenamefont {Taborda}, \citenamefont {Tanaka}, \citenamefont {Tomé}, \citenamefont {Toshito}, \citenamefont {Tran}, \citenamefont {Truscott}, \citenamefont {Urban}, \citenamefont {Uzhinsky}, \citenamefont {Verbeke}, \citenamefont {Verderi}, \citenamefont {Wendt}, \citenamefont {Wenzel}, \citenamefont {Wright}, \citenamefont {Wright}, \citenamefont {Yamashita}, \citenamefont {Yarba},\ and\ \citenamefont {Yoshida}}]{Allison2016-zr}%
  \BibitemOpen
  \bibfield  {author} {\bibinfo {author} {\bibfnamefont {J.}~\bibnamefont {Allison}}, \bibinfo {author} {\bibfnamefont {K.}~\bibnamefont {Amako}}, \bibinfo {author} {\bibfnamefont {J.}~\bibnamefont {Apostolakis}}, \bibinfo {author} {\bibfnamefont {P.}~\bibnamefont {Arce}}, \bibinfo {author} {\bibfnamefont {M.}~\bibnamefont {Asai}}, \bibinfo {author} {\bibfnamefont {T.}~\bibnamefont {Aso}}, \bibinfo {author} {\bibfnamefont {E.}~\bibnamefont {Bagli}}, \bibinfo {author} {\bibfnamefont {A.}~\bibnamefont {Bagulya}}, \bibinfo {author} {\bibfnamefont {S.}~\bibnamefont {Banerjee}}, \bibinfo {author} {\bibfnamefont {G.}~\bibnamefont {Barrand}}, \bibinfo {author} {\bibfnamefont {B.~R.}\ \bibnamefont {Beck}}, \bibinfo {author} {\bibfnamefont {A.~G.}\ \bibnamefont {Bogdanov}}, \bibinfo {author} {\bibfnamefont {D.}~\bibnamefont {Brandt}}, \bibinfo {author} {\bibfnamefont {J.~M.~C.}\ \bibnamefont {Brown}}, \bibinfo {author} {\bibfnamefont {H.}~\bibnamefont {Burkhardt}}, \bibinfo {author} {\bibfnamefont {P.}~\bibnamefont
  {Canal}}, \bibinfo {author} {\bibfnamefont {D.}~\bibnamefont {Cano-Ott}}, \bibinfo {author} {\bibfnamefont {S.}~\bibnamefont {Chauvie}}, \bibinfo {author} {\bibfnamefont {K.}~\bibnamefont {Cho}}, \bibinfo {author} {\bibfnamefont {G.~A.~P.}\ \bibnamefont {Cirrone}}, \bibinfo {author} {\bibfnamefont {G.}~\bibnamefont {Cooperman}}, \bibinfo {author} {\bibfnamefont {M.~A.}\ \bibnamefont {Cortés-Giraldo}}, \bibinfo {author} {\bibfnamefont {G.}~\bibnamefont {Cosmo}}, \bibinfo {author} {\bibfnamefont {G.}~\bibnamefont {Cuttone}}, \bibinfo {author} {\bibfnamefont {G.}~\bibnamefont {Depaola}}, \bibinfo {author} {\bibfnamefont {L.}~\bibnamefont {Desorgher}}, \bibinfo {author} {\bibfnamefont {X.}~\bibnamefont {Dong}}, \bibinfo {author} {\bibfnamefont {A.}~\bibnamefont {Dotti}}, \bibinfo {author} {\bibfnamefont {V.~D.}\ \bibnamefont {Elvira}}, \bibinfo {author} {\bibfnamefont {G.}~\bibnamefont {Folger}}, \bibinfo {author} {\bibfnamefont {Z.}~\bibnamefont {Francis}}, \bibinfo {author} {\bibfnamefont {A.}~\bibnamefont
  {Galoyan}}, \bibinfo {author} {\bibfnamefont {L.}~\bibnamefont {Garnier}}, \bibinfo {author} {\bibfnamefont {M.}~\bibnamefont {Gayer}}, \bibinfo {author} {\bibfnamefont {K.~L.}\ \bibnamefont {Genser}}, \bibinfo {author} {\bibfnamefont {V.~M.}\ \bibnamefont {Grichine}}, \bibinfo {author} {\bibfnamefont {S.}~\bibnamefont {Guatelli}}, \bibinfo {author} {\bibfnamefont {P.}~\bibnamefont {Guèye}}, \bibinfo {author} {\bibfnamefont {P.}~\bibnamefont {Gumplinger}}, \bibinfo {author} {\bibfnamefont {A.~S.}\ \bibnamefont {Howard}}, \bibinfo {author} {\bibfnamefont {I.}~\bibnamefont {Hřivnáčová}}, \bibinfo {author} {\bibfnamefont {S.}~\bibnamefont {Hwang}}, \bibinfo {author} {\bibfnamefont {S.}~\bibnamefont {Incerti}}, \bibinfo {author} {\bibfnamefont {A.}~\bibnamefont {Ivanchenko}}, \bibinfo {author} {\bibfnamefont {V.~N.}\ \bibnamefont {Ivanchenko}}, \bibinfo {author} {\bibfnamefont {F.~W.}\ \bibnamefont {Jones}}, \bibinfo {author} {\bibfnamefont {S.~Y.}\ \bibnamefont {Jun}}, \bibinfo {author} {\bibfnamefont
  {P.}~\bibnamefont {Kaitaniemi}}, \bibinfo {author} {\bibfnamefont {N.}~\bibnamefont {Karakatsanis}}, \bibinfo {author} {\bibfnamefont {M.}~\bibnamefont {Karamitros}}, \bibinfo {author} {\bibfnamefont {M.}~\bibnamefont {Kelsey}}, \bibinfo {author} {\bibfnamefont {A.}~\bibnamefont {Kimura}}, \bibinfo {author} {\bibfnamefont {T.}~\bibnamefont {Koi}}, \bibinfo {author} {\bibfnamefont {H.}~\bibnamefont {Kurashige}}, \bibinfo {author} {\bibfnamefont {A.}~\bibnamefont {Lechner}}, \bibinfo {author} {\bibfnamefont {S.~B.}\ \bibnamefont {Lee}}, \bibinfo {author} {\bibfnamefont {F.}~\bibnamefont {Longo}}, \bibinfo {author} {\bibfnamefont {M.}~\bibnamefont {Maire}}, \bibinfo {author} {\bibfnamefont {D.}~\bibnamefont {Mancusi}}, \bibinfo {author} {\bibfnamefont {A.}~\bibnamefont {Mantero}}, \bibinfo {author} {\bibfnamefont {E.}~\bibnamefont {Mendoza}}, \bibinfo {author} {\bibfnamefont {B.}~\bibnamefont {Morgan}}, \bibinfo {author} {\bibfnamefont {K.}~\bibnamefont {Murakami}}, \bibinfo {author} {\bibfnamefont
  {T.}~\bibnamefont {Nikitina}}, \bibinfo {author} {\bibfnamefont {L.}~\bibnamefont {Pandola}}, \bibinfo {author} {\bibfnamefont {P.}~\bibnamefont {Paprocki}}, \bibinfo {author} {\bibfnamefont {J.}~\bibnamefont {Perl}}, \bibinfo {author} {\bibfnamefont {I.}~\bibnamefont {Petrović}}, \bibinfo {author} {\bibfnamefont {M.~G.}\ \bibnamefont {Pia}}, \bibinfo {author} {\bibfnamefont {W.}~\bibnamefont {Pokorski}}, \bibinfo {author} {\bibfnamefont {J.~M.}\ \bibnamefont {Quesada}}, \bibinfo {author} {\bibfnamefont {M.}~\bibnamefont {Raine}}, \bibinfo {author} {\bibfnamefont {M.~A.}\ \bibnamefont {Reis}}, \bibinfo {author} {\bibfnamefont {A.}~\bibnamefont {Ribon}}, \bibinfo {author} {\bibfnamefont {A.}~\bibnamefont {Ristić~Fira}}, \bibinfo {author} {\bibfnamefont {F.}~\bibnamefont {Romano}}, \bibinfo {author} {\bibfnamefont {G.}~\bibnamefont {Russo}}, \bibinfo {author} {\bibfnamefont {G.}~\bibnamefont {Santin}}, \bibinfo {author} {\bibfnamefont {T.}~\bibnamefont {Sasaki}}, \bibinfo {author} {\bibfnamefont
  {D.}~\bibnamefont {Sawkey}}, \bibinfo {author} {\bibfnamefont {J.~I.}\ \bibnamefont {Shin}}, \bibinfo {author} {\bibfnamefont {I.~I.}\ \bibnamefont {Strakovsky}}, \bibinfo {author} {\bibfnamefont {A.}~\bibnamefont {Taborda}}, \bibinfo {author} {\bibfnamefont {S.}~\bibnamefont {Tanaka}}, \bibinfo {author} {\bibfnamefont {B.}~\bibnamefont {Tomé}}, \bibinfo {author} {\bibfnamefont {T.}~\bibnamefont {Toshito}}, \bibinfo {author} {\bibfnamefont {H.~N.}\ \bibnamefont {Tran}}, \bibinfo {author} {\bibfnamefont {P.~R.}\ \bibnamefont {Truscott}}, \bibinfo {author} {\bibfnamefont {L.}~\bibnamefont {Urban}}, \bibinfo {author} {\bibfnamefont {V.}~\bibnamefont {Uzhinsky}}, \bibinfo {author} {\bibfnamefont {J.~M.}\ \bibnamefont {Verbeke}}, \bibinfo {author} {\bibfnamefont {M.}~\bibnamefont {Verderi}}, \bibinfo {author} {\bibfnamefont {B.~L.}\ \bibnamefont {Wendt}}, \bibinfo {author} {\bibfnamefont {H.}~\bibnamefont {Wenzel}}, \bibinfo {author} {\bibfnamefont {D.~H.}\ \bibnamefont {Wright}}, \bibinfo {author}
  {\bibfnamefont {D.~M.}\ \bibnamefont {Wright}}, \bibinfo {author} {\bibfnamefont {T.}~\bibnamefont {Yamashita}}, \bibinfo {author} {\bibfnamefont {J.}~\bibnamefont {Yarba}},\ and\ \bibinfo {author} {\bibfnamefont {H.}~\bibnamefont {Yoshida}},\ }\href@noop {} {\bibfield  {journal} {\bibinfo  {journal} {Nucl. Instrum. Methods Phys. Res. A}\ }\textbf {\bibinfo {volume} {835}},\ \bibinfo {pages} {186} (\bibinfo {year} {2016})}\BibitemShut {NoStop}%
\bibitem [{\citenamefont {Shamsuzzoha~Basunia}(2012)}]{Shamsuzzoha-Basunia2012-jo}%
  \BibitemOpen
  \bibfield  {author} {\bibinfo {author} {\bibfnamefont {M.}~\bibnamefont {Shamsuzzoha~Basunia}},\ }\href@noop {} {\bibfield  {journal} {\bibinfo  {journal} {Nucl. Data Sheets}\ }\textbf {\bibinfo {volume} {113}},\ \bibinfo {pages} {909} (\bibinfo {year} {2012})}\BibitemShut {NoStop}%
\bibitem [{\citenamefont {Shamsuzzoha~Basunia}(2013)}]{Shamsuzzoha-Basunia2013-wi}%
  \BibitemOpen
  \bibfield  {author} {\bibinfo {author} {\bibfnamefont {M.}~\bibnamefont {Shamsuzzoha~Basunia}},\ }\href@noop {} {\bibfield  {journal} {\bibinfo  {journal} {Nucl. Data Sheets}\ }\textbf {\bibinfo {volume} {114}},\ \bibinfo {pages} {1189} (\bibinfo {year} {2013})}\BibitemShut {NoStop}%
\bibitem [{\citenamefont {Shamsuzzoha~Basunia}(2011)}]{Shamsuzzoha-Basunia2011-nq}%
  \BibitemOpen
  \bibfield  {author} {\bibinfo {author} {\bibfnamefont {M.}~\bibnamefont {Shamsuzzoha~Basunia}},\ }\href@noop {} {\bibfield  {journal} {\bibinfo  {journal} {Nucl. Data Sheets}\ }\textbf {\bibinfo {volume} {112}},\ \bibinfo {pages} {1875} (\bibinfo {year} {2011})}\BibitemShut {NoStop}%
\bibitem [{\citenamefont {Basunia}\ and\ \citenamefont {Hurst}(2016)}]{Basunia2016-mn}%
  \BibitemOpen
  \bibfield  {author} {\bibinfo {author} {\bibfnamefont {M.~S.}\ \bibnamefont {Basunia}}\ and\ \bibinfo {author} {\bibfnamefont {A.~M.}\ \bibnamefont {Hurst}},\ }\href@noop {} {\bibfield  {journal} {\bibinfo  {journal} {Nucl. Data Sheets}\ }\textbf {\bibinfo {volume} {134}},\ \bibinfo {pages} {1} (\bibinfo {year} {2016})}\BibitemShut {NoStop}%
\bibitem [{\citenamefont {Firestone}(2009)}]{Firestone2009-aw}%
  \BibitemOpen
  \bibfield  {author} {\bibinfo {author} {\bibfnamefont {R.~B.}\ \bibnamefont {Firestone}},\ }\href@noop {} {\bibfield  {journal} {\bibinfo  {journal} {Nucl. Data Sheets}\ }\textbf {\bibinfo {volume} {110}},\ \bibinfo {pages} {1691} (\bibinfo {year} {2009})}\BibitemShut {NoStop}%
\bibitem [{\citenamefont {Basunia}\ and\ \citenamefont {Chakraborty}(2022)}]{Basunia2022-zw}%
  \BibitemOpen
  \bibfield  {author} {\bibinfo {author} {\bibfnamefont {M.~S.}\ \bibnamefont {Basunia}}\ and\ \bibinfo {author} {\bibfnamefont {A.}~\bibnamefont {Chakraborty}},\ }\href@noop {} {\bibfield  {journal} {\bibinfo  {journal} {Nucl. Data Sheets}\ }\textbf {\bibinfo {volume} {186}},\ \bibinfo {pages} {3} (\bibinfo {year} {2022})}\BibitemShut {NoStop}%
\bibitem [{\citenamefont {Shamsuzzoha~Basunia}\ and\ \citenamefont {Chakraborty}(2021)}]{Shamsuzzoha-Basunia2021-sj}%
  \BibitemOpen
  \bibfield  {author} {\bibinfo {author} {\bibfnamefont {M.}~\bibnamefont {Shamsuzzoha~Basunia}}\ and\ \bibinfo {author} {\bibfnamefont {A.}~\bibnamefont {Chakraborty}},\ }\href@noop {} {\bibfield  {journal} {\bibinfo  {journal} {Nucl. Data Sheets}\ }\textbf {\bibinfo {volume} {171}},\ \bibinfo {pages} {1} (\bibinfo {year} {2021})}\BibitemShut {NoStop}%
\bibitem [{\citenamefont {Basunia}(2015)}]{Basunia2015-ab}%
  \BibitemOpen
  \bibfield  {author} {\bibinfo {author} {\bibfnamefont {M.~S.}\ \bibnamefont {Basunia}},\ }\href@noop {} {\bibfield  {journal} {\bibinfo  {journal} {Nucl. Data Sheets}\ }\textbf {\bibinfo {volume} {127}},\ \bibinfo {pages} {69} (\bibinfo {year} {2015})}\BibitemShut {NoStop}%
\bibitem [{\citenamefont {Firestone}(2015)}]{Firestone2015-jd}%
  \BibitemOpen
  \bibfield  {author} {\bibinfo {author} {\bibfnamefont {R.~B.}\ \bibnamefont {Firestone}},\ }\href@noop {} {\bibfield  {journal} {\bibinfo  {journal} {Nucl. Data Sheets}\ }\textbf {\bibinfo {volume} {127}},\ \bibinfo {pages} {1} (\bibinfo {year} {2015})}\BibitemShut {NoStop}%
\bibitem [{\citenamefont {Tilley}\ \emph {et~al.}(1998)\citenamefont {Tilley}, \citenamefont {Cheves}, \citenamefont {Kelley}, \citenamefont {Raman},\ and\ \citenamefont {Weller}}]{Tilley1998-sf}%
  \BibitemOpen
  \bibfield  {author} {\bibinfo {author} {\bibfnamefont {D.~R.}\ \bibnamefont {Tilley}}, \bibinfo {author} {\bibfnamefont {C.~M.}\ \bibnamefont {Cheves}}, \bibinfo {author} {\bibfnamefont {J.~H.}\ \bibnamefont {Kelley}}, \bibinfo {author} {\bibfnamefont {S.}~\bibnamefont {Raman}},\ and\ \bibinfo {author} {\bibfnamefont {H.~R.}\ \bibnamefont {Weller}},\ }\href@noop {} {\bibfield  {journal} {\bibinfo  {journal} {Nucl. Phys. A}\ }\textbf {\bibinfo {volume} {636}},\ \bibinfo {pages} {249} (\bibinfo {year} {1998})}\BibitemShut {NoStop}%
\bibitem [{\citenamefont {Tilley}\ \emph {et~al.}(1995)\citenamefont {Tilley}, \citenamefont {Weller}, \citenamefont {Cheves},\ and\ \citenamefont {Chasteler}}]{Tilley1995-zp}%
  \BibitemOpen
  \bibfield  {author} {\bibinfo {author} {\bibfnamefont {D.~R.}\ \bibnamefont {Tilley}}, \bibinfo {author} {\bibfnamefont {H.~R.}\ \bibnamefont {Weller}}, \bibinfo {author} {\bibfnamefont {C.~M.}\ \bibnamefont {Cheves}},\ and\ \bibinfo {author} {\bibfnamefont {R.~M.}\ \bibnamefont {Chasteler}},\ }\href@noop {} {\bibfield  {journal} {\bibinfo  {journal} {Nucl. Phys. A}\ }\textbf {\bibinfo {volume} {595}},\ \bibinfo {pages} {1} (\bibinfo {year} {1995})}\BibitemShut {NoStop}%
\bibitem [{\citenamefont {Alburger}\ and\ \citenamefont {Harris}(1969)}]{Alburger1969-ic}%
  \BibitemOpen
  \bibfield  {author} {\bibinfo {author} {\bibfnamefont {D.~E.}\ \bibnamefont {Alburger}}\ and\ \bibinfo {author} {\bibfnamefont {W.~R.}\ \bibnamefont {Harris}},\ }\href@noop {} {\bibfield  {journal} {\bibinfo  {journal} {Phys. Rev.}\ }\textbf {\bibinfo {volume} {185}},\ \bibinfo {pages} {1495} (\bibinfo {year} {1969})}\BibitemShut {NoStop}%
\bibitem [{\citenamefont {Bateman}(1910)}]{Bateman1910-sl}%
  \BibitemOpen
  \bibfield  {author} {\bibinfo {author} {\bibfnamefont {H.}~\bibnamefont {Bateman}},\ }\href@noop {} {\bibfield  {journal} {\bibinfo  {journal} {Proc. Cambridge Philos. Soc.}\ }\textbf {\bibinfo {volume} {15}},\ \bibinfo {pages} {423} (\bibinfo {year} {1910})}\BibitemShut {NoStop}%
\bibitem [{\citenamefont {Endt}\ and\ \citenamefont {Van Der~Leun}(1974)}]{Endt1974-tz}%
  \BibitemOpen
  \bibfield  {author} {\bibinfo {author} {\bibfnamefont {P.~M.}\ \bibnamefont {Endt}}\ and\ \bibinfo {author} {\bibfnamefont {C.}~\bibnamefont {Van Der~Leun}},\ }\href@noop {} {\bibfield  {journal} {\bibinfo  {journal} {At. Data Nucl. Data Tables}\ }\textbf {\bibinfo {volume} {13}},\ \bibinfo {pages} {67} (\bibinfo {year} {1974})}\BibitemShut {NoStop}%
\bibitem [{\citenamefont {Klotz}\ and\ \citenamefont {Walter}(1974)}]{Klotz1974-mi}%
  \BibitemOpen
  \bibfield  {author} {\bibinfo {author} {\bibfnamefont {G.}~\bibnamefont {Klotz}}\ and\ \bibinfo {author} {\bibfnamefont {G.}~\bibnamefont {Walter}},\ }\href@noop {} {\bibfield  {journal} {\bibinfo  {journal} {Nucl. Phys. A}\ }\textbf {\bibinfo {volume} {227}},\ \bibinfo {pages} {341} (\bibinfo {year} {1974})}\BibitemShut {NoStop}%
\bibitem [{\citenamefont {Alburger}\ and\ \citenamefont {Goosman}(1974)}]{Alburger1974-ie}%
  \BibitemOpen
  \bibfield  {author} {\bibinfo {author} {\bibfnamefont {D.~E.}\ \bibnamefont {Alburger}}\ and\ \bibinfo {author} {\bibfnamefont {D.~R.}\ \bibnamefont {Goosman}},\ }\href@noop {} {\bibfield  {journal} {\bibinfo  {journal} {Phys. Rev. C}\ }\textbf {\bibinfo {volume} {9}},\ \bibinfo {pages} {2236} (\bibinfo {year} {1974})}\BibitemShut {NoStop}%
\bibitem [{\citenamefont {Bunatyan}\ and\ \citenamefont {Ecseev}(1970)}]{Bunatyan1970-mo}%
  \BibitemOpen
  \bibfield  {author} {\bibinfo {author} {\bibfnamefont {G.~G.}\ \bibnamefont {Bunatyan}}\ and\ \bibinfo {author} {\bibfnamefont {V.~S.}\ \bibnamefont {Ecseev}},\ }\href@noop {} {\bibfield  {journal} {\bibinfo  {journal} {Soviet Journal of Nuclear Physics}\ }\textbf {\bibinfo {volume} {11}},\ \bibinfo {pages} {445} (\bibinfo {year} {1970})}\BibitemShut {NoStop}%
\bibitem [{\citenamefont {Vil'gel'mona}\ and\ \citenamefont {Evseev}(1970)}]{Vilgelmona1970-ec}%
  \BibitemOpen
  \bibfield  {author} {\bibinfo {author} {\bibfnamefont {L.}~\bibnamefont {Vil'gel'mona}}\ and\ \bibinfo {author} {\bibfnamefont {V.~S.}\ \bibnamefont {Evseev}},\ }\href@noop {} {\bibfield  {journal} {\bibinfo  {journal} {Soviet Journal of Nuclear Physics}\ }\textbf {\bibinfo {volume} {13}},\ \bibinfo {pages} {310} (\bibinfo {year} {1970})}\BibitemShut {NoStop}%
\bibitem [{\citenamefont {Heusser}\ and\ \citenamefont {Kirsten}(1972)}]{Heusser1972-vs}%
  \BibitemOpen
  \bibfield  {author} {\bibinfo {author} {\bibfnamefont {G.}~\bibnamefont {Heusser}}\ and\ \bibinfo {author} {\bibfnamefont {T.}~\bibnamefont {Kirsten}},\ }\href@noop {} {\bibfield  {journal} {\bibinfo  {journal} {Nucl. Phys. A}\ }\textbf {\bibinfo {volume} {195}},\ \bibinfo {pages} {369} (\bibinfo {year} {1972})}\BibitemShut {NoStop}%
\bibitem [{\citenamefont {Wyttenbach}\ \emph {et~al.}(1978)\citenamefont {Wyttenbach}, \citenamefont {Baertschi}, \citenamefont {Bajo}, \citenamefont {Hadermann}, \citenamefont {Junker}, \citenamefont {Katcoff}, \citenamefont {Hermes},\ and\ \citenamefont {Pruys}}]{Wyttenbach1978-ks}%
  \BibitemOpen
  \bibfield  {author} {\bibinfo {author} {\bibfnamefont {A.}~\bibnamefont {Wyttenbach}}, \bibinfo {author} {\bibfnamefont {P.}~\bibnamefont {Baertschi}}, \bibinfo {author} {\bibfnamefont {S.}~\bibnamefont {Bajo}}, \bibinfo {author} {\bibfnamefont {J.}~\bibnamefont {Hadermann}}, \bibinfo {author} {\bibfnamefont {K.}~\bibnamefont {Junker}}, \bibinfo {author} {\bibfnamefont {S.}~\bibnamefont {Katcoff}}, \bibinfo {author} {\bibfnamefont {E.~A.}\ \bibnamefont {Hermes}},\ and\ \bibinfo {author} {\bibfnamefont {H.~S.}\ \bibnamefont {Pruys}},\ }\href@noop {} {\bibfield  {journal} {\bibinfo  {journal} {Nucl. Phys. A}\ }\textbf {\bibinfo {volume} {294}},\ \bibinfo {pages} {278} (\bibinfo {year} {1978})}\BibitemShut {NoStop}%
\bibitem [{\citenamefont {Heisinger}\ \emph {et~al.}(2002)\citenamefont {Heisinger}, \citenamefont {Lal}, \citenamefont {Jull}, \citenamefont {Kubik}, \citenamefont {Ivy-Ochs}, \citenamefont {Knie},\ and\ \citenamefont {Nolte}}]{Heisinger2002-nm}%
  \BibitemOpen
  \bibfield  {author} {\bibinfo {author} {\bibfnamefont {B.}~\bibnamefont {Heisinger}}, \bibinfo {author} {\bibfnamefont {D.}~\bibnamefont {Lal}}, \bibinfo {author} {\bibfnamefont {A.~J.~T.}\ \bibnamefont {Jull}}, \bibinfo {author} {\bibfnamefont {P.}~\bibnamefont {Kubik}}, \bibinfo {author} {\bibfnamefont {S.}~\bibnamefont {Ivy-Ochs}}, \bibinfo {author} {\bibfnamefont {K.}~\bibnamefont {Knie}},\ and\ \bibinfo {author} {\bibfnamefont {E.}~\bibnamefont {Nolte}},\ }\href@noop {} {\bibfield  {journal} {\bibinfo  {journal} {Earth Planet. Sci. Lett.}\ }\textbf {\bibinfo {volume} {200}},\ \bibinfo {pages} {357} (\bibinfo {year} {2002})}\BibitemShut {NoStop}%
\bibitem [{\citenamefont {Miller}\ \emph {et~al.}(1972)\citenamefont {Miller}, \citenamefont {Eckhause}, \citenamefont {Martin},\ and\ \citenamefont {Welsh}}]{Miller1972-wm}%
  \BibitemOpen
  \bibfield  {author} {\bibinfo {author} {\bibfnamefont {G.~H.}\ \bibnamefont {Miller}}, \bibinfo {author} {\bibfnamefont {M.}~\bibnamefont {Eckhause}}, \bibinfo {author} {\bibfnamefont {P.}~\bibnamefont {Martin}},\ and\ \bibinfo {author} {\bibfnamefont {R.~E.}\ \bibnamefont {Welsh}},\ }\href@noop {} {\bibfield  {journal} {\bibinfo  {journal} {Phys. Rev. C}\ }\textbf {\bibinfo {volume} {6}},\ \bibinfo {pages} {487} (\bibinfo {year} {1972})}\BibitemShut {NoStop}%
\bibitem [{\citenamefont {Pratt}(1969)}]{Pratt1969-sk}%
  \BibitemOpen
  \bibfield  {author} {\bibinfo {author} {\bibfnamefont {T.~A. E.~C.}\ \bibnamefont {Pratt}},\ }\href@noop {} {\bibfield  {journal} {\bibinfo  {journal} {Il Nuovo Cimento B (1965-1970)}\ }\textbf {\bibinfo {volume} {61}},\ \bibinfo {pages} {119} (\bibinfo {year} {1969})}\BibitemShut {NoStop}%
\bibitem [{\citenamefont {{L. E. Temple, S. N. Kaplan, R. V. Pyle, and G. F. Valby}}(1971)}]{L_E_Temple_S_N_Kaplan_R_V_Pyle_and_G_F_Valby1971-mu}%
  \BibitemOpen
  \bibfield  {author} {\bibinfo {author} {\bibnamefont {{L. E. Temple, S. N. Kaplan, R. V. Pyle, and G. F. Valby}}},\ }\href@noop {} {\bibinfo {title} {{NUCLEAR} {EXCITATION} {OF} al, si, ca and co {PRODUCED} {BY} mu- {CAPTURE}}},\ \bibinfo {howpublished} {\url{https://escholarship.org/content/qt3w73563k/qt3w73563k.pdf}} (\bibinfo {year} {1971}),\ \bibinfo {note} {accessed: 2023-6-27}\BibitemShut {NoStop}%
\bibitem [{\citenamefont {Gorringe}\ \emph {et~al.}(1999)\citenamefont {Gorringe}, \citenamefont {Armstrong}, \citenamefont {Arole}, \citenamefont {Boleman}, \citenamefont {Gete}, \citenamefont {Kuzmin}, \citenamefont {Moftah}, \citenamefont {Sedlar}, \citenamefont {Stocki},\ and\ \citenamefont {Tetereva}}]{Gorringe1999-nv}%
  \BibitemOpen
  \bibfield  {author} {\bibinfo {author} {\bibfnamefont {T.~P.}\ \bibnamefont {Gorringe}}, \bibinfo {author} {\bibfnamefont {D.~S.}\ \bibnamefont {Armstrong}}, \bibinfo {author} {\bibfnamefont {S.}~\bibnamefont {Arole}}, \bibinfo {author} {\bibfnamefont {M.}~\bibnamefont {Boleman}}, \bibinfo {author} {\bibfnamefont {E.}~\bibnamefont {Gete}}, \bibinfo {author} {\bibfnamefont {V.}~\bibnamefont {Kuzmin}}, \bibinfo {author} {\bibfnamefont {B.~A.}\ \bibnamefont {Moftah}}, \bibinfo {author} {\bibfnamefont {R.}~\bibnamefont {Sedlar}}, \bibinfo {author} {\bibfnamefont {T.~J.}\ \bibnamefont {Stocki}},\ and\ \bibinfo {author} {\bibfnamefont {T.}~\bibnamefont {Tetereva}},\ }\href@noop {} {\bibfield  {journal} {\bibinfo  {journal} {Phys. Rev. C}\ }\textbf {\bibinfo {volume} {60}},\ \bibinfo {pages} {055501} (\bibinfo {year} {1999})}\BibitemShut {NoStop}%
\bibitem [{\citenamefont {Measday}\ \emph {et~al.}(2007{\natexlab{a}})\citenamefont {Measday}, \citenamefont {Stocki}, \citenamefont {Moftah},\ and\ \citenamefont {Tam}}]{Measday2007-os}%
  \BibitemOpen
  \bibfield  {author} {\bibinfo {author} {\bibfnamefont {D.~F.}\ \bibnamefont {Measday}}, \bibinfo {author} {\bibfnamefont {T.~J.}\ \bibnamefont {Stocki}}, \bibinfo {author} {\bibfnamefont {B.~A.}\ \bibnamefont {Moftah}},\ and\ \bibinfo {author} {\bibfnamefont {H.}~\bibnamefont {Tam}},\ }\href@noop {} {\bibfield  {journal} {\bibinfo  {journal} {Phys. Rev. C}\ }\textbf {\bibinfo {volume} {76}} (\bibinfo {year} {2007}{\natexlab{a}})}\BibitemShut {NoStop}%
\bibitem [{\citenamefont {Sobottka}\ and\ \citenamefont {Wills}(1968)}]{Sobottka1968-px}%
  \BibitemOpen
  \bibfield  {author} {\bibinfo {author} {\bibfnamefont {S.~E.}\ \bibnamefont {Sobottka}}\ and\ \bibinfo {author} {\bibfnamefont {E.~L.}\ \bibnamefont {Wills}},\ }\href@noop {} {\bibfield  {journal} {\bibinfo  {journal} {Phys. Rev. Lett.}\ }\textbf {\bibinfo {volume} {20}},\ \bibinfo {pages} {596} (\bibinfo {year} {1968})}\BibitemShut {NoStop}%
\bibitem [{\citenamefont {Budyashov}\ \emph {et~al.}(1971)\citenamefont {Budyashov}, \citenamefont {Zinov}, \citenamefont {Konin}, \citenamefont {Mukhin},\ and\ \citenamefont {Chatrchyan}}]{Budyashov1971-wc}%
  \BibitemOpen
  \bibfield  {author} {\bibinfo {author} {\bibfnamefont {Y.~G.}\ \bibnamefont {Budyashov}}, \bibinfo {author} {\bibfnamefont {V.~G.}\ \bibnamefont {Zinov}}, \bibinfo {author} {\bibfnamefont {A.~D.}\ \bibnamefont {Konin}}, \bibinfo {author} {\bibfnamefont {A.~I.}\ \bibnamefont {Mukhin}},\ and\ \bibinfo {author} {\bibfnamefont {A.~M.}\ \bibnamefont {Chatrchyan}},\ }\href@noop {} {\bibfield  {journal} {\bibinfo  {journal} {Soviet Journal of Experimental and Theoretical Physics}\ }\textbf {\bibinfo {volume} {33}},\ \bibinfo {pages} {11} (\bibinfo {year} {1971})}\BibitemShut {NoStop}%
\bibitem [{\citenamefont {Krane}\ \emph {et~al.}(1979)\citenamefont {Krane}, \citenamefont {Sharma}, \citenamefont {Swenson}, \citenamefont {McDaniels}, \citenamefont {Varghese}, \citenamefont {Wood}, \citenamefont {Silbar}, \citenamefont {Wohlfahrt},\ and\ \citenamefont {Goulding}}]{Krane1979-hc}%
  \BibitemOpen
  \bibfield  {author} {\bibinfo {author} {\bibfnamefont {K.~S.}\ \bibnamefont {Krane}}, \bibinfo {author} {\bibfnamefont {T.~C.}\ \bibnamefont {Sharma}}, \bibinfo {author} {\bibfnamefont {L.~W.}\ \bibnamefont {Swenson}}, \bibinfo {author} {\bibfnamefont {D.~K.}\ \bibnamefont {McDaniels}}, \bibinfo {author} {\bibfnamefont {P.}~\bibnamefont {Varghese}}, \bibinfo {author} {\bibfnamefont {B.~E.}\ \bibnamefont {Wood}}, \bibinfo {author} {\bibfnamefont {R.~R.}\ \bibnamefont {Silbar}}, \bibinfo {author} {\bibfnamefont {H.~D.}\ \bibnamefont {Wohlfahrt}},\ and\ \bibinfo {author} {\bibfnamefont {C.~A.}\ \bibnamefont {Goulding}},\ }\href@noop {} {\bibfield  {journal} {\bibinfo  {journal} {Phys. Rev. C}\ }\textbf {\bibinfo {volume} {20}},\ \bibinfo {pages} {1873} (\bibinfo {year} {1979})}\BibitemShut {NoStop}%
\bibitem [{\citenamefont {Gaponenko}\ \emph {et~al.}(2020)\citenamefont {Gaponenko}, \citenamefont {Grossheim}, \citenamefont {Hillairet}, \citenamefont {Marshall}, \citenamefont {Mischke},\ and\ \citenamefont {Olin}}]{Gaponenko2020-io}%
  \BibitemOpen
  \bibfield  {author} {\bibinfo {author} {\bibfnamefont {A.}~\bibnamefont {Gaponenko}}, \bibinfo {author} {\bibfnamefont {A.}~\bibnamefont {Grossheim}}, \bibinfo {author} {\bibfnamefont {A.}~\bibnamefont {Hillairet}}, \bibinfo {author} {\bibfnamefont {G.~M.}\ \bibnamefont {Marshall}}, \bibinfo {author} {\bibfnamefont {R.~E.}\ \bibnamefont {Mischke}},\ and\ \bibinfo {author} {\bibfnamefont {A.}~\bibnamefont {Olin}},\ }\href@noop {} {\bibfield  {journal} {\bibinfo  {journal} {Phys. Rev. C}\ }\textbf {\bibinfo {volume} {101}},\ \bibinfo {pages} {035502} (\bibinfo {year} {2020})}\BibitemShut {NoStop}%
\bibitem [{\citenamefont {Sundelin}\ and\ \citenamefont {Edelstein}(1973)}]{Sundelin1973-gw}%
  \BibitemOpen
  \bibfield  {author} {\bibinfo {author} {\bibfnamefont {R.~M.}\ \bibnamefont {Sundelin}}\ and\ \bibinfo {author} {\bibfnamefont {R.~M.}\ \bibnamefont {Edelstein}},\ }\href@noop {} {\bibfield  {journal} {\bibinfo  {journal} {Phys. Rev. C}\ }\textbf {\bibinfo {volume} {7}},\ \bibinfo {pages} {1037} (\bibinfo {year} {1973})}\BibitemShut {NoStop}%
\bibitem [{\citenamefont {Kozlowski}\ \emph {et~al.}(1985)\citenamefont {Kozlowski}, \citenamefont {Bertl}, \citenamefont {Povel}, \citenamefont {Sennhauser}, \citenamefont {Walter}, \citenamefont {Zglinski}, \citenamefont {Engfer}, \citenamefont {Grab}, \citenamefont {Hermes}, \citenamefont {Isaak}, \citenamefont {Van Der~Schaaf}, \citenamefont {Van Der~Pluym},\ and\ \citenamefont {Hesselink}}]{Kozlowski1985-tf}%
  \BibitemOpen
  \bibfield  {author} {\bibinfo {author} {\bibfnamefont {T.}~\bibnamefont {Kozlowski}}, \bibinfo {author} {\bibfnamefont {W.}~\bibnamefont {Bertl}}, \bibinfo {author} {\bibfnamefont {H.~P.}\ \bibnamefont {Povel}}, \bibinfo {author} {\bibfnamefont {U.}~\bibnamefont {Sennhauser}}, \bibinfo {author} {\bibfnamefont {H.~K.}\ \bibnamefont {Walter}}, \bibinfo {author} {\bibfnamefont {A.}~\bibnamefont {Zglinski}}, \bibinfo {author} {\bibfnamefont {R.}~\bibnamefont {Engfer}}, \bibinfo {author} {\bibfnamefont {C.~H.}\ \bibnamefont {Grab}}, \bibinfo {author} {\bibfnamefont {E.~A.}\ \bibnamefont {Hermes}}, \bibinfo {author} {\bibfnamefont {H.~P.}\ \bibnamefont {Isaak}}, \bibinfo {author} {\bibfnamefont {A.}~\bibnamefont {Van Der~Schaaf}}, \bibinfo {author} {\bibfnamefont {J.}~\bibnamefont {Van Der~Pluym}},\ and\ \bibinfo {author} {\bibfnamefont {W.~H.~A.}\ \bibnamefont {Hesselink}},\ }\href@noop {} {\bibfield  {journal} {\bibinfo  {journal} {Nucl. Phys. A}\ }\textbf {\bibinfo {volume} {436}},\ \bibinfo {pages} {717}
  (\bibinfo {year} {1985})}\BibitemShut {NoStop}%
\bibitem [{\citenamefont {Macdonald}\ \emph {et~al.}(1965)\citenamefont {Macdonald}, \citenamefont {Diaz}, \citenamefont {Kaplan},\ and\ \citenamefont {Pyle}}]{Macdonald1965-ab}%
  \BibitemOpen
  \bibfield  {author} {\bibinfo {author} {\bibfnamefont {B.}~\bibnamefont {Macdonald}}, \bibinfo {author} {\bibfnamefont {J.~A.}\ \bibnamefont {Diaz}}, \bibinfo {author} {\bibfnamefont {S.~N.}\ \bibnamefont {Kaplan}},\ and\ \bibinfo {author} {\bibfnamefont {R.~V.}\ \bibnamefont {Pyle}},\ }\href@noop {} {\bibfield  {journal} {\bibinfo  {journal} {Physical Review}\ }\textbf {\bibinfo {volume} {139}},\ \bibinfo {pages} {B1253} (\bibinfo {year} {1965})}\BibitemShut {NoStop}%
\bibitem [{\citenamefont {Singer}(1962)}]{Singer1962-lj}%
  \BibitemOpen
  \bibfield  {author} {\bibinfo {author} {\bibfnamefont {P.}~\bibnamefont {Singer}},\ }\href@noop {} {\bibfield  {journal} {\bibinfo  {journal} {Il Nuovo Cimento (1955-1965)}\ }\textbf {\bibinfo {volume} {23}},\ \bibinfo {pages} {669} (\bibinfo {year} {1962})}\BibitemShut {NoStop}%
\bibitem [{\citenamefont {Singer}(1974)}]{Singer1974-zv}%
  \BibitemOpen
  \bibfield  {author} {\bibinfo {author} {\bibfnamefont {P.}~\bibnamefont {Singer}},\ }in\ \href@noop {} {\emph {\bibinfo {booktitle} {Nuclear Physics}}},\ \bibinfo {editor} {edited by\ \bibinfo {editor} {\bibfnamefont {G.}~\bibnamefont {Höhler}}, \bibinfo {editor} {\bibfnamefont {A.}~\bibnamefont {Fujimori}}, \bibinfo {editor} {\bibfnamefont {J.}~\bibnamefont {Kühn}}, \bibinfo {editor} {\bibfnamefont {T.}~\bibnamefont {Müller}}, \bibinfo {editor} {\bibfnamefont {F.}~\bibnamefont {Steiner}}, \bibinfo {editor} {\bibfnamefont {W.~C.}\ \bibnamefont {Stwalley}}, \bibinfo {editor} {\bibfnamefont {J.~E.}\ \bibnamefont {Trümper}}, \bibinfo {editor} {\bibfnamefont {P.}~\bibnamefont {Wölfle}},\ and\ \bibinfo {editor} {\bibfnamefont {U.}~\bibnamefont {Woggon}}}\ (\bibinfo  {publisher} {Springer Berlin Heidelberg},\ \bibinfo {address} {Berlin, Heidelberg},\ \bibinfo {year} {1974})\ pp.\ \bibinfo {pages} {39--87}\BibitemShut {NoStop}%
\bibitem [{\citenamefont {Lifshitz}\ and\ \citenamefont {Singer}(1978)}]{Lifshitz1978-zz}%
  \BibitemOpen
  \bibfield  {author} {\bibinfo {author} {\bibfnamefont {M.}~\bibnamefont {Lifshitz}}\ and\ \bibinfo {author} {\bibfnamefont {P.}~\bibnamefont {Singer}},\ }\href@noop {} {\bibfield  {journal} {\bibinfo  {journal} {Phys. Rev. Lett.}\ }\textbf {\bibinfo {volume} {41}},\ \bibinfo {pages} {18} (\bibinfo {year} {1978})}\BibitemShut {NoStop}%
\bibitem [{\citenamefont {Lifshitz}\ and\ \citenamefont {Singer}(1980)}]{Lifshitz1980-bu}%
  \BibitemOpen
  \bibfield  {author} {\bibinfo {author} {\bibfnamefont {M.}~\bibnamefont {Lifshitz}}\ and\ \bibinfo {author} {\bibfnamefont {P.}~\bibnamefont {Singer}},\ }\href@noop {} {\bibfield  {journal} {\bibinfo  {journal} {Phys. Rev. C}\ }\textbf {\bibinfo {volume} {22}},\ \bibinfo {pages} {2135} (\bibinfo {year} {1980})}\BibitemShut {NoStop}%
\bibitem [{\citenamefont {Balashov}\ and\ \citenamefont {Eramzhyan}(1967)}]{Balashov1967-ml}%
  \BibitemOpen
  \bibfield  {author} {\bibinfo {author} {\bibfnamefont {V.~V.}\ \bibnamefont {Balashov}}\ and\ \bibinfo {author} {\bibfnamefont {R.~A.}\ \bibnamefont {Eramzhyan}},\ }\href@noop {} {\bibfield  {journal} {\bibinfo  {journal} {At. Energy Rev.}\ }\textbf {\bibinfo {volume} {5}},\ \bibinfo {pages} {3} (\bibinfo {year} {1967})}\BibitemShut {NoStop}%
\bibitem [{\citenamefont {Foldy}\ and\ \citenamefont {Walecka}(1964)}]{Foldy1964-ea}%
  \BibitemOpen
  \bibfield  {author} {\bibinfo {author} {\bibfnamefont {L.~L.}\ \bibnamefont {Foldy}}\ and\ \bibinfo {author} {\bibfnamefont {J.~D.}\ \bibnamefont {Walecka}},\ }\href@noop {} {\bibfield  {journal} {\bibinfo  {journal} {Nuovo Cimento: C: Geophys. Space Phys.}\ }\textbf {\bibinfo {volume} {34}},\ \bibinfo {pages} {1026} (\bibinfo {year} {1964})}\BibitemShut {NoStop}%
\bibitem [{\citenamefont {Überall}(1974)}]{Uberall1974-tk}%
  \BibitemOpen
  \bibfield  {author} {\bibinfo {author} {\bibfnamefont {H.}~\bibnamefont {Überall}},\ }in\ \href@noop {} {\emph {\bibinfo {booktitle} {Nuclear Physics}}},\ \bibinfo {editor} {edited by\ \bibinfo {editor} {\bibfnamefont {G.}~\bibnamefont {Höhler}}, \bibinfo {editor} {\bibfnamefont {A.}~\bibnamefont {Fujimori}}, \bibinfo {editor} {\bibfnamefont {J.}~\bibnamefont {Kühn}}, \bibinfo {editor} {\bibfnamefont {T.}~\bibnamefont {Müller}}, \bibinfo {editor} {\bibfnamefont {F.}~\bibnamefont {Steiner}}, \bibinfo {editor} {\bibfnamefont {W.~C.}\ \bibnamefont {Stwalley}}, \bibinfo {editor} {\bibfnamefont {J.~E.}\ \bibnamefont {Trümper}}, \bibinfo {editor} {\bibfnamefont {P.}~\bibnamefont {Wölfle}},\ and\ \bibinfo {editor} {\bibfnamefont {U.}~\bibnamefont {Woggon}}}\ (\bibinfo  {publisher} {Springer Berlin Heidelberg},\ \bibinfo {address} {Berlin, Heidelberg},\ \bibinfo {year} {1974})\ pp.\ \bibinfo {pages} {1--38}\BibitemShut {NoStop}%
\bibitem [{\citenamefont {Mukhopadhyay}(1977)}]{Mukhopadhyay1977-qf}%
  \BibitemOpen
  \bibfield  {author} {\bibinfo {author} {\bibfnamefont {N.~C.}\ \bibnamefont {Mukhopadhyay}},\ }\href@noop {} {\bibfield  {journal} {\bibinfo  {journal} {Phys. Rep.}\ }\textbf {\bibinfo {volume} {30}},\ \bibinfo {pages} {1} (\bibinfo {year} {1977})}\BibitemShut {NoStop}%
\bibitem [{\citenamefont {Kozłowski}\ and\ \citenamefont {Zgliński}(1978)}]{Kozlowski1978-di}%
  \BibitemOpen
  \bibfield  {author} {\bibinfo {author} {\bibfnamefont {T.}~\bibnamefont {Kozłowski}}\ and\ \bibinfo {author} {\bibfnamefont {A.}~\bibnamefont {Zgliński}},\ }\href@noop {} {\bibfield  {journal} {\bibinfo  {journal} {Nucl. Phys. A}\ }\textbf {\bibinfo {volume} {305}},\ \bibinfo {pages} {368} (\bibinfo {year} {1978})}\BibitemShut {NoStop}%
\bibitem [{\citenamefont {Auerbach}\ and\ \citenamefont {Klein}(1984)}]{Auerbach1984-gf}%
  \BibitemOpen
  \bibfield  {author} {\bibinfo {author} {\bibfnamefont {N.}~\bibnamefont {Auerbach}}\ and\ \bibinfo {author} {\bibfnamefont {A.}~\bibnamefont {Klein}},\ }\href@noop {} {\bibfield  {journal} {\bibinfo  {journal} {Nuclear Physics}\ }\textbf {\bibinfo {volume} {422}},\ \bibinfo {pages} {480} (\bibinfo {year} {1984})}\BibitemShut {NoStop}%
\bibitem [{\citenamefont {Kolbe}\ \emph {et~al.}(1994)\citenamefont {Kolbe}, \citenamefont {Langanke},\ and\ \citenamefont {Vogel}}]{Kolbe1994-zz}%
  \BibitemOpen
  \bibfield  {author} {\bibinfo {author} {\bibfnamefont {E.}~\bibnamefont {Kolbe}}, \bibinfo {author} {\bibfnamefont {K.}~\bibnamefont {Langanke}},\ and\ \bibinfo {author} {\bibfnamefont {P.}~\bibnamefont {Vogel}},\ }\href@noop {} {\bibfield  {journal} {\bibinfo  {journal} {Phys. Rev. C}\ }\textbf {\bibinfo {volume} {50}},\ \bibinfo {pages} {2576} (\bibinfo {year} {1994})}\BibitemShut {NoStop}%
\bibitem [{\citenamefont {Wang}\ \emph {et~al.}(2021)\citenamefont {Wang}, \citenamefont {Huang}, \citenamefont {Kondev}, \citenamefont {Audi},\ and\ \citenamefont {Naimi}}]{Wang2021-uw}%
  \BibitemOpen
  \bibfield  {author} {\bibinfo {author} {\bibfnamefont {M.}~\bibnamefont {Wang}}, \bibinfo {author} {\bibfnamefont {W.~J.}\ \bibnamefont {Huang}}, \bibinfo {author} {\bibfnamefont {F.~G.}\ \bibnamefont {Kondev}}, \bibinfo {author} {\bibfnamefont {G.}~\bibnamefont {Audi}},\ and\ \bibinfo {author} {\bibfnamefont {S.}~\bibnamefont {Naimi}},\ }\href@noop {} {\bibfield  {journal} {\bibinfo  {journal} {Chin. Phys. C}\ }\textbf {\bibinfo {volume} {45}},\ \bibinfo {pages} {030003} (\bibinfo {year} {2021})}\BibitemShut {NoStop}%
\bibitem [{\citenamefont {Watanabe}\ \emph {et~al.}(1987)\citenamefont {Watanabe}, \citenamefont {Kumabe}, \citenamefont {Hyakutake}, \citenamefont {Koori}, \citenamefont {Ogawa}, \citenamefont {Orito}, \citenamefont {Akagi},\ and\ \citenamefont {Oda}}]{Watanabe1987-jr}%
  \BibitemOpen
  \bibfield  {author} {\bibinfo {author} {\bibfnamefont {Y.}~\bibnamefont {Watanabe}}, \bibinfo {author} {\bibfnamefont {I.}~\bibnamefont {Kumabe}, \bibfnamefont {I}}, \bibinfo {author} {\bibfnamefont {M.}~\bibnamefont {Hyakutake}}, \bibinfo {author} {\bibfnamefont {N.}~\bibnamefont {Koori}}, \bibinfo {author} {\bibfnamefont {K.}~\bibnamefont {Ogawa}}, \bibinfo {author} {\bibfnamefont {K.}~\bibnamefont {Orito}}, \bibinfo {author} {\bibfnamefont {K.}~\bibnamefont {Akagi}},\ and\ \bibinfo {author} {\bibfnamefont {N.}~\bibnamefont {Oda}},\ }\href@noop {} {\bibfield  {journal} {\bibinfo  {journal} {Phys. Rev. C}\ }\textbf {\bibinfo {volume} {36}},\ \bibinfo {pages} {1325} (\bibinfo {year} {1987})}\BibitemShut {NoStop}%
\bibitem [{\citenamefont {Cannata}\ \emph {et~al.}(1977)\citenamefont {Cannata}, \citenamefont {Graves},\ and\ \citenamefont {Überall}}]{Cannata1977-vm}%
  \BibitemOpen
  \bibfield  {author} {\bibinfo {author} {\bibfnamefont {F.}~\bibnamefont {Cannata}}, \bibinfo {author} {\bibfnamefont {R.}~\bibnamefont {Graves}},\ and\ \bibinfo {author} {\bibfnamefont {H.}~\bibnamefont {Überall}},\ }\href@noop {} {\bibfield  {journal} {\bibinfo  {journal} {Riv. Nuovo Cimento}\ }\textbf {\bibinfo {volume} {7}},\ \bibinfo {pages} {133} (\bibinfo {year} {1977})}\BibitemShut {NoStop}%
\bibitem [{\citenamefont {Yoshida}(2013)}]{Yoshida2013-vv}%
  \BibitemOpen
  \bibfield  {author} {\bibinfo {author} {\bibfnamefont {K.}~\bibnamefont {Yoshida}},\ }\href@noop {} {\bibfield  {journal} {\bibinfo  {journal} {Prog. Theor. Exp. Phys.}\ }\textbf {\bibinfo {volume} {2013}},\ \bibinfo {pages} {113D02} (\bibinfo {year} {2013})}\BibitemShut {NoStop}%
\bibitem [{\citenamefont {Bai}\ \emph {et~al.}(2014)\citenamefont {Bai}, \citenamefont {Sagawa}, \citenamefont {Colò}, \citenamefont {Fujita}, \citenamefont {Zhang}, \citenamefont {Zhang},\ and\ \citenamefont {Xu}}]{Bai2014-au}%
  \BibitemOpen
  \bibfield  {author} {\bibinfo {author} {\bibfnamefont {C.~L.}\ \bibnamefont {Bai}}, \bibinfo {author} {\bibfnamefont {H.}~\bibnamefont {Sagawa}}, \bibinfo {author} {\bibfnamefont {G.}~\bibnamefont {Colò}}, \bibinfo {author} {\bibfnamefont {Y.}~\bibnamefont {Fujita}}, \bibinfo {author} {\bibfnamefont {H.~Q.}\ \bibnamefont {Zhang}}, \bibinfo {author} {\bibfnamefont {X.~Z.}\ \bibnamefont {Zhang}},\ and\ \bibinfo {author} {\bibfnamefont {F.~R.}\ \bibnamefont {Xu}},\ }\href@noop {} {\bibfield  {journal} {\bibinfo  {journal} {Phys. Rev. C}\ }\textbf {\bibinfo {volume} {90}},\ \bibinfo {pages} {054335} (\bibinfo {year} {2014})}\BibitemShut {NoStop}%
\bibitem [{\citenamefont {Sagawa}\ \emph {et~al.}(2016)\citenamefont {Sagawa}, \citenamefont {Bai},\ and\ \citenamefont {Colò}}]{Sagawa2016-zx}%
  \BibitemOpen
  \bibfield  {author} {\bibinfo {author} {\bibfnamefont {H.}~\bibnamefont {Sagawa}}, \bibinfo {author} {\bibfnamefont {C.~L.}\ \bibnamefont {Bai}},\ and\ \bibinfo {author} {\bibfnamefont {G.}~\bibnamefont {Colò}},\ }\href@noop {} {\bibfield  {journal} {\bibinfo  {journal} {Phys. Scr.}\ }\textbf {\bibinfo {volume} {91}},\ \bibinfo {pages} {083011} (\bibinfo {year} {2016})}\BibitemShut {NoStop}%
\bibitem [{\citenamefont {Yüksel}\ \emph {et~al.}(2020)\citenamefont {Yüksel}, \citenamefont {Paar}, \citenamefont {Colò}, \citenamefont {Khan},\ and\ \citenamefont {Niu}}]{Yuksel2020-se}%
  \BibitemOpen
  \bibfield  {author} {\bibinfo {author} {\bibfnamefont {E.}~\bibnamefont {Yüksel}}, \bibinfo {author} {\bibfnamefont {N.}~\bibnamefont {Paar}}, \bibinfo {author} {\bibfnamefont {G.}~\bibnamefont {Colò}}, \bibinfo {author} {\bibfnamefont {E.}~\bibnamefont {Khan}},\ and\ \bibinfo {author} {\bibfnamefont {Y.~F.}\ \bibnamefont {Niu}},\ }\href@noop {} {\bibfield  {journal} {\bibinfo  {journal} {Phys. Rev. C}\ }\textbf {\bibinfo {volume} {101}},\ \bibinfo {pages} {044305} (\bibinfo {year} {2020})}\BibitemShut {NoStop}%
\bibitem [{\citenamefont {Fujita}\ \emph {et~al.}(2014)\citenamefont {Fujita}, \citenamefont {Fujita}, \citenamefont {Adachi}, \citenamefont {Bai}, \citenamefont {Algora}, \citenamefont {Berg}, \citenamefont {von Brentano}, \citenamefont {Colò}, \citenamefont {Csatlós}, \citenamefont {Deaven}, \citenamefont {Estevez-Aguado}, \citenamefont {Fransen}, \citenamefont {De~Frenne}, \citenamefont {Fujita}, \citenamefont {Ganioğlu}, \citenamefont {Guess}, \citenamefont {Gulyás}, \citenamefont {Hatanaka}, \citenamefont {Hirota}, \citenamefont {Honma}, \citenamefont {Ishikawa}, \citenamefont {Jacobs}, \citenamefont {Krasznahorkay}, \citenamefont {Matsubara}, \citenamefont {Matsuyanagi}, \citenamefont {Meharchand}, \citenamefont {Molina}, \citenamefont {Muto}, \citenamefont {Nakanishi}, \citenamefont {Negret}, \citenamefont {Okamura}, \citenamefont {Ong}, \citenamefont {Otsuka}, \citenamefont {Pietralla}, \citenamefont {Perdikakis}, \citenamefont {Popescu}, \citenamefont {Rubio}, \citenamefont {Sagawa},
  \citenamefont {Sarriguren}, \citenamefont {Scholl}, \citenamefont {Shimbara}, \citenamefont {Shimizu}, \citenamefont {Susoy}, \citenamefont {Suzuki}, \citenamefont {Tameshige}, \citenamefont {Tamii}, \citenamefont {Thies}, \citenamefont {Uchida}, \citenamefont {Wakasa}, \citenamefont {Yosoi}, \citenamefont {Zegers}, \citenamefont {Zell},\ and\ \citenamefont {Zenihiro}}]{Fujita2014-qz}%
  \BibitemOpen
  \bibfield  {author} {\bibinfo {author} {\bibfnamefont {Y.}~\bibnamefont {Fujita}}, \bibinfo {author} {\bibfnamefont {H.}~\bibnamefont {Fujita}}, \bibinfo {author} {\bibfnamefont {T.}~\bibnamefont {Adachi}}, \bibinfo {author} {\bibfnamefont {C.~L.}\ \bibnamefont {Bai}}, \bibinfo {author} {\bibfnamefont {A.}~\bibnamefont {Algora}}, \bibinfo {author} {\bibfnamefont {G.~P.~A.}\ \bibnamefont {Berg}}, \bibinfo {author} {\bibfnamefont {P.}~\bibnamefont {von Brentano}}, \bibinfo {author} {\bibfnamefont {G.}~\bibnamefont {Colò}}, \bibinfo {author} {\bibfnamefont {M.}~\bibnamefont {Csatlós}}, \bibinfo {author} {\bibfnamefont {J.~M.}\ \bibnamefont {Deaven}}, \bibinfo {author} {\bibfnamefont {E.}~\bibnamefont {Estevez-Aguado}}, \bibinfo {author} {\bibfnamefont {C.}~\bibnamefont {Fransen}}, \bibinfo {author} {\bibfnamefont {D.}~\bibnamefont {De~Frenne}}, \bibinfo {author} {\bibfnamefont {K.}~\bibnamefont {Fujita}}, \bibinfo {author} {\bibfnamefont {E.}~\bibnamefont {Ganioğlu}}, \bibinfo {author} {\bibfnamefont
  {C.~J.}\ \bibnamefont {Guess}}, \bibinfo {author} {\bibfnamefont {J.}~\bibnamefont {Gulyás}}, \bibinfo {author} {\bibfnamefont {K.}~\bibnamefont {Hatanaka}}, \bibinfo {author} {\bibfnamefont {K.}~\bibnamefont {Hirota}}, \bibinfo {author} {\bibfnamefont {M.}~\bibnamefont {Honma}}, \bibinfo {author} {\bibfnamefont {D.}~\bibnamefont {Ishikawa}}, \bibinfo {author} {\bibfnamefont {E.}~\bibnamefont {Jacobs}}, \bibinfo {author} {\bibfnamefont {A.}~\bibnamefont {Krasznahorkay}}, \bibinfo {author} {\bibfnamefont {H.}~\bibnamefont {Matsubara}}, \bibinfo {author} {\bibfnamefont {K.}~\bibnamefont {Matsuyanagi}}, \bibinfo {author} {\bibfnamefont {R.}~\bibnamefont {Meharchand}}, \bibinfo {author} {\bibfnamefont {F.}~\bibnamefont {Molina}}, \bibinfo {author} {\bibfnamefont {K.}~\bibnamefont {Muto}}, \bibinfo {author} {\bibfnamefont {K.}~\bibnamefont {Nakanishi}}, \bibinfo {author} {\bibfnamefont {A.}~\bibnamefont {Negret}}, \bibinfo {author} {\bibfnamefont {H.}~\bibnamefont {Okamura}}, \bibinfo {author} {\bibfnamefont
  {H.~J.}\ \bibnamefont {Ong}}, \bibinfo {author} {\bibfnamefont {T.}~\bibnamefont {Otsuka}}, \bibinfo {author} {\bibfnamefont {N.}~\bibnamefont {Pietralla}}, \bibinfo {author} {\bibfnamefont {G.}~\bibnamefont {Perdikakis}}, \bibinfo {author} {\bibfnamefont {L.}~\bibnamefont {Popescu}}, \bibinfo {author} {\bibfnamefont {B.}~\bibnamefont {Rubio}}, \bibinfo {author} {\bibfnamefont {H.}~\bibnamefont {Sagawa}}, \bibinfo {author} {\bibfnamefont {P.}~\bibnamefont {Sarriguren}}, \bibinfo {author} {\bibfnamefont {C.}~\bibnamefont {Scholl}}, \bibinfo {author} {\bibfnamefont {Y.}~\bibnamefont {Shimbara}}, \bibinfo {author} {\bibfnamefont {Y.}~\bibnamefont {Shimizu}}, \bibinfo {author} {\bibfnamefont {G.}~\bibnamefont {Susoy}}, \bibinfo {author} {\bibfnamefont {T.}~\bibnamefont {Suzuki}}, \bibinfo {author} {\bibfnamefont {Y.}~\bibnamefont {Tameshige}}, \bibinfo {author} {\bibfnamefont {A.}~\bibnamefont {Tamii}}, \bibinfo {author} {\bibfnamefont {J.~H.}\ \bibnamefont {Thies}}, \bibinfo {author} {\bibfnamefont
  {M.}~\bibnamefont {Uchida}}, \bibinfo {author} {\bibfnamefont {T.}~\bibnamefont {Wakasa}}, \bibinfo {author} {\bibfnamefont {M.}~\bibnamefont {Yosoi}}, \bibinfo {author} {\bibfnamefont {R.~G.~T.}\ \bibnamefont {Zegers}}, \bibinfo {author} {\bibfnamefont {K.~O.}\ \bibnamefont {Zell}},\ and\ \bibinfo {author} {\bibfnamefont {J.}~\bibnamefont {Zenihiro}},\ }\href@noop {} {\bibfield  {journal} {\bibinfo  {journal} {Phys. Rev. Lett.}\ }\textbf {\bibinfo {volume} {112}},\ \bibinfo {pages} {112502} (\bibinfo {year} {2014})}\BibitemShut {NoStop}%
\bibitem [{\citenamefont {Matsubara}\ and\ \citenamefont {Tamii}(2021)}]{Matsubara2021-hl}%
  \BibitemOpen
  \bibfield  {author} {\bibinfo {author} {\bibfnamefont {H.}~\bibnamefont {Matsubara}}\ and\ \bibinfo {author} {\bibfnamefont {A.}~\bibnamefont {Tamii}},\ }\href@noop {} {\bibfield  {journal} {\bibinfo  {journal} {Front. Astron. Space Sci.}\ }\textbf {\bibinfo {volume} {8}},\ \bibinfo {pages} {667058} (\bibinfo {year} {2021})}\BibitemShut {NoStop}%
\bibitem [{\citenamefont {Yoshida}\ and\ \citenamefont {Tanimura}(2021)}]{Yoshida2021-fp}%
  \BibitemOpen
  \bibfield  {author} {\bibinfo {author} {\bibfnamefont {K.}~\bibnamefont {Yoshida}}\ and\ \bibinfo {author} {\bibfnamefont {Y.}~\bibnamefont {Tanimura}},\ }\href@noop {} {\bibfield  {journal} {\bibinfo  {journal} {Phys. Rev. C.}\ }\textbf {\bibinfo {volume} {104}},\ \bibinfo {pages} {014319} (\bibinfo {year} {2021})}\BibitemShut {NoStop}%
\bibitem [{\citenamefont {Shimizu}\ \emph {et~al.}(2019)\citenamefont {Shimizu}, \citenamefont {Mizusaki}, \citenamefont {Utsuno},\ and\ \citenamefont {Tsunoda}}]{Shimizu2019-va}%
  \BibitemOpen
  \bibfield  {author} {\bibinfo {author} {\bibfnamefont {N.}~\bibnamefont {Shimizu}}, \bibinfo {author} {\bibfnamefont {T.}~\bibnamefont {Mizusaki}}, \bibinfo {author} {\bibfnamefont {Y.}~\bibnamefont {Utsuno}},\ and\ \bibinfo {author} {\bibfnamefont {Y.}~\bibnamefont {Tsunoda}},\ }\href@noop {} {\bibfield  {journal} {\bibinfo  {journal} {Comput. Phys. Commun.}\ }\textbf {\bibinfo {volume} {244}},\ \bibinfo {pages} {372} (\bibinfo {year} {2019})}\BibitemShut {NoStop}%
\bibitem [{\citenamefont {Schröder}\ \emph {et~al.}(1974)\citenamefont {Schröder}, \citenamefont {Jahnke}, \citenamefont {Lindenberger}, \citenamefont {Röschert}, \citenamefont {Engfer},\ and\ \citenamefont {Walter}}]{Schroder1974-us}%
  \BibitemOpen
  \bibfield  {author} {\bibinfo {author} {\bibfnamefont {W.~U.}\ \bibnamefont {Schröder}}, \bibinfo {author} {\bibfnamefont {U.}~\bibnamefont {Jahnke}}, \bibinfo {author} {\bibfnamefont {K.~H.}\ \bibnamefont {Lindenberger}}, \bibinfo {author} {\bibfnamefont {G.}~\bibnamefont {Röschert}}, \bibinfo {author} {\bibfnamefont {R.}~\bibnamefont {Engfer}},\ and\ \bibinfo {author} {\bibfnamefont {H.~K.}\ \bibnamefont {Walter}},\ }\href@noop {} {\bibfield  {journal} {\bibinfo  {journal} {Eur. Phys. J. A}\ }\textbf {\bibinfo {volume} {268}},\ \bibinfo {pages} {57} (\bibinfo {year} {1974})}\BibitemShut {NoStop}%
\bibitem [{\citenamefont {Measday}\ and\ \citenamefont {Stocki}(2006)}]{Measday2006-kb}%
  \BibitemOpen
  \bibfield  {author} {\bibinfo {author} {\bibfnamefont {D.~F.}\ \bibnamefont {Measday}}\ and\ \bibinfo {author} {\bibfnamefont {T.~J.}\ \bibnamefont {Stocki}},\ }\href@noop {} {\bibfield  {journal} {\bibinfo  {journal} {Phys. Rev. C}\ }\textbf {\bibinfo {volume} {73}},\ \bibinfo {pages} {045501} (\bibinfo {year} {2006})}\BibitemShut {NoStop}%
\bibitem [{\citenamefont {Stocki}\ \emph {et~al.}(2002)\citenamefont {Stocki}, \citenamefont {Measday}, \citenamefont {Gete}, \citenamefont {Saliba}, \citenamefont {Moftah},\ and\ \citenamefont {Gorringe}}]{Stocki2002-bw}%
  \BibitemOpen
  \bibfield  {author} {\bibinfo {author} {\bibfnamefont {T.~J.}\ \bibnamefont {Stocki}}, \bibinfo {author} {\bibfnamefont {D.~F.}\ \bibnamefont {Measday}}, \bibinfo {author} {\bibfnamefont {E.}~\bibnamefont {Gete}}, \bibinfo {author} {\bibfnamefont {M.~A.}\ \bibnamefont {Saliba}}, \bibinfo {author} {\bibfnamefont {B.~A.}\ \bibnamefont {Moftah}},\ and\ \bibinfo {author} {\bibfnamefont {T.~P.}\ \bibnamefont {Gorringe}},\ }\href@noop {} {\bibfield  {journal} {\bibinfo  {journal} {Nucl. Phys. A}\ }\textbf {\bibinfo {volume} {697}},\ \bibinfo {pages} {55} (\bibinfo {year} {2002})}\BibitemShut {NoStop}%
\bibitem [{\citenamefont {Winsberg}(1954)}]{Winsberg1954-lj}%
  \BibitemOpen
  \bibfield  {author} {\bibinfo {author} {\bibfnamefont {L.}~\bibnamefont {Winsberg}},\ }\href@noop {} {\bibfield  {journal} {\bibinfo  {journal} {Phys. Rev.}\ }\textbf {\bibinfo {volume} {95}},\ \bibinfo {pages} {205} (\bibinfo {year} {1954})}\BibitemShut {NoStop}%
\bibitem [{\citenamefont {Measday}\ \emph {et~al.}(2007{\natexlab{b}})\citenamefont {Measday}, \citenamefont {Stocki},\ and\ \citenamefont {Tam}}]{Measday2007-zh}%
  \BibitemOpen
  \bibfield  {author} {\bibinfo {author} {\bibfnamefont {D.~F.}\ \bibnamefont {Measday}}, \bibinfo {author} {\bibfnamefont {T.~J.}\ \bibnamefont {Stocki}},\ and\ \bibinfo {author} {\bibfnamefont {H.}~\bibnamefont {Tam}},\ }\href@noop {} {\bibfield  {journal} {\bibinfo  {journal} {Phys. Rev. C}\ }\textbf {\bibinfo {volume} {75}},\ \bibinfo {pages} {045501} (\bibinfo {year} {2007}{\natexlab{b}})}\BibitemShut {NoStop}%
\bibitem [{\citenamefont {Sato}\ \emph {et~al.}(2013)\citenamefont {Sato}, \citenamefont {Niita}, \citenamefont {Matsuda}, \citenamefont {Hashimoto}, \citenamefont {Iwamoto}, \citenamefont {Noda}, \citenamefont {Ogawa}, \citenamefont {Iwase}, \citenamefont {Nakashima}, \citenamefont {Fukahori}, \citenamefont {Okumura}, \citenamefont {Kai}, \citenamefont {Chiba}, \citenamefont {Furuta},\ and\ \citenamefont {Sihver}}]{Sato2013-ym}%
  \BibitemOpen
  \bibfield  {author} {\bibinfo {author} {\bibfnamefont {T.}~\bibnamefont {Sato}}, \bibinfo {author} {\bibfnamefont {K.}~\bibnamefont {Niita}}, \bibinfo {author} {\bibfnamefont {N.}~\bibnamefont {Matsuda}}, \bibinfo {author} {\bibfnamefont {S.}~\bibnamefont {Hashimoto}}, \bibinfo {author} {\bibfnamefont {Y.}~\bibnamefont {Iwamoto}}, \bibinfo {author} {\bibfnamefont {S.}~\bibnamefont {Noda}}, \bibinfo {author} {\bibfnamefont {T.}~\bibnamefont {Ogawa}}, \bibinfo {author} {\bibfnamefont {H.}~\bibnamefont {Iwase}}, \bibinfo {author} {\bibfnamefont {H.}~\bibnamefont {Nakashima}}, \bibinfo {author} {\bibfnamefont {T.}~\bibnamefont {Fukahori}}, \bibinfo {author} {\bibfnamefont {K.}~\bibnamefont {Okumura}}, \bibinfo {author} {\bibfnamefont {T.}~\bibnamefont {Kai}}, \bibinfo {author} {\bibfnamefont {S.}~\bibnamefont {Chiba}}, \bibinfo {author} {\bibfnamefont {T.}~\bibnamefont {Furuta}},\ and\ \bibinfo {author} {\bibfnamefont {L.}~\bibnamefont {Sihver}},\ }\href@noop {} {\bibfield  {journal} {\bibinfo  {journal} {J.
  Nucl. Sci. Technol.}\ }\textbf {\bibinfo {volume} {50}},\ \bibinfo {pages} {913} (\bibinfo {year} {2013})}\BibitemShut {NoStop}%
\bibitem [{\citenamefont {Abe}\ and\ \citenamefont {Sato}(2017)}]{Abe2017-ol}%
  \BibitemOpen
  \bibfield  {author} {\bibinfo {author} {\bibfnamefont {S.-I.}\ \bibnamefont {Abe}}\ and\ \bibinfo {author} {\bibfnamefont {T.}~\bibnamefont {Sato}},\ }\href@noop {} {\bibfield  {journal} {\bibinfo  {journal} {J. Nucl. Sci. Technol.}\ }\textbf {\bibinfo {volume} {54}},\ \bibinfo {pages} {101} (\bibinfo {year} {2017})}\BibitemShut {NoStop}%
\bibitem [{\citenamefont {Amado}(1976)}]{Amado1976-nd}%
  \BibitemOpen
  \bibfield  {author} {\bibinfo {author} {\bibfnamefont {R.~D.}\ \bibnamefont {Amado}},\ }\href@noop {} {\bibfield  {journal} {\bibinfo  {journal} {Phys. Rev. C}\ }\textbf {\bibinfo {volume} {14}},\ \bibinfo {pages} {1264} (\bibinfo {year} {1976})}\BibitemShut {NoStop}%
\bibitem [{\citenamefont {Niita}\ \emph {et~al.}(1995)\citenamefont {Niita}, \citenamefont {Chiba}, \citenamefont {Maruyama}, \citenamefont {Takada}, \citenamefont {Fukahori}, \citenamefont {Nakahara},\ and\ \citenamefont {Iwamoto}}]{Niita1995-yt}%
  \BibitemOpen
  \bibfield  {author} {\bibinfo {author} {\bibfnamefont {K.}~\bibnamefont {Niita}}, \bibinfo {author} {\bibfnamefont {S.}~\bibnamefont {Chiba}}, \bibinfo {author} {\bibfnamefont {T.}~\bibnamefont {Maruyama}}, \bibinfo {author} {\bibfnamefont {H.}~\bibnamefont {Takada}}, \bibinfo {author} {\bibfnamefont {T.}~\bibnamefont {Fukahori}}, \bibinfo {author} {\bibfnamefont {Y.}~\bibnamefont {Nakahara}},\ and\ \bibinfo {author} {\bibfnamefont {A.}~\bibnamefont {Iwamoto}},\ }\href@noop {} {\bibfield  {journal} {\bibinfo  {journal} {Phys. Rev. C}\ }\textbf {\bibinfo {volume} {52}},\ \bibinfo {pages} {2620} (\bibinfo {year} {1995})}\BibitemShut {NoStop}%
\bibitem [{\citenamefont {Ogawa}\ \emph {et~al.}(2015)\citenamefont {Ogawa}, \citenamefont {Sato}, \citenamefont {Hashimoto}, \citenamefont {Satoh}, \citenamefont {Tsuda},\ and\ \citenamefont {Niita}}]{Ogawa2015-qq}%
  \BibitemOpen
  \bibfield  {author} {\bibinfo {author} {\bibfnamefont {T.}~\bibnamefont {Ogawa}}, \bibinfo {author} {\bibfnamefont {T.}~\bibnamefont {Sato}}, \bibinfo {author} {\bibfnamefont {S.}~\bibnamefont {Hashimoto}}, \bibinfo {author} {\bibfnamefont {D.}~\bibnamefont {Satoh}}, \bibinfo {author} {\bibfnamefont {S.}~\bibnamefont {Tsuda}},\ and\ \bibinfo {author} {\bibfnamefont {K.}~\bibnamefont {Niita}},\ }\href@noop {} {\bibfield  {journal} {\bibinfo  {journal} {Phys. Rev. C}\ }\textbf {\bibinfo {volume} {92}},\ \bibinfo {pages} {024614} (\bibinfo {year} {2015})}\BibitemShut {NoStop}%
\bibitem [{\citenamefont {Furihata}(2000)}]{Furihata2000-da}%
  \BibitemOpen
  \bibfield  {author} {\bibinfo {author} {\bibfnamefont {S.}~\bibnamefont {Furihata}},\ }\href@noop {} {\bibfield  {journal} {\bibinfo  {journal} {Nucl. Instrum. Methods Phys. Res. B}\ }\textbf {\bibinfo {volume} {171}},\ \bibinfo {pages} {251} (\bibinfo {year} {2000})}\BibitemShut {NoStop}%
\bibitem [{\citenamefont {Watanabe}\ and\ \citenamefont {Kadrev}(2007)}]{Watanabe2007-cq}%
  \BibitemOpen
  \bibfield  {author} {\bibinfo {author} {\bibfnamefont {Y.}~\bibnamefont {Watanabe}}\ and\ \bibinfo {author} {\bibfnamefont {D.~N.}\ \bibnamefont {Kadrev}},\ }in\ \href@noop {} {\emph {\bibinfo {booktitle} {ND2007}}}\ (\bibinfo  {publisher} {EDP Sciences},\ \bibinfo {address} {Les Ulis, France},\ \bibinfo {year} {2007})\ pp.\ \bibinfo {pages} {1121--1124}\BibitemShut {NoStop}%
\bibitem [{\citenamefont {Angeli}\ and\ \citenamefont {Marinova}(2013)}]{Angeli2013-to}%
  \BibitemOpen
  \bibfield  {author} {\bibinfo {author} {\bibfnamefont {I.}~\bibnamefont {Angeli}}\ and\ \bibinfo {author} {\bibfnamefont {K.~P.}\ \bibnamefont {Marinova}},\ }\href@noop {} {\bibfield  {journal} {\bibinfo  {journal} {At. Data Nucl. Data Tables}\ }\textbf {\bibinfo {volume} {99}},\ \bibinfo {pages} {69} (\bibinfo {year} {2013})}\BibitemShut {NoStop}%
\bibitem [{\citenamefont {Saito}(2022)}]{Saito2022-vc}%
  \BibitemOpen
  \bibfield  {author} {\bibinfo {author} {\bibfnamefont {T.~Y.}\ \bibnamefont {Saito}},\ }\emph {\bibinfo {title} {Study of muonic {X}-ray spectroscopy and nuclear muon capture reaction}},\ \href@noop {} {Ph.D. thesis},\ \bibinfo  {school} {Univ. Tokyo} (\bibinfo {year} {2022})\BibitemShut {NoStop}%
\bibitem [{\citenamefont {Da~Providência}(1965)}]{Da_Providencia1965-cb}%
  \BibitemOpen
  \bibfield  {author} {\bibinfo {author} {\bibfnamefont {J.}~\bibnamefont {Da~Providência}},\ }\href {https://doi.org/10.1016/0029-5582(65)90937-5} {\bibfield  {journal} {\bibinfo  {journal} {Phys. Rev. C Nucl. Phys.}\ }\textbf {\bibinfo {volume} {61}},\ \bibinfo {pages} {87} (\bibinfo {year} {1965})}\BibitemShut {NoStop}%
\bibitem [{\citenamefont {Minato}(2016)}]{Minato2016-qm}%
  \BibitemOpen
  \bibfield  {author} {\bibinfo {author} {\bibfnamefont {F.}~\bibnamefont {Minato}},\ }\href {https://doi.org/10.1103/PhysRevC.93.044319} {\bibfield  {journal} {\bibinfo  {journal} {Phys. Rev. C Nucl. Phys.}\ }\textbf {\bibinfo {volume} {93}},\ \bibinfo {pages} {044319} (\bibinfo {year} {2016})}\BibitemShut {NoStop}%
\bibitem [{\citenamefont {Reinhard}\ \emph {et~al.}(1999)\citenamefont {Reinhard}, \citenamefont {Dean}, \citenamefont {Nazarewicz}, \citenamefont {Dobaczewski}, \citenamefont {Maruhn},\ and\ \citenamefont {Strayer}}]{sko2}%
  \BibitemOpen
  \bibfield  {author} {\bibinfo {author} {\bibfnamefont {P.-G.}\ \bibnamefont {Reinhard}}, \bibinfo {author} {\bibfnamefont {D.~J.}\ \bibnamefont {Dean}}, \bibinfo {author} {\bibfnamefont {W.}~\bibnamefont {Nazarewicz}}, \bibinfo {author} {\bibfnamefont {J.}~\bibnamefont {Dobaczewski}}, \bibinfo {author} {\bibfnamefont {J.~A.}\ \bibnamefont {Maruhn}},\ and\ \bibinfo {author} {\bibfnamefont {M.~R.}\ \bibnamefont {Strayer}},\ }\href {https://doi.org/10.1103/physrevc.60.014316} {\bibfield  {journal} {\bibinfo  {journal} {Phys. Rev. C Nucl. Phys.}\ }\textbf {\bibinfo {volume} {60}},\ \bibinfo {pages} {014316} (\bibinfo {year} {1999})}\BibitemShut {NoStop}%
\bibitem [{\citenamefont {Van~Giai}\ and\ \citenamefont {Sagawa}(1981)}]{sg2}%
  \BibitemOpen
  \bibfield  {author} {\bibinfo {author} {\bibfnamefont {N.}~\bibnamefont {Van~Giai}}\ and\ \bibinfo {author} {\bibfnamefont {H.}~\bibnamefont {Sagawa}},\ }\href {https://doi.org/10.1016/0370-2693(81)90646-8} {\bibfield  {journal} {\bibinfo  {journal} {Phys. Lett. B}\ }\textbf {\bibinfo {volume} {106}},\ \bibinfo {pages} {379} (\bibinfo {year} {1981})}\BibitemShut {NoStop}%
\bibitem [{\citenamefont {Vautherin}\ and\ \citenamefont {Brink}(1972)}]{Vautherin1972Phys.Rev.C5_626}%
  \BibitemOpen
  \bibfield  {author} {\bibinfo {author} {\bibfnamefont {D.}~\bibnamefont {Vautherin}}\ and\ \bibinfo {author} {\bibfnamefont {D.~M.}\ \bibnamefont {Brink}},\ }\href {https://doi.org/10.1103/PhysRevC.5.626} {\bibfield  {journal} {\bibinfo  {journal} {Phys. Rev. C}\ }\textbf {\bibinfo {volume} {5}},\ \bibinfo {pages} {626} (\bibinfo {year} {1972})}\BibitemShut {NoStop}%
\bibitem [{\citenamefont {Koning}\ and\ \citenamefont {Duijvestijn}(2004)}]{Koning2004-bl}%
  \BibitemOpen
  \bibfield  {author} {\bibinfo {author} {\bibfnamefont {A.~J.}\ \bibnamefont {Koning}}\ and\ \bibinfo {author} {\bibfnamefont {M.~C.}\ \bibnamefont {Duijvestijn}},\ }\href@noop {} {\bibfield  {journal} {\bibinfo  {journal} {Nucl. Phys. A}\ }\textbf {\bibinfo {volume} {744}},\ \bibinfo {pages} {15} (\bibinfo {year} {2004})}\BibitemShut {NoStop}%
\bibitem [{\citenamefont {Iwamoto}\ \emph {et~al.}(2016)\citenamefont {Iwamoto}, \citenamefont {Iwamoto}, \citenamefont {Kunieda}, \citenamefont {Minato},\ and\ \citenamefont {Shibata}}]{Iwamoto2016-pl}%
  \BibitemOpen
  \bibfield  {author} {\bibinfo {author} {\bibfnamefont {O.}~\bibnamefont {Iwamoto}}, \bibinfo {author} {\bibfnamefont {N.}~\bibnamefont {Iwamoto}}, \bibinfo {author} {\bibfnamefont {S.}~\bibnamefont {Kunieda}}, \bibinfo {author} {\bibfnamefont {F.}~\bibnamefont {Minato}},\ and\ \bibinfo {author} {\bibfnamefont {K.}~\bibnamefont {Shibata}},\ }\href@noop {} {\bibfield  {journal} {\bibinfo  {journal} {Nucl. Data Sheets}\ }\textbf {\bibinfo {volume} {131}},\ \bibinfo {pages} {259} (\bibinfo {year} {2016})}\BibitemShut {NoStop}%
\bibitem [{\citenamefont {Hauser}\ and\ \citenamefont {Feshbach}(1952)}]{Hauser1952-cs}%
  \BibitemOpen
  \bibfield  {author} {\bibinfo {author} {\bibfnamefont {W.}~\bibnamefont {Hauser}}\ and\ \bibinfo {author} {\bibfnamefont {H.}~\bibnamefont {Feshbach}},\ }\href {https://doi.org/10.1103/PhysRev.87.366} {\bibfield  {journal} {\bibinfo  {journal} {Phys. Rev.}\ }\textbf {\bibinfo {volume} {87}},\ \bibinfo {pages} {366} (\bibinfo {year} {1952})}\BibitemShut {NoStop}%
\bibitem [{\citenamefont {Iwamoto}(2007)}]{Iwamoto2007-xf}%
  \BibitemOpen
  \bibfield  {author} {\bibinfo {author} {\bibfnamefont {O.}~\bibnamefont {Iwamoto}},\ }\href@noop {} {\bibfield  {journal} {\bibinfo  {journal} {J. Nucl. Sci. Technol.}\ }\textbf {\bibinfo {volume} {44}},\ \bibinfo {pages} {687} (\bibinfo {year} {2007})}\BibitemShut {NoStop}%
\end{thebibliography}%

\end{document}